\documentclass{optica-article}
\journal{opticajournal} 
\articletype{Roadmap}

\usepackage{lineno}
\usepackage{capt-of}
\usepackage{tabularx}
\usepackage{makecell}
\usepackage{gensymb}
\usepackage{enumitem}
\usepackage{wrapfig}
\usepackage{braket}
\newcommand{\qo}[1]{``#1''}
\usepackage[many]{tcolorbox}
\usepackage[export]{adjustbox}[2011/08/13]
\usepackage{titlesec}
\usepackage{pgfplots}
\usepackage{amsmath}
\usepackage{dsfont}
\usepackage[table]{xcolor}
\usepackage{enumitem}
\usepackage{soul}
\usepackage{graphicx}
\usepackage{verbatim, float}
\usepackage{psfrag}
\usepackage[normalem]{ulem}
\usepackage{physics}		

\renewcommand{\t}[1]{\mathrm{#1}}
\newcommand{\be}{\begin{equation}}
\newcommand{\ee}{\end{equation}}

\def\<{\langle} 
\def\>{\rangle}

\newcommand{\subtiny}[3]{\ensuremath{_{\hspace{#1 pt}\protect\raisebox{#2 pt}{\tiny{$ #3$}}}}}
\newcommand{\suptiny}[3]{\ensuremath{^{\hspace{#1 pt}\protect\raisebox{#2 pt}{\tiny{$ #3$}}}}}

\newcommand{\nr}{\ensuremath{\hspace*{0.5pt}}}

\pgfplotsset{compat=1.18}
\usetikzlibrary{calc}
\pgfplotsset{
  FreeSpace/.style={only marks, mark=*, mark size=2.6pt},
  MultiMode/.style={only marks, mark=triangle*, mark size=4.0pt},
  MultiCore/.style={only marks, mark=square*, mark size=2.8pt},
}
\NewTColorBox{NewBox}{ s O{!htbp} }{
  IfBooleanTF={#1}{float*,width=\textwidth}{float},
  colframe=teal!20, colback=teal!5, coltitle=black,
  title= {#2}}

\titleformat*{\subsection}{\bfseries\large}
\titleformat*{\subsubsection}{}

\begin{document}
\title{High-Dimensional Quantum Photonics: Roadmap}

\author{
Mehul Malik\authormark{1},
Micheal Kues\authormark{2},
Takuya Ikuta\authormark{3}, 
Hiroki Takesue\authormark{3},
Daniele Bajoni\authormark{4},
David J. Moss\authormark{5},
Roberto Morandotti\authormark{6},
Andrew Forbes\authormark{7}, 
Stephen Walborn\authormark{8,9},
Ebrahim Karimi\authormark{10,11},
Yunhong Ding\authormark{12},
Stefano Paesani\authormark{13},
Caterina Vigliar\authormark{12},
Benjamin Brecht\authormark{14},
Christine Silberhorn\authormark{14},
Fr\'ed\'eric Bouchard\authormark{15},
Micha\l{} Karpi\'nski\authormark{16}, 
Benjamin Sussman\authormark{15}, 
Joseph M. Lukens\authormark{17,18},
Yaron Bromberg\authormark{19}, 
Robert Fickler\authormark{20}, 
Taira Giordani\authormark{21}, 
Fabio Sciarrino\authormark{21}, 
Yun Zheng\authormark{22}, 
Jianwei Wang\authormark{22},
Marcus Huber\authormark{23,24},
Armin Tavakoli\authormark{25},
Roope Uola\authormark{26,27},
Nicolas Brunner\authormark{28},
Nicolai Friis\authormark{23},
Natalia Herrera Valencia\authormark{1},
Jacquiline Romero\authormark{29,30},
Will McCutcheon\authormark{1}
}

\address{\authormark{1}Institute of Photonics and Quantum Sciences (IPAQS), Heriot-Watt University, Edinburgh, EH14 4AS, United Kingdom}
\address{\authormark{2}Institute of Photonics and PhoenixD Cluster of Excellence, Leibniz University Hannover, Hannover, Germany}
\address{\authormark{3}Basic Research Laboratories, NTT, Inc., 3-1 Morinosato Wakamiya, Atsugi, Kanagawa 243-0198, Japan}
\address{\authormark{4}Dipartimento di Ingegneria Industriale e dell'Informazione, Università di Pavia, 27100 Pavia, Italy}
\address{\authormark{5}Optical Sciences Centre, Swinburne University of Technology, Hawthorn, Victoria, Australia 3122}
\address{\authormark{6}Institut national de la recherche scientifique, INRS-EMT, Varennes (QC), Canada}
\address{\authormark{7}School of Physics, University of the Witwatersrand, Johannesburg, South Africa}
\address{\authormark{8}Departamento de Física, Universidad de Concepción, 160-C Concepción, Chile}
\address{\authormark{9}Millennium Institute for Research in Optics, Universidad de Concepción, 160-C Concepción, Chile}
\address{\authormark{10}Institute for Quantum Studies, Chapman University, Orange, California 92866, USA}
\address{\authormark{11}Nexus for Quantum Technologies, University of Ottawa, K1N 5N6 Ottawa, Ontario, Canada}
\address{\authormark{12}Department of Electrical and Photonics Engineering, Denmark
Technical University, Ørsteds Plads, Lyngby, 2800, Hovedstaden, Denmark}
\address{\authormark{13}NNF Quantum Computing Programme, Niels Bohr Institute, University of Copenhagen, Blegdamsvej 17, 2100 Copenhagen, Denmark.}
\address{\authormark{14}Paderborn University, Integrated Quantum Optics, Institute for Photonic Quantum Systems (PhoQS), 33095 Paderborn, Germany}
\address{\authormark{15}National Research Council of Canada, 100 Sussex Drive, Ottawa, Ontario K1A 0R6, Canada.}
\address{\authormark{16}Faculty of Physics, University of Warsaw, Pasteura 5, 02-093 Warszawa, Poland}
\address{\authormark{17}Elmore Family School of Electrical and Computer Engineering and Purdue Quantum Science and Engineering Institute, Purdue University, West Lafayette, Indiana 47907, USA}
\address{\authormark{18}Quantum Information Science Section, Oak Ridge National Laboratory, Oak Ridge, Tennessee 37831, USA}
\address{\authormark{19}Racah Institute of Physics, The Hebrew University of Jerusalem, Jerusalem, 91904, Israel}
\address{\authormark{20}Photonics Laboratory, Physics Unit, Tampere University, Tampere, FI-33720, Finland}
\address{\authormark{21}Dipartimento di Fisica, Sapienza Universit\`{a} di Roma,
Piazzale Aldo Moro 5, I-00185 Roma, Italy}
\address{\authormark{22} State Key Laboratory for Mesoscopic Physics, School of Physics, Peking University, Beijing, 100871, China}
\address{\authormark{23} Atominstitut, Technische Universit\"at Wien, Stadionallee 2, 1020 Vienna, Austria}
\address{\authormark{24} Institute for Quantum Optics and Quantum Information (IQOQI), Austrian Academy of Sciences, Boltzmanngasse 3, 1090 Vienna, Austria}
\address{\authormark{25}Physics Department and NanoLund, Lund University, Box 118, 22100 Lund, Sweden.}
\address{\authormark{26}Department of Physics and Astronomy, Uppsala University, Box 516, 751 20 Uppsala, Sweden}
\address{\authormark{27}Nordita, KTH Royal Institute of Technology and Stockholm University, 10691 Stockholm, Sweden}
\address{\authormark{28}Department of Applied Physics University of Geneva, 1211 Geneva, Switzerland}
\address{\authormark{29}School of Mathematics and Physics, University of Queensland, Brisbane, 4072, Australia}
\address{\authormark{30}Australian Research Council Training Centre for Current and Emergent Quantum Technologies (QuTech), Brisbane, 4072, Australia}

\begin{abstract*} 
The field of high-dimensional quantum photonics involves the use of multimode photonic degrees-of-freedom such as the spatial, temporal, or spectral structure of light to encode multi-level quantum states. Recent years have seen rapid progress in the development of methods to generate, manipulate, and distribute such quantum states of light and their use in a range of quantum technology applications that offer practical advantages over conventional qubit-based approaches. High-dimensional quantum states of light encoded in photonic time-bins, frequency-bins, transverse-spatial modes, waveguide paths, and temporal modes have enabled noise-robust fundamental tests of quantum mechanics, error-resilient and high-capacity quantum communication protocols, as well as efficient approaches for quantum information processing, to name just a few examples. However, research in this field has progressed fairly independently, with little exchange across different photonic degrees-of-freedom or between experiment and theory and no comprehensive comparison between degrees-of-freedom. This roadmap aims to bridge this gap by surveying progress in each area and identifying shared challenges and opportunities that cut across two or more photonic degrees-of-freedoms. We review early work and state-of-the-art experimental techniques under development for high-dimensional quantum states encoded in single and entangled photons, as well as theoretical tools for their measurement and certification. We outline the main outstanding challenges for theory and each experimental degree-of-freedom, identifying promising future directions of research that may enable these to be overcome. We end by discussing interconnections between degrees-of-freedom and shared challenges centered around their distribution, measurement, and manipulation, with a view towards their integration into next-generation quantum technology platforms for applications in communications, sensing and computing. 
\end{abstract*}

\setcounter{tocdepth}{2}
\tableofcontents

\section{Introduction}

From vibrant rainbows in the sky to images on our screens, the multi-modal nature of light makes itself evident in our lives every day. Light can carry a vast amount of information on its many different properties—ranging from its structure in time, frequency, position, momentum and polarization. Classical communication technologies have long harnessed these properties of light in the form of wavelength and space-division-multiplexing to push their information capacity limits. In contrast, commercial quantum technologies have primarily been limited to the use of two-level systems or qubits. However, the last two decades have seen a surge of research on the multi-modal properties of quantum light, giving birth to the field of high-dimensional (HD) quantum photonics.

HD quantum photonics involves the study of multimode-capable photonic degrees-of-freedom such as frequency, time, transverse-position, and integrated waveguide path to encode quantum states of light carrying more than two levels of information, also known as qudits. In recent years, photonic qudits have enabled fundamental tests of key quantum mechanical concepts such as contextuality and entanglement. In parallel, qudits provide the advantage of carrying significantly more information per particle than qubits, enabling high-capacity quantum communication and computing systems. Qudits additionally offer robustness to noise and loss, which is important for quantum protocol operation in realistic environments. Finally, through control over multiple properties of quantum light, qudits open a pathway towards complete control of matter-based quantum systems.

In their simplest application, qudits can be used to combine, or multiplex, multiple channels of quantum information, significantly increasing system information capacity. For example, three independent polarization qubits ($d=2$) can be encoded on three different spatial modes ($d=3$), resulting in a composite dimension of $d=2^3=8$. However, this does not require the three spatial modes to be coherent with respect to one another, as the multiplexed channels carry independent streams of information. On the other hand, encoding an arbitrary eight-dimensional quantum state on a qudit ($d=8$) does require the presence of quantum coherence between its eight levels. General transformations of a qudit also need to preserve quantum coherence, making them particularly challenging to implement. Similarly, the entanglement of two or more qudits poses many interesting problems, from both a theoretical and experimental perspective.

Recent developments in experimental capabilities for controlling the HD quantum properties of light have led to rapid advances in their use for quantum technologies, ranging from entanglement-based quantum communication, cluster-state quantum computing, to quantum-enhanced sensing. However, each photonic degree-of-freedom comes with its own benefits and challenges, making it suitable for certain quantum technology applications while hindering others. In addition, research in each of these areas has been fairly siloed, with little exchange across different HD quantum properties. In this roadmap, we aim to initiate a dialogue across different photonic degrees-of-freedom as well as their application areas, allowing the cross-fertilisation of ideas and techniques from one area to another. We survey state-of-the-art techniques for generating, manipulating, and measuring quantum light (single and entangled photons) in every multi-mode degree-of-freedom and discuss their applications in next-generation quantum technologies for communication, computing, and sensing.

High-dimensional photonic quantum states are driving several applications in quantum technologies. In quantum communication and networking, high-dimensional encoding enables increased information capacity per photon, improved tolerance to noise and loss, and enhanced resilience against certain classes of eavesdropping attacks. These features are particularly relevant for advancing long-distance fiber and free-space links, satellite-based quantum communication, and scalable multi-user quantum networks. In quantum information processing, high-dimensional systems provide access to richer Hilbert spaces that can reduce circuit depth, simplify certain algorithms, and enable more compact implementations of quantum protocols. Beyond information processing, high-dimensional photonic states offer new opportunities in quantum sensing and metrology, where increased dimensionality can improve sensitivity, enable multiplexed measurements, and enhance robustness against technical imperfections. Collectively, these application drivers motivate sustained efforts to develop scalable, well-controlled, and interconnected high-dimensional photonic platforms and out-of-lab systems.

Despite rapid experimental and theoretical progress, several cross-cutting challenges remain before high-dimensional photonic systems can be deployed in practical quantum technologies. Key challenges include scalable state generation with high purity and brightness, low-loss and reconfigurable high-dimensional unitary transformations, and efficient, high-fidelity projective measurements across large modal spaces. From a theoretical perspective, the certification, characterization, and benchmarking of high-dimensional quantum states and processes present nontrivial challenges, as resource requirements often scale unfavorably with dimension. Future directions therefore include the development of dimension-independent or dimension-efficient certification tools, integrated and hybrid photonic platforms, and standardized interfaces between different encoding schemes. Addressing these issues will be essential for translating high-dimensional concepts from laboratory demonstrations to deployable systems.

Alongside light, recent years have also seen a surge in experimental research on matter-based qudits, exploring platforms such as ions and superconducting circuits with an emphasis on quantum computing applications. A roadmap on high-dimensional quantum photonics will be remiss without pointing out the subtle difference between 
photonic and matter-based qubits and qudits.  Photonic qubits are not stationary, while matter-based qubits are. A photon carries within it all its qubits and qudits encoded on its various internal degrees of freedom. This is not the case for matter-based qudits, wherein adding another level often requires adding another transition, and thus more complexity (e.g., differing transition times). In contrast, photonic qudits can be realized by just increasing the number of modes or combining different photonic internal degrees-of-freedom, which entails different experimental challenges from matter-based qudits. 

This roadmap is organized to reflect the natural evolution of HD photonic quantum technologies, progressing from foundational concepts to emerging applications and open challenges. We first focus on the generation and measurement of HD single photons and entangled states across different degrees-of-freedom (time-bins, frequency-bins, transverse-spatial modes, path encoding, and temporal modes). We then address the distribution, manipulation, and applications of photonic qudits in each degree-of-freedom. Subsequently, we discuss theoretical tools for HD systems and protocols. Each thematic section follows a common narrative, beginning with early proof-of-principle demonstrations, followed by recent experimental and theoretical advances, and concluding with current limitations and future outlooks. The roadmap concludes by highlighting interconnections between degrees-of-freedom and outlining promising directions for future research and necessary steps---with the ultimate aim of providing researchers, both new to the field and experts, with a coherent overview of the field’s current status and future trajectory.

\section{Generation and Measurement}
\subsection{Time-bin} \label{sec:GenAndMeas:Time_Bin}
\author{Takuya Ikuta\authormark{3}, Hiroki Takesue\authormark{3}}

\address{\authormark{3}Basic Research Laboratories, NTT, Inc., 3-1 Morinosato Wakamiya, Atsugi, Kanagawa 243-0198, Japan}

\subsubsection{Early work} \label{sec:GenAndMeas_TimeBin_EarlyWork}

\begin{figure}[ht!]
    \centering
    \includegraphics[width=\linewidth]{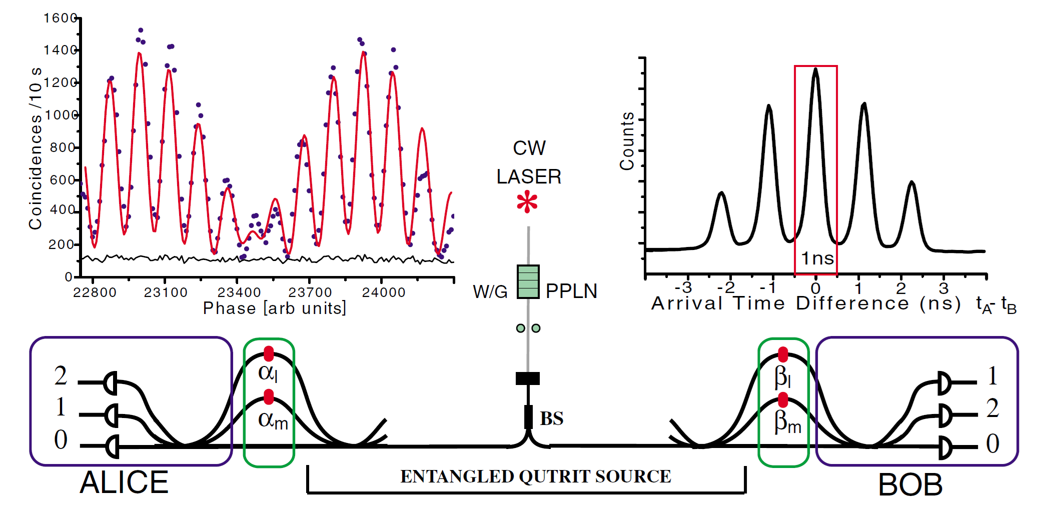}
    \caption{Generation and measurement of time-energy entangled qutrits using three-arm interferometers \cite{Thew2004PRL_Qutrits}. Reprinted figure with permission from \href{https://dx.doi.org/10.1103/PhysRevLett.93.010503}{[R. T. Thew, A. Ac\'{i}n, H. Zbinden, and N. Gisin, Phys. Rev. Lett. \textbf{93}, 010503 (2004).]} Copyright (2004) by the American Physical Society.}
    \label{fig:Thew2004}
\end{figure}

The temporal degree-of-freedom or the arrival time of a photon is an important physical property of light,
having paved the way for stable and robust quantum communication over optical fibers.
Time-energy entanglement \cite{fransonBellInequalityPosition1989} and time-bin entanglement \cite{brendelPulsedEnergyTimeEntangled1999} have been used in many fundamental research areas and applications since their proposal.
Even at a stage when a qubit was the main research target,
it was already recognized that the temporal degree of freedom can be used to generate a high-dimensional quantum state \cite{brendelPulsedEnergyTimeEntangled1999}.

In early work on high-dimensional quantum states in the time domain, time–energy uncertainty was mainly employed instead of discrete time bins to demonstrate high-dimensional entanglement. A central measurement tool in such experiments is the Franson interferometer \cite{fransonBellInequalityPosition1989}, which consists of two unbalanced Mach–Zehnder interferometers with identical path-length differences placed at spatially separated measurement stations. When the path-length imbalance exceeds the coherence time of the individual photons but remains shorter than that of the pump, two-photon interference arises from the indistinguishability between the short–short and long–long propagation amplitudes. By controlling the relative phases of the interferometers, projections onto superposed states of different emission times can be realized, enabling phase-sensitive measurements of time-energy entanglement. While this configuration relies on post-selection, it provides the fundamental mechanism for analyzing superpositions in the time domain and forms the conceptual basis for later time-bin measurement schemes.

High-dimensional time-energy entanglement was first realized by a group at the University of Geneva using three-arm interferometers \cite{Thew2004PRL_Qutrits}. In this experiment, they controlled interference between three time bins using the interferometers (Fig.~\ref{fig:Thew2004}) and observed a violation of Bell’s inequality for high-dimensional entanglement, which is known as the Collins--Gisin--Linden--Massar--Popescu (CGLMP) inequality \cite{Collins2002}.

Richart et al. found that an exact mutually unbiased bases (MUB) measurement for time-bins can be implemented by cascading delay interferometers with delay times of $\Delta T$, $2\Delta T$, $4\Delta T$, $\cdots$, where $\Delta T$ is the separation between time-bins \cite{LampertRichart2014PhD},
and experimentally confirmed the generation of four-dimensional time-energy entanglement \cite{Richart2012ApplPhysB}. Brougham and Barnett conceived of another implementation of MUB measurements for time-bins that uses a delay-loop configuration \cite{Brougham2013_MUB_TimeBinQudits}. However, due to the intrinsic attenuation of the optical field in the delay-loop, the amplitudes of successive time bins are not equal, preventing the realization of an exact MUB. As a result, this configuration can realize only approximate MUB measurements.

Alongside the developments described above, de Reidmatten et al. demonstrated pulsed time-energy entanglement (or sequential time-bin entanglement) using coherent pump pulses \cite{deriedmattenTailoringPhotonicEntanglement2004}. Takesue and Inoue proposed that time-bin entanglement can be modulated by intensity and phase modulation of the pump light \cite{takesueQuantumSecretSharing2006}. An alternative approach for certifying time-energy entanglement based on geometric phase was proposed and demonstrated by Jha et al \cite{Jha2008}. These advances in concepts and technologies paved the way for various theoretical and experimental works that we discuss in Subsection~\ref{timebin-recent}. 

\subsubsection{Recent developments}
 \label{timebin-recent}

\begin{figure}[htb]
    \centering
    \includegraphics[width=.9 \linewidth]{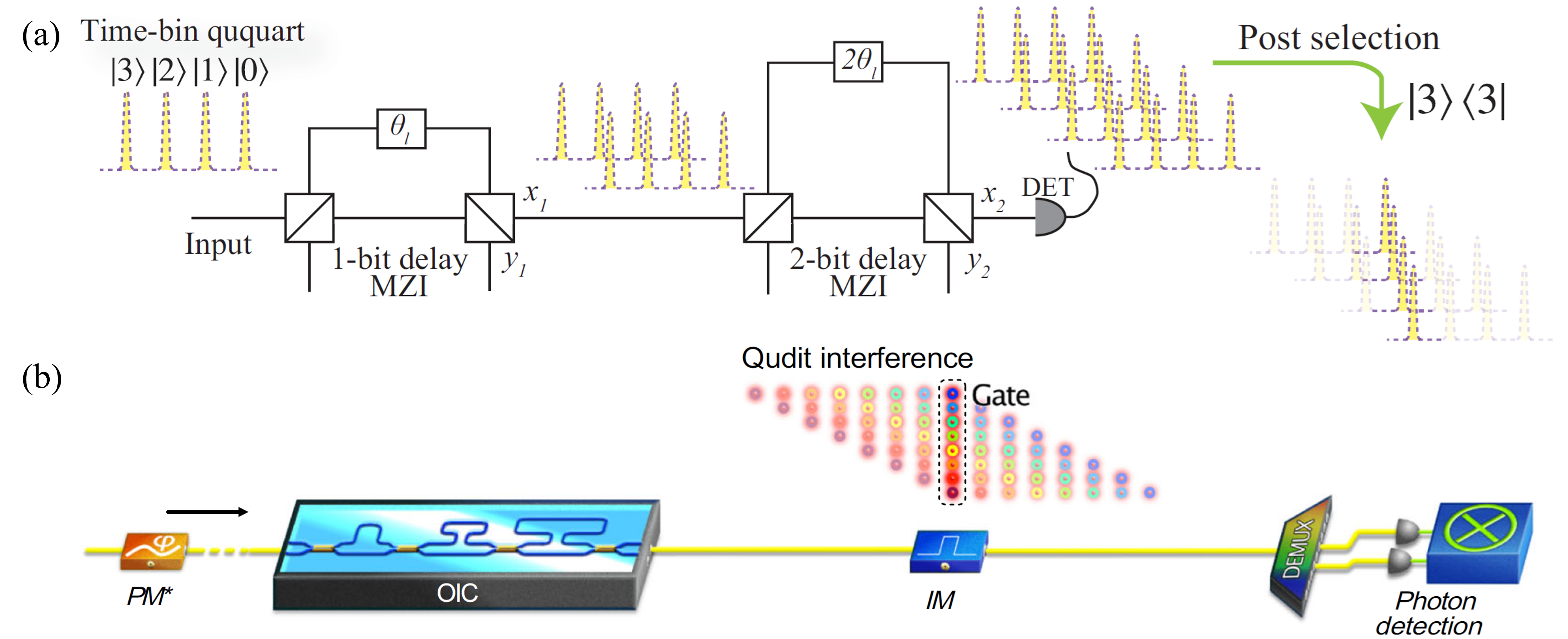}
    \caption{Measurement of a high-dimensional time-bin state using (a) cascaded delay interferomters (Reprinted figure with permission from \href{https://dx.doi.org/10.1103/PhysRevA.93.022307}{[T. Ikuta and H. Takesue, Phys. Rev. A, \textbf{93}, 022307 (2016).]} Copyright (2016) by the American Physical Society.) \cite{Ikuta2016PRA_CGLMP_ququart} and (b) additional intensity modulator for gate operation (reproduced from Ref. \cite{yuQuantumKeyDistribution2025}, licensed under CC BY 4.0).}
    \label{fig:Ikuta2016Yu2025}
\end{figure}

A high-dimensional time-bin state is conventionally generated as sequential pulses using an intensity modulator and a phase modulator.
High-dimensional time-bin entanglement can also be generated via spontaneous parametric downconversion (SPDC) by launching these modulated pulses into a nonlinear medium.
In many cases, more elaborate techniques correspond to the inverse of measurement operations.
For example, a delay interferometer can be used for not only measurements but also preparations of a superposed state by launching a single pulse into it.
Therefore,
how to measure the high-dimensional time-bin state is the most fundamental part.
With this in mind,
we focus on the recent developments of measurements for a high-dimensional time-bin state.

Similar to their application for certifying time-energy entanglement \cite{LampertRichart2014PhD, Richart2012ApplPhysB},
cascaded delay interferometers can be used to implement Fourier-basis measurements for a high-dimensional time-bin state \cite{Ikuta2016PRA_CGLMP_ququart}.
The Fourier basis is an important measurement basis that appears in many quantum applications such as the CGLMP inequality test \cite{Collins2002} and high-dimensional quantum key distribution \cite{Islam2017_a}.
Let us assume the case of dimension $d=4$, where four optical pulses are equally separated by an interval of $\Delta T$.
If we launch the four-dimensional state into two cascaded delay interferometers having delay times of $\Delta T$ and $2\Delta T$,
we can observe the interference of all input pulses in the central peak at the output (see Fig.~\ref{fig:Ikuta2016Yu2025} (a)). 
By setting the relative phases between the short and long arms to be $\theta$ and $2\theta$,
this interference corresponds to the projective measurement onto the Fourier basis state $\ket{f_\theta}$, which is given by,
\begin{equation}
  \ket{f_\theta} = \frac{1}{\sqrt{d}} \sum_{k=0}^{d-1} e^{ik\theta} \ket{k} .
\end{equation}
Therefore, we can project the high-dimensional time-bin state onto the Fourier-basis state having an arbitrary phase parameter $\theta$.
This method can be easily extended to an arbitrary dimension $d$ by cascading $\log d$ delay interferometers.
While it looks simple, it has significant potential.
High-dimensional time-bin entanglement with $\Delta T = 32$ ps was measured using three cascaded delay interferometers recently \cite{yuQuantumKeyDistribution2025},
where an additional phase modulator set the relative phases between time-bins and an intensity modulator removed unnecessary interference (Fig.~\ref{fig:Ikuta2016Yu2025} (b)).
However, methods based on cascaded interferometers implement only one single-outcome projective measurement for each measurement setting. Therefore, to realize a complete multi-outcome measurement, we need to change the phase parameter $\theta$ many times, which can cause unnecessary fluctuations in the measurement.

\begin{figure}[htb]
    \centering
    \includegraphics[width=.7 \linewidth]{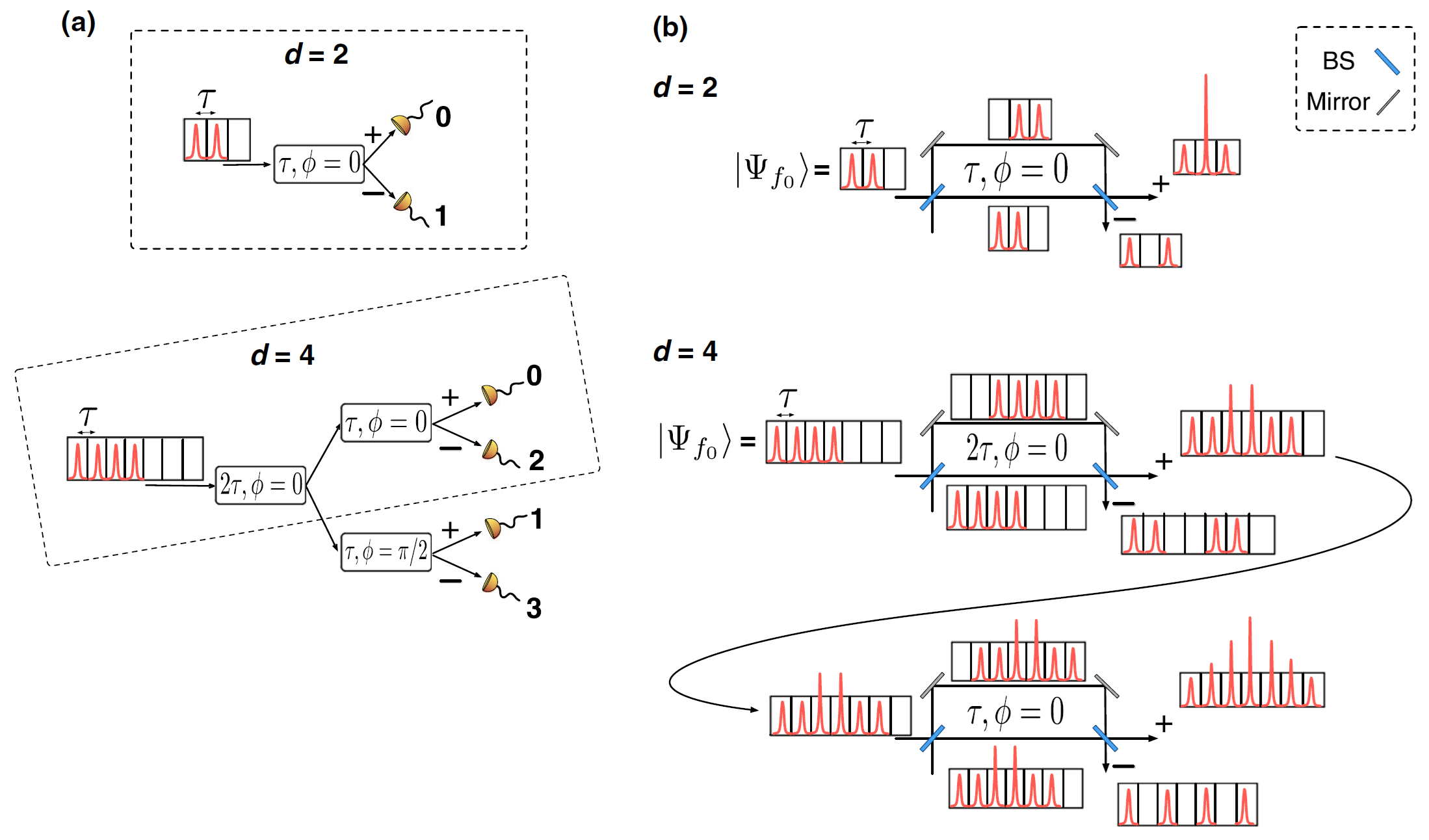}
    \caption{All projective measurements in the Fourier basis. (a) Diagram of the tree configuration of interferometers. (b) Details of the cascaded interferometers. \cite{islam2017robust}. Reprinted figure with permission from \href{https://dx.doi.org/10.1103/PhysRevApplied.7.044010}{[N. T. Islam et al., Phys. Rev. Applied \textbf{7}, 044010 (2017).]} Copyright (2017) by the American Physical Society. }
    \label{fig:Islam2017}
\end{figure}

Although changing the phase is necessary for some applications such as the CGLMP inequality test \cite{Collins2002},
a high-dimensional QKD protocol using two MUBs requires only a limited set of states \cite{Cerf2002, Sheridan2010}.
For example,
$\left\{\ket{f_\theta} | \theta \in \{0, \pi/2, \pi, 3\pi/2\} \right\}$ forms a basis that is mutually unbiased to
the time-bin basis $\left\{ \ket{k} | k \in \{0, 1, 2, 3\} \right\}$.
Projective measurements onto the four states in this basis can be implemented simultaneously \cite{broughamSecurityHighdimensionalQuantum2013}.
For example, the simultaneous implementation of four projective measurements in the Fourier basis using cascaded delay interferometers in a tree configuration (Fig.~\ref{fig:Islam2017}) was demonstrated by Islam et al \cite{islam2017robust, Islam2017_a}.
In this configuration,
two interferometers with a delay time $\Delta T$ are connected to the output ports of an interferometer with a delay time $2\Delta T$.
Although the two interferometers have the same delay times,
the relative phases between their short and long arms are different---one has a phase of 0, and the other has a phase of $\pi/2$.
Similar to the cascaded interferometers in \cite{Ikuta2016PRA_CGLMP_ququart},
we can observe the interference of all input pulses in the central peak of the output.
The four possible output ports correspond to the four projective measurements in the Fourier-basis.
These works demonstrated the case of $d=4$. In general,
this method can be extended to $d=2^n$ by increasing the tree depth \cite{broughamSecurityHighdimensionalQuantum2013}.
The advantage of the tree configuration is that it enables us to implement all projective measurements simultaneously.
It is not only efficient, but also more stable as it does not require the relative phase to be changed once set.
On the other hand,
$(d-1)$ interferometers are required for measuring a $d$-dimensional state.
Thus, the setup can become large quite quickly, dramatically increasing the complexity of implementation.
In addition, the smallest relative phase among the interferometers is $2\pi/d$.
Therefore, the initial calibration becomes difficult when the number of dimensions increases.
Unfortunately,
the difficulty of the small phase stems from the characteristic of the Fourier basis itself.
Therefore,
a different basis could be used to circumvent these problems.

Alternatively, propagation in a medium with group velocity dispersion can be used to approximately implement the Fourier basis measurement. Here, for sufficiently short time bins chirping of the pulses in the dispersive medium for high enough dispersion and low time-bin separation leads to interference between the time bins. In particular, Widomski et al. \cite{widomski2024efficient} used the temporal Talbot effect to approximately detect 4-dimensional time-bin superpositions from the Fourier basis. 

\begin{figure}[htb]
    \centering
    \includegraphics[width= \linewidth]{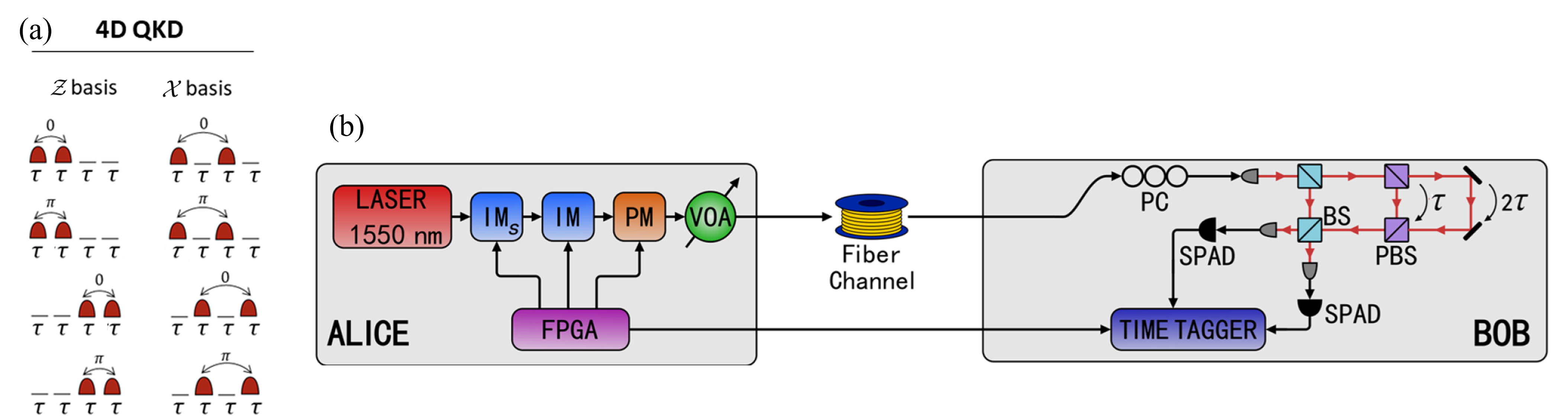}
    \caption{(a) Practical phase encoding in the subspace of a four-dimensional time-bin state. (b) Measurement setup using a delay interferometer with a variable delay time (reproduced from Ref. \cite{Vagniluca2020PRAppl_d4QKD_2det}, licensed under CC BY 4.0). }
    \label{fig:Vagniluca2020}
\end{figure}

If the number of dimensions is fixed,
one may find a practically useful implementation of two time-bin MUBs.
A good example is the interferometers multiplexed by using polarization \cite{Vagniluca2020PRAppl_d4QKD_2det}.
The key idea is to use a phase encoding for a qubit,
whose Hilbert space is spanned by a different combination of the four-dimensional time-bin basis states (Fig.~\ref{fig:Vagniluca2020} (a)). 
This results in two four-dimensional mutually unbiased bases where the basis states are composed of only qubits.
These states can be measured using delay interferometers with different delay times.
This allows one to switch between the two bases by changing the delay time using the polarization degree of freedom as shown in Fig.~\ref{fig:Vagniluca2020} (b).
Although this is a solution specific to the case of $d=4$,
the measurement setup is practical because it can be easily calibrated in a similar manner to the conventional implementation for a time-bin qubit.

\begin{figure}[htb]
    \centering
    \includegraphics[width= .6\linewidth]{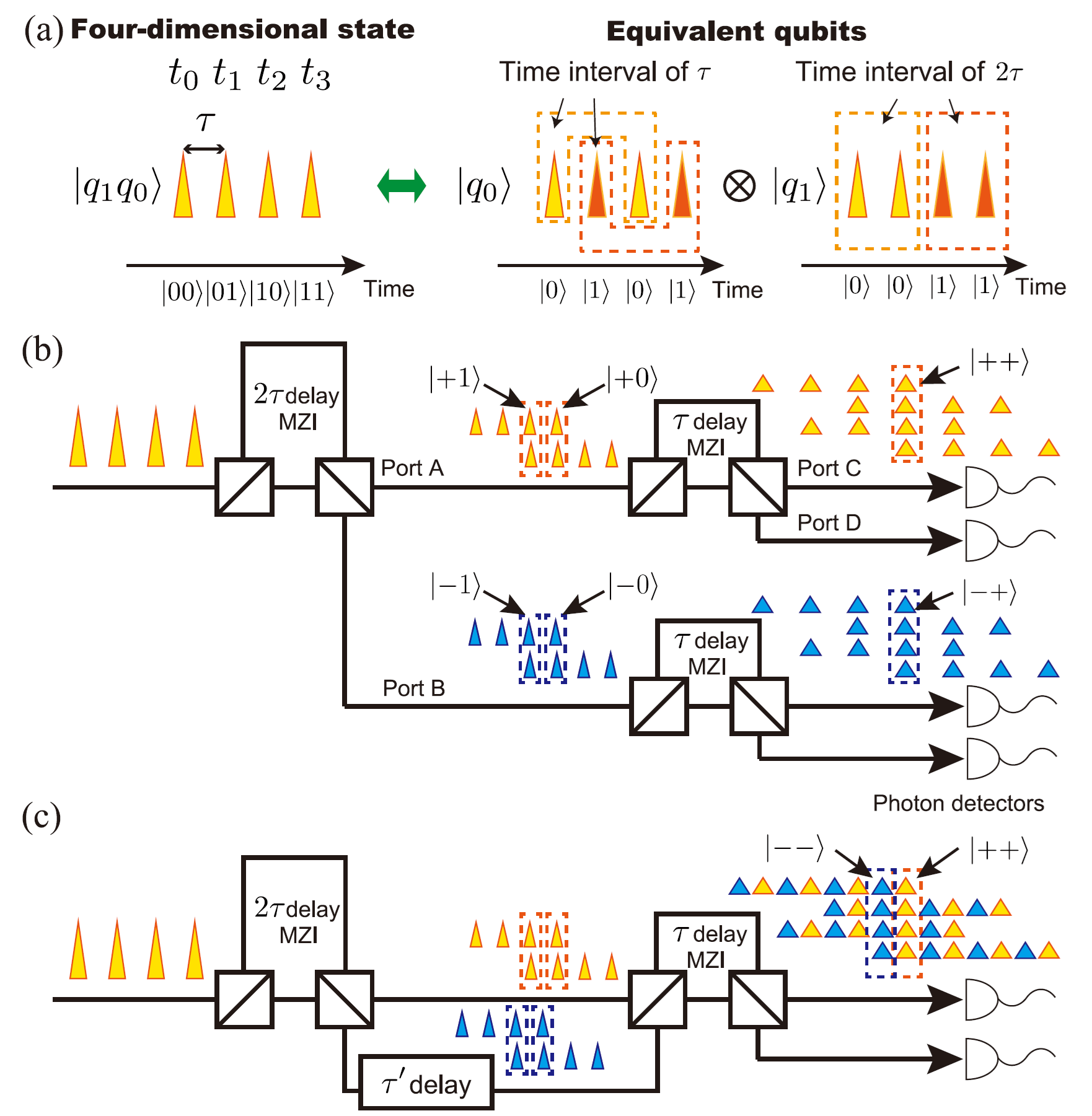}
    \caption{(a) The concept of how a four-dimensional time-bin state is informationally equivalent to two qubits. Measurement setup for the Hadamard basis using (b) a tree structure and (c) time-division multiplexing (reproduced from Ref. \cite{Ikuta2022PhysRevRes_MUB}, licensed under CC BY 4.0). }
    \label{fig:Ikuta2022_1}
\end{figure}

\begin{figure}[b!]
    \centering
    \includegraphics[width= \linewidth]{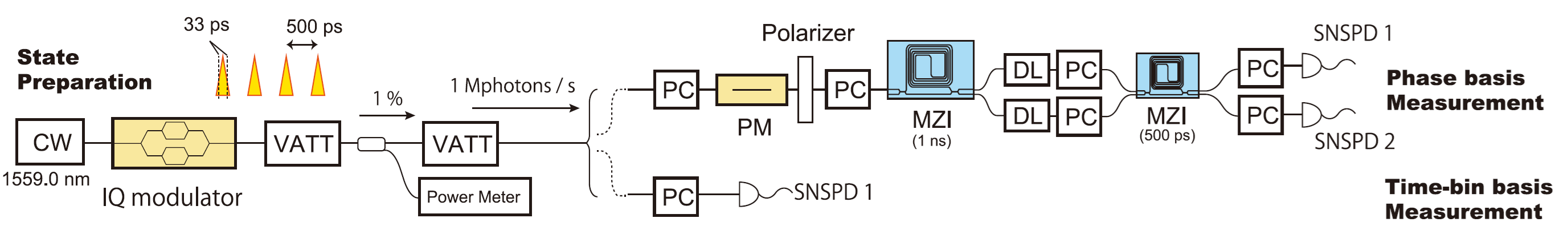}
    \caption{Experimental setup for measuring the complete set of MUBs for a four-dimensional time-bin state (reproduced from Ref. \cite{Ikuta2022PhysRevRes_MUB}, licensed under CC BY 4.0). }
    \label{fig:Ikuta2022_2}
\end{figure}

The Hadamard basis is another basis that is mutually unbiased to the time-bin basis.
For the four-dimensional states $\{\ket{0}, \ket{1}, \ket{2}, \ket{3}\}$,
we can consider an equivalent two-qubit representation
$\{\ket{00}, \ket{01}, \ket{10}, \ket{11}\}$ (Fig.~\ref{fig:Ikuta2022_1} (a)).
The $d$-dimensional Hadamard basis is
obtained by applying the two-dimensional Hadamard transform to all the equivalent qubits.
A nice feature of the Hadamard basis is that all the phases are $0$ or $\pi$
and it has a tensor product form in the equivalent qubit representation.
Because of this,
we can make projective measurements of time-bin Hadamard basis states
by using cascaded interferometers with zero relative phases only (Fig.~\ref{fig:Ikuta2022_1} (b)) \cite{Ikuta2022PhysRevRes_MUB}.
In addition,
we can reduce the number of interferometers from $(d-1)$ to $\log d$
by using time-division multiplexing instead of the tree configuration (Fig.~\ref{fig:Ikuta2022_1} (c)).
We can easily calibrate the interferometers to have a zero relative phase by maximizing the extinction ratio at the output of each interferometer.
Thus, this configuration reduces the difficulties associated with encoding small relative phases and the number of cascaded interferometers required.

In addition,
the use of the Hadamard basis enables us to employ the complete set of MUBs in $d$ dimensions by
only adding a phase modulator in front of the interferometers (Fig.~\ref{fig:Ikuta2022_2}) \cite{Ikuta2022PhysRevRes_MUB}.
In a $d$-dimensional Hilbert space,
there are at most $(d+1)$ mutually unbiased bases (MUBs) \cite{Wootters1989, DURT2010}.
Such a complete set of MUBs have been found in prime-power dimensions,
but their existence in other dimensions is an open problem.
A typical example is the eigenstates of the Pauli operators $\sigma_x, \sigma_y, \sigma_z$ for a qubit.
In this case,
two eigenstates chosen from different Pauli operators are mutually unbiased in all combinations.
The complete set of MUBs can be used to perform quantum state tomography \cite{DURT2010}
and make a QKD protocol more noise tolerant such as the six-state protocol \cite{Brus1998}. 
The configuration in \cite{Ikuta2022PhysRevRes_MUB} can switch between all different MUBs by changing the modulation signal applied to the phase modulator except for the time-bin basis.
This method was demonstrated for $d=2^n$ and it was shown that it can be extended to more general $d=p^n$ for odd prime numbers $p$,
which are known to be the dimensions for which a complete set of MUBs exists.

A recent alternative to interferometric setups for measurements of HD superpositions of time-bin was demonstrated by Danese et al.~\cite{danese2026} that uses spatial-mode dispersion in multi-mode fibers. By harnessing the coupling between spatial and temporal information in a commercial multi-mode fiber, they showed how large, multi-mode Franson-type interferometers can be programmed inside the fiber itself. This experiment used a 40m-long fiber to implement generalized quantum measurements of time-bins in up to dimension 11. In contrast with the cascaded interferometer approach, this method does not require any active interferometric stabilization and can implement arbitrary HD measurements. The use of a single fiber demonstrates potential for scalability to larger time-bin dimensions, as the achievable measurement dimension and time-scales are directly related to the fiber parameters such as its core size and length.

In many cases,
we discard the side peaks in the signal at the output of the interferometers.
However, these side peaks also provide useful information that can have some applications.
For example, in two dimensions, they correspond to the projective measurements in the $\sigma_z$ basis for a time-bin qubit.
Therefore, we need only two phase settings to perform quantum state tomography (QST) \cite{Takesue2009}.
This idea can be extended to a higher-dimensional state \cite{Ikuta2017NJP_Tomography, Ikuta2018SciRep_100km_d4}.
By using information contained in all the side peaks,
we can reduce the number of parameters required in the setup.
Although the analysis of the measurement data becomes complex,
the setup is more user-friendly to experimentalists.

An optical switch is another key device used for the measurement of a high-dimensional time-bin state.
The first important work in this regard was demonstrated by Nowierski et al \cite{Nowierski2016}.
They mapped a time-bin qubit in a subspace of a high-dimensional time-bin state onto a polarization qubit using a delay interferometer and a fast optical switch.
The optical switch enabled the selection of short pulses with an interval of 100 ps.
By combining this method with conventional polarization-qubit measurements,
they could perform QST for the entire state in the original high-dimensional Hilbert space.
The key device---the optical switch---used cross-phase modulation (XPM) for its operation \cite{hallAllopticalSwitchingPhotonic2011, ozaEntanglementPreservingPhotonicSwitching2014}.
Such a fast optical phenomenon opens the way to overcoming the limitations due to slow electrical devices, e.g., a single-photon detector,
although it introduces extra noise due to other nonlinear effects such as Raman scattering.

\begin{figure}[htb]
    \centering
    \includegraphics[width= \linewidth]{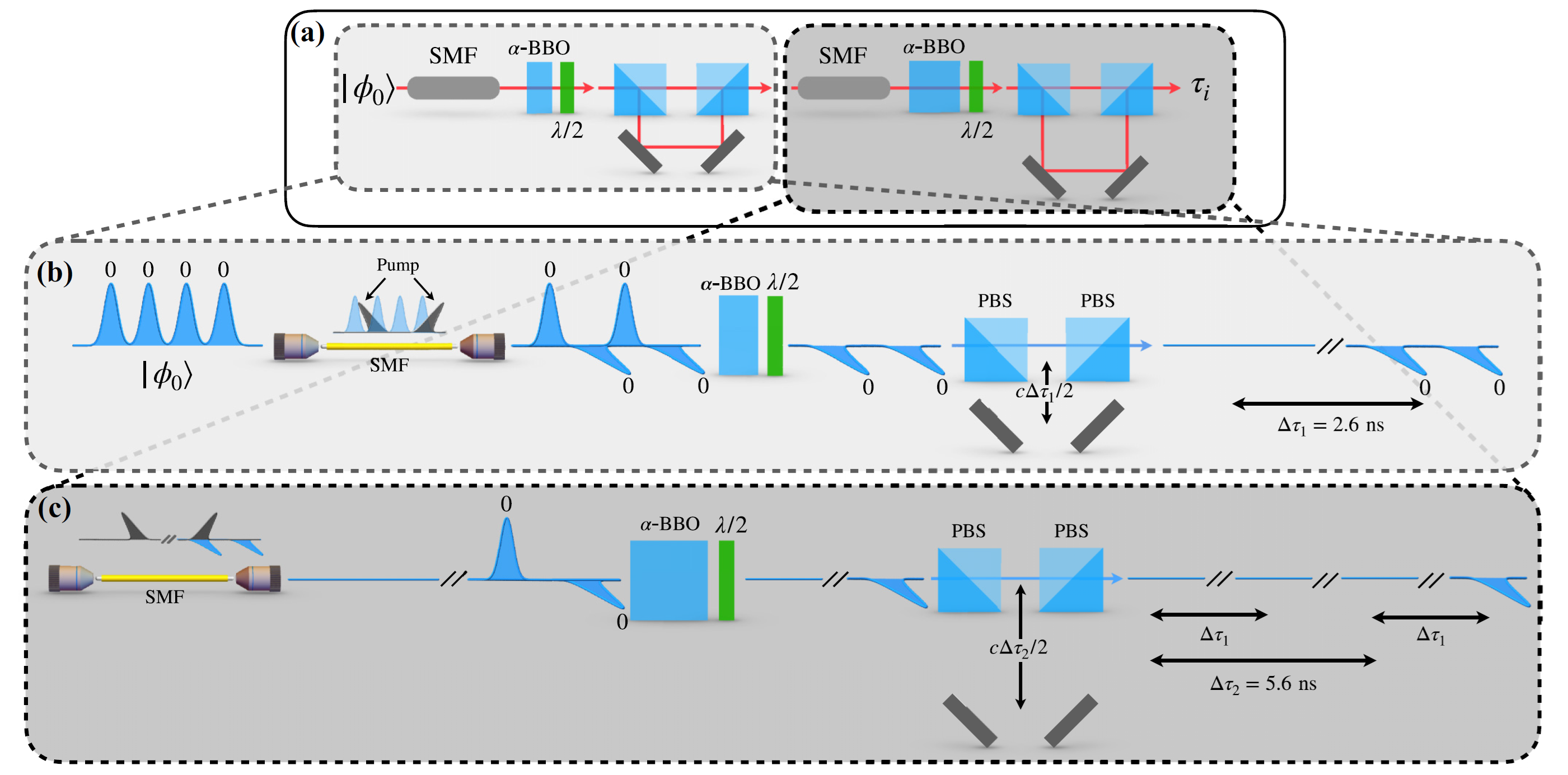}
    \caption{Deterministic measurement of a high-dimensional time-bin state in the Hadamard basis using ultra-fast optical switch \cite{Bouchard2023}. Reprinted figure with permission from \href{https://dx.doi.org/10.1103/PhysRevA.107.022618}{[F. Bouchard et al., Phys. Rev. A \textbf{107}, 022618 (2023).]} Copyright (2023) by the American Physical Society.
    }
    \label{fig:Bouchard2023}
\end{figure}
\label{sec:GenAndMeasTimeBinXPMSwitch}
Soon after, the technique using XPM was elaborated for the measurement of an ultra-fast time-bin qubit \cite{kupchakTimebintopolarizationConversionUltrafast2017, kupchakTerahertzbandwidthSwitchingHeralded2019, bouchardQuantumCommunicationUltrafast2022},
and further extended to an ultra-fast high-dimensional time-bin state by Bouchard et al \cite{Bouchard2023}.
They demonstrated the deterministic measurement in the time-bin basis and the Hadamard basis,
where the pulse interval of a four-dimensional state is only 2.25 ps.
They used XPM to change the polarization of specific pulses
and implemented delay interferometers using the large birefringence of an $\alpha$-barium borate (BBO) crystal and polarising beam splitter (as shown in Fig.~\ref{fig:Bouchard2023}).
This method is not only ultra-fast but also deterministic.
Namely,
the output of their interferometers do not have side peaks
and perform the measurement in the selected basis deterministically,
which is an important demonstration of an efficient measurement for a high-dimensional time-bin state.

A fundamental phenomenon in quantum mechanics is two-photon interference,
also known as {Hong--Ou--Mandel} (HOM) interference.
Using HOM interference,
Islam et al.~demonstrated a measurement for determining whether an input state is a high-dimensional target state or not \cite{Islam2019}.
Similarly to HOM interference for a time-bin qubit,
an input and target high-dimensional time-bin state are launched into a 50:50 beam splitter.
If the input and target states are the same,
there are no coincidence counts at the output of the beam splitter due to the HOM effect.
The difference between a qubit and qudit is that this measurement does not correspond to the Bell measurement,
which requires ancilla photons for a qudit \cite{calsamigliaGeneralizedMeasurementsLinear2002, Goyal2014}.
However,
this measurement is still powerful if one only needs to monitor the closeness to a chosen target state.

\begin{figure}[htb]
    \centering
    \includegraphics[width= .6\linewidth]{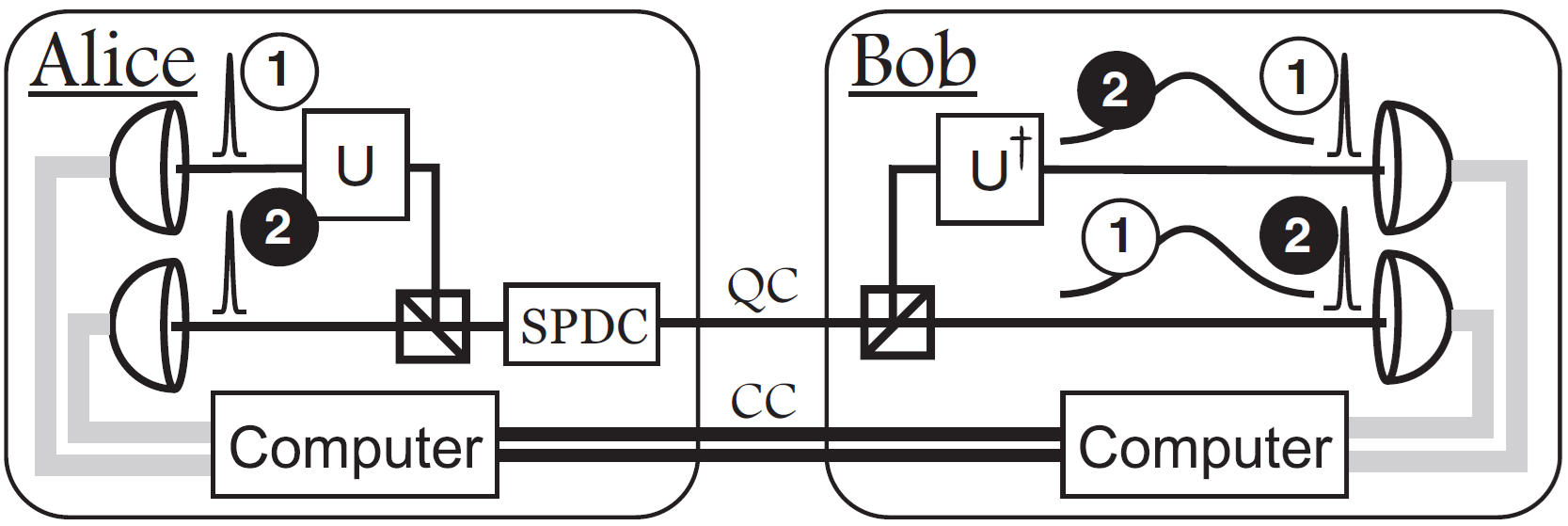}
    \caption{Measurement of a high-dimensional time-bin state using dispersive optics \cite{mowerHighdimensionalQuantumKey2013}. Reprinted figure with permission from \href{https://dx.doi.org/10.1103/PhysRevA.87.062322}{[J. Mower et al., Phys. Rev. A \textbf{87}, 062322 (2013).]} Copyright (2023) by the American Physical Society.}
    \label{fig:Mower}
\end{figure}

As time-energy entanglement and the Fourier basis indicate,
the frequency of a photon is a physical quantity complementary to time.
The dispersive optics approach can be used to convert the information on optical frequencies into the information on temporal delays by means of chirping, provided the temporal far-field condition is met \cite{Torres:2011spacetime}. This approach is equivalent or closely related to the classical pulsed dispersive Fourier transformation technique \cite{Goda:2013}. It has been pioneered for spectrally-resolved single photon detection by Avenhaus et al. using long optical fibers \cite{Avenhaus:2009}, while Davis et al.~increased the resolution of the technique using chirped fiber Bragg gratings as the dispersive medium \cite{Davis:2017}.  Therefore, for the case of time-energy entanglement,
we can measure the two complementary quantities by observing the arrival time with or without dispersive optics respectively.
This technique was used for performing QKD with the time-frequency degrees-of-freedom by several groups (Fig.~\ref{fig:Mower}) \cite{mowerHighdimensionalQuantumKey2013, Liu2019, liuHighdimensionalQuantumKey2024, ogrodnik2025high}.
It is important to note that this method involves measurements of continuous variable properties of time and frequency, to be contrasted with discretized time-bins.
This limits one to using theoretical tools developed for continuous variables, and/or places additional security considerations arising from the implementation, such as detection efficiency mismatch \cite{Grasselli:2025}. This approach is nevertheless appealing because the setup is simple, scalable in dimension, and exhibits low loss.

\begin{figure}[b!]
    \centering
    \includegraphics[width= .7\linewidth]{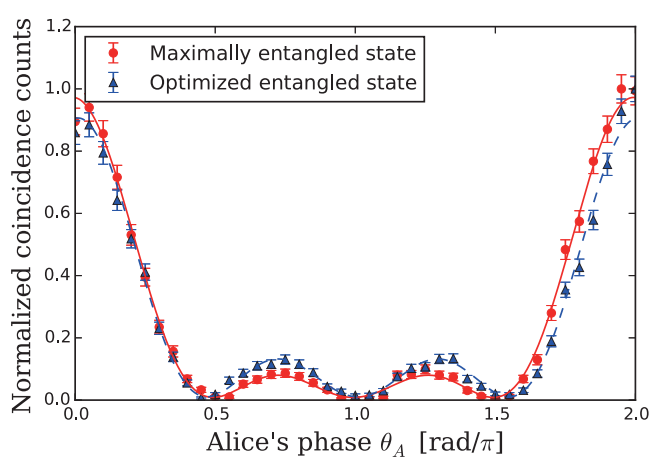}
    \caption{Difference between the coincidence fringes of four-dimensional time-bin maximally entangled state and optimized entangled state \cite{Ikuta2016PRA_CGLMP_ququart}. Reprinted figure with permission from \href{https://dx.doi.org/10.1103/PhysRevA.93.022307}{[T. Ikuta and H. Takesue, Phys. Rev. A, \textbf{93}, 022307 (2016).]} Copyright (2016) by the American Physical Society.}
    \label{fig:Ikuta_fringe}
\end{figure}

\textit{Measurement of HD time-bin entanglement}---Recent work on time-bin entanglement has moved towards quantifying and precisely characterizing HD entanglement and exploring its noise-robustness. It is known that maximally entangled states do not maximize the violation of the CGLMP inequality for entangled states with 
$d>2$. Although an experiment attempted to maximize the violation by optimizing an energy–time entangled state, a clear enhancement of the violation was not confirmed \cite{Schwarz2014IJQI_EnergyTimeQutrits}.
In \cite{Ikuta2016PRA_CGLMP_ququart}, Ikuta and Takesue optimized a four-dimensional time-bin state by varying the intensity of the pump pulses for each slot, and experimentally observed a clear difference from the maximally entangled state in a two-photon coincidence fringe (Fig.~\ref{fig:Ikuta_fringe}). They also performed the CGLMP inequality test and confirmed an enhanced violation ($I=2.913\pm0.023$)
that clearly exceeded results obtained with maximally entangled states ($I=2.774 \pm 0.025$). 

The same team also developed a scalable quantum state tomography (QST) scheme for HD time-bin states, in which cascaded delay interferometers simultaneously implement multiple projective measurements and reduce the number of measurement settings to scale linearly with the dimension \cite{Ikuta2017NJP_Tomography}. Using this scheme, they reconstructed the density matrix of a four-dimensional time-bin maximally entangled state with only 16 settings, achieving an average fidelity of about 0.95. Alternatively, a dispersion-based approach has been proposed by Czerwi\'nski et al. to perform multidimensional time-bin state tomography \cite{czerwinski2021phase}.

\begin{figure}[t!]
    \centering
    \includegraphics[width= \linewidth]{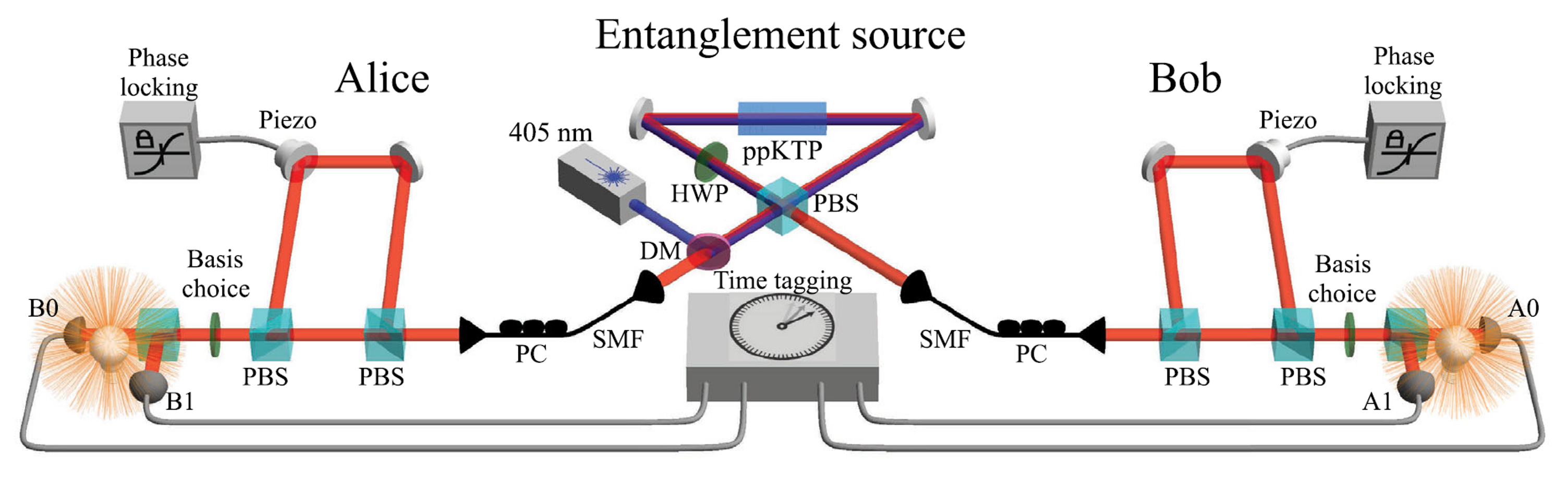}
    \caption{Experiment certifying the noise-robustness of a high-dimensional time-bin entanglement (reproduced from Ref. \cite{EckerHuber2019}, licensed under CC BY 4.0).}
    \label{fig:Ecker_highnoise}
\end{figure}

Another interesting application of high-dimensional encoding is certifying the existence of entanglement in a noisy environment.
While it was well-known that an advantage of HD entanglement is its large noise robustness,
it had not been confirmed in an experiment.
In \cite{EckerHuber2019},
Ecker et al.~confirmed the increased robustness of HD time-bin entanglement by intentionally adding noise using light emitting diodes (LEDs) in front of their single-photon detectors (Fig.~\ref{fig:Ecker_highnoise}).
They changed the number of dimensions simply by changing the resolution of the detection time
while the duration determining the state was fixed.
The use of a Franson-type interferometer with a fixed delay time limited their measurements to time-bin qubit spaces.
Even with this limitation,
they showed that time-energy entanglement with larger dimensions had greater robustness to background noise.

\subsubsection{Challenges and outlook}
\label{sec:GenAndMeas_TimeBinChallengesAndOutlook}

While passive cascaded interferometers have played a central role in the generation and measurement of high-dimensional time-bin states,
active interferometers, such as the recent implementation using XPM \cite{Bouchard2023} are an important direction in development.
As the efficiency of passive interferometers for implementing the Fourier and Hadamard bases decreases by 3 dB per interferometer,
the deterministic operation of active interferometers is a key requirement to further increase the dimensions of a high-dimensional time-bin measurement.
The interferometer in \cite{Bouchard2023} used an $\alpha$-BBO crystal to induce the mode-dependent delay.
Therefore,
the minimum and maximum thickness of the crystal determine the available number of the dimensions as long as the optical pulses are short enough (at least less than a ps).
To enlarge the dimensions more than this limit,
an actively stabilized delay interferometer using a PBS would be required.
In addition,
the effect of the extra noise should be carefully managed by considering the accumulation at each stage of the interferometer.

An optical switch using an electro-optic (EO) modulator is also a promising candidate for realizing active interferometers.
Recently,
Vedovato et al. implemented such an active interferometer for time-bin qubits to demonstrate a postselection-loophole-free Bell test \cite{vedovatoPostselectionLoopholeFreeBellViolation2018}.
This implementation can be extended to a high-dimensional time-bin state \cite{Bouchard2023, Ikuta2022PhysRevRes_MUB},
and takes advantage of the fact that it introduces no extra noise in principle.
Although the bandwidth of conventional EO modulators is limited to several tens of GHz,
much faster operation is possible by using thin-film lithium niobate (TFLN) \cite{bacchiPostselectionFreeTimebin2025},
for which commercial modulators with bandwidths exceeding 100 GHz are already available \cite{hyperlightProductPage}.
It would enable low-noise operations on time-bin qudits with intervals of several picoseconds.
Challenges in this direction include stabilizing the bias drift for reliable operation,
precisely modulating the pulses to achieve a high extinction ratio,
and reducing the insertion loss to realize a truly efficient active interferometer.
Specifically,
an insertion loss less than 3 dB per interferometer is an important criterion.
Besides improving the fabrication process,
a more elaborate configuration of EO modulators such as a Sagnac interferometer \cite{scalconLowerrorEncoderTimebin2025, vijayadharanSagnacbasedArbitraryTimebin2025} would be beneficial for a more stable and practical implementation.

\label{sec:GenTimeDetectorDeadTime}Unfortunately,
the capacity of a photon increases logarithmically with the number of dimensions $d$.
Since increasing the dimension of a time-bin state consumes multiple time slots,
the capacity per time $(\log_2 d )/ d$ is the same at $d=2$ and $4$ and decreases for larger $d$ in the noiseless case \cite{Islam2017_a}.
However, in many cases
the current bottleneck for quantum communication is the dead time of single-photon detectors.
By combining active measurement methods and increasing $d$,
there is a pathway to enabling ultra-fast quantum communications as demonstrated in \cite{Islam2017_a, yuQuantumKeyDistribution2025}.
A postselection-loophole-free experiment for the CGLMP inequality is also an interesting fundamental challenge, which could be significant for device-independent quantum protocols.
In addition,
optical switches and cascaded interferometers are key elements for realizing an arbitrary unitary transformation for time-bin qudits \cite{Bussieres2006}, where a scalable implementation poses a significant challenge.

Finally,
the noise-robustness of a high-dimensional time-bin entangled state is an important research topic as recently demonstrated \cite{EckerHuber2019}.
However, noise involves many aspects of a given experiment or implementation and can be complex to precisely quantify.
For example,
the total amount of noise also increases when we increase the dimensions by adding extra time slots,
while it is fixed when we increase the dimensions by dividing the original temporal frame.
If we use optical switches based on XPM, the extra noise induced by nonlinear effects depends on the dimension, as the number of optical switches required increases with dimensions.
In addition,
the noise affects different physical quantities---such as the state fidelity, violation of Bell's inequality, other entanglement witnesses, and QKD error rate---in different ways.
We should keep in mind that there is no free lunch,
and it is important to identify which combinations of noise sources and measurements genuinely offer an advantage for high-dimensional time-bin states.

\subsection{Frequency-bin} \label{sec:GenAndMeas_Freq_Bin}

\author{Daniele Bajoni\authormark{4}, David J. Moss\authormark{5}, Roberto Morandotti\authormark{6}}

\address{\authormark{4}Dipartimento di Ingegneria Industriale e dell'Informazione, Università di Pavia, 27100 Pavia, Italy. \email{daniele.bajoni@unipv.it}}

\address{\authormark{5}Optical Sciences Centre, Swinburne University of Technology, Hawthorn, Victoria, Australia 3122. \email{dmoss@swin.edu.au}}

\address{\authormark{6}Institut national de la recherche scientifique, INRS-EMT, Varennes (QC), Canada. \email{roberto.morandotti@inrs.ca}}

\subsubsection{Early work}
\label{sec:GenAndMeasFreqBins}

Frequency-bin qubits are being studied as an alternative DoF with respect to more established path and time-bin encodings. In the case of frequency bins, the spectral degree of freedom of photons is leveraged to encode coherent superpositions of discrete optical frequencies. A generic frequency-bin qubit is defined as
\begin{equation}
|\psi\rangle = \alpha |\nu_0\rangle + \beta |\nu_1\rangle, \qquad |\alpha|^2 + |\beta|^2 = 1,
\end{equation}
where the basis states $|\nu_i\rangle$ correspond to photons that occupy distinct frequency modes, typically separated by an integer multiple of the free spectral range (FSR) of a resonator or the fundamental drive frequency of a modulator. 
Frequency-bin states offer intrinsic compatibility with telecom networks, integrated photonics in telecom bands, and wavelength-division multiplexing. The theoretical foundations of manipulating quantum information in the frequency domain were developed through electro-optic phase modulation, Fourier-transform pulse shaping, and parametric processes \cite{Lukens2017, brecht_photon_2015}. These techniques laid the foundation for using the frequency domain to implement arbitrary single-qubit operations, beamsplitter-like transformations, and scalable unitaries in high-dimensional Hilbert spaces.

\begin{figure*}[t]
    \centering
    \includegraphics[width=\textwidth]{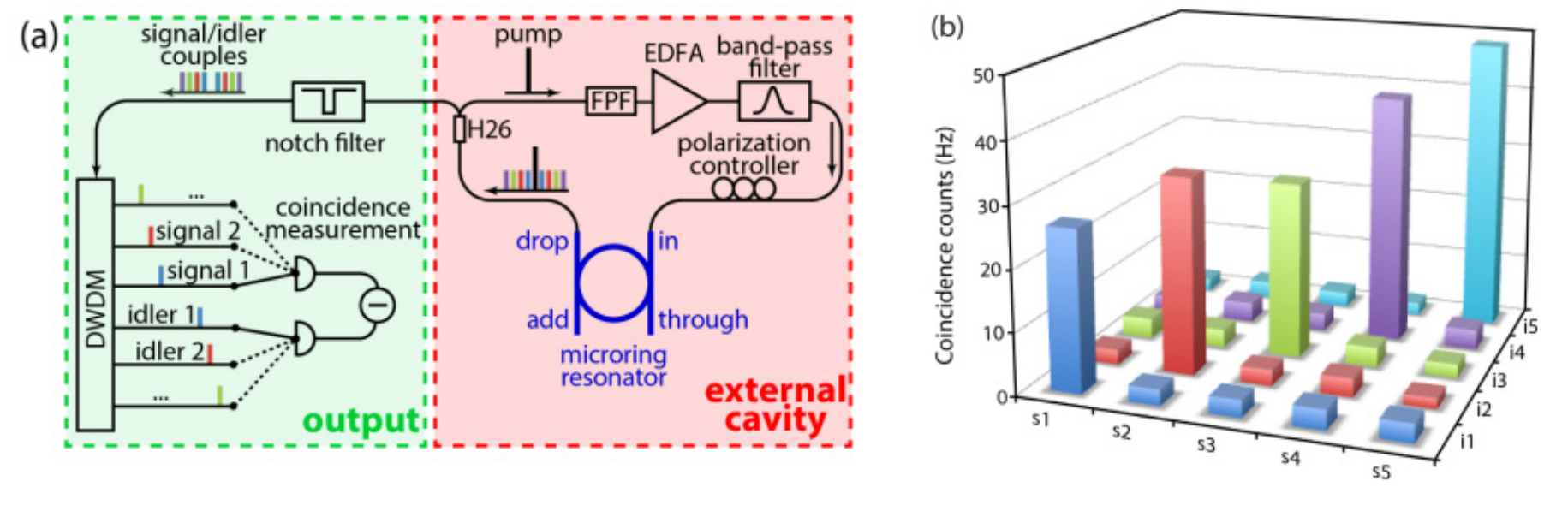}
    \caption{(a) Scheme for the generation of frequency-bin-encoded heralded single photons taken from Ref. \cite{Reimer2014}. The source is a microring embedded in an external cavity including a gain medium (EDFA), a band-pass filter (centered at the pump wavelength), and a polarization controller. The signal/idler photon pairs exiting the drop port are transmitted through the high-isolation notch filter, separated by a commercial DWDM filter, and then characterized by coincidence detection. (b) Coincidence count rates measured for several signal/idler combinations, showing that significant coincidence counts are only visible between symmetric channels. Images reproduced with permission from an Optica Publishing Group Open Access License.}
    \label{fig_fbin_1}
\end{figure*}

Early and influential demonstrations of the generation of integrated frequency-bin photon pairs  relied on dissipative Kerr soliton microcombs in silicon nitride or Hydex microring resonators \cite{Reimer2014, Kues2017, Kippenberg2018}. In Ref.~\cite{Reimer2014}, Reimer \textit{et al.} reported a CMOS-compatible microring source generating signal and idler photon pairs on adjacent frequency bins of an optical frequency comb (shown in Fig.~\ref{fig_fbin_1}). These early results provided spectrally narrow, intrinsically phase-locked frequency modes, forming the basis for entangled qubit and qudit states. Kues \textit{et al.} \cite{Kues2017} also extended this approach to generate high-dimensional entangled states with coherent control across multiple frequency bins, demonstrating on-chip quantum state manipulation using pulse shaping.

The characterization of frequency-bin states has also been an important early topic of research. Clemmen \textit{et al.} demonstrated Ramsey interference with single photons by employing Bragg-scattering four-wave mixing \cite{Clemmen2016AUTO} instead of electro-optic modulators. The resulting interference fringes directly probed the relative phase between frequency-bin amplitudes. Imany \textit{et al.} introduced a complementary electro-optic phase modulation technique for characterizing biphoton frequency combs, using a combination of pulse shapers and electro-optic modulators to measure both the amplitude and phase of entangled frequency-bin states \cite{Imany2018a}. These optical tools provided the foundations for frequency-domain quantum state tomography but also for the universal manipulation of frequency-bin states.

The realization of coherent optical operations in the frequency domain is essential to transform frequency-bin encoding into a universal photonic platform. Lu \textit{et al.} demonstrated the first frequency beamsplitter and tritter based on cascaded electro-optic modulators and pulse shapers, implementing the unitary transformations
\begin{equation}
\frac{1}{\sqrt{2}}
\begin{pmatrix}
1 & 1 \\
1 & -1
\end{pmatrix}
\text{ and }
\frac{1}{\sqrt{3}}
\begin{pmatrix}
1 & 1 & 1 \\
1 & e^{2\pi i/3} & e^{4\pi i/3} \\
1 & e^{4\pi i/3} & e^{2\pi i/3} \\
\end{pmatrix},\
\label{eq:freq_bs}
\end{equation}
fully analogous to their spatial counterparts but operating between spectral modes \cite{Lu2018a}. Using the same architecture, Lu \textit{et al.} later extended the approach of in-line pulse shapers and modulators to implement a coincidence-basis frequency-bin controlled-NOT (CNOT) gate \cite{Lu2019a}. This work established the first two-qubit logic operation in the spectral domain.

During the same years the frequency-domain analogue of Hong-Ou-Mandel (HOM) interference was reported by Kobayashi \textit{et al.} \cite{Kobayashi2016AUTO} and Imany \textit{et al.} \cite{Imany2018c}, where two photons occupying frequency modes $\omega_0$ and $\omega_1$ were combined through the transformation in Eq.~(\ref{eq:freq_bs}), leading to a frequency-domain HOM dip in the joint spectral correlations. Subsequent work by Lu \textit{et al.} demonstrated full control of quantum interference between two frequency-bin qubits using programmable electro-optic modulation and measured frequency-resolved coincidence fringes \cite{Lu2018b}. Lingaraju \textit{et al.} conducted an experimental investigation of the role of spectral phase coherence in traditional path-encoded HOM interference, showing that high-visibility oscillations at the bin spacing are surprisingly not a direct signature of inter-comb-line coherence, thus reinforcing the need for coherent active operations for frequency-bin characterization \cite{Lingaraju2019}. 

The logical extension of such works is the generation and manipulation of high-dimensional states. Imany \textit{et al.} demonstrated 50-GHz-spaced, high-dimensional frequency-bin entangled photon pairs from a silicon-nitride resonator \cite{Imany2018b}, while MacLellan \textit{et al.} reported coherent control over pulsed quantum frequency combs using a nested cavity configuration\cite{MacLellan2018}. Imany \textit{et al.} also implemented high-dimensional state manipulation in large Hilbert spaces, demonstrating controlled multi-bin interference and generalized Pauli operations \cite{Imany2019}. These experiments pave the way for scalable cluster states and might open the possibility of frequency-multiplexed quantum repeaters.

When taken together, the early works described in this section established frequency-bin qubits and qudits as a leading platform for integrated quantum photonics. The combination of microresonator-based biphoton generation \cite{Reimer2014, Kues2017, Imany2018b}, electro-optic modulation-based gates \cite{Lu2018a, Lu2019a}, and precision frequency-domain interferometry \cite{Clemmen2016AUTO, Imany2018c} form the basis for quantum information processing in the spectral domain. With the theoretical framework of frequency-bin encoding \cite{Lukens2017, brecht_photon_2015} and advances in dissipative Kerr soliton microcombs \cite{Kippenberg2018}, the field has rapidly evolved toward programmable, high-dimensional photonic processors operating in the frequency domain.

\subsubsection{Recent developments}

The early demonstrations of integrated frequency-bin photon-pair generation described in the previous subsection have led to the development of fully programmable quantum frequency processors and increasingly complex applications of frequency-bin qubits and qudits. Most of these advances are based on the architecture of universal unitary transformations using cascaded pulse-shapers and electro-optical modulators. This architecture has been employed to demonstrate fully arbitrary control over frequency-bin qubits \cite{Lu2020_ArbitraryControl} with fidelities exceeding 98\%.

This synthetic-frequency approach has been employed in a series of experiments on the quantum frequency processor (QFP). In Ref. \cite{Lu2019_Simulations}, the authors used the QFP to simulate subatomic many-body physics, mapping effective Hamiltonians for scattering processes onto frequency-bin qubits and implementing time evolution via programmable spectral transformations. In related work by the same group,  quantum phase estimation was implemented using time-frequency qudits encoded in a single photon \cite{Lu2020_QPE}. High-dimensional frequency-bin entanglement has also been used to probe quantum walks on synthetic spectral lattices \cite{Imany2020}. By programming specific spectral phase patterns, tunable coin operations were realized and  signatures of nontrivial walk dynamics were observed, illustrating the suitability of frequency bins for studying quantum transport and topological features in synthetic dimensions.

\begin{figure*}[t]
    \centering
    \includegraphics[width=\textwidth]{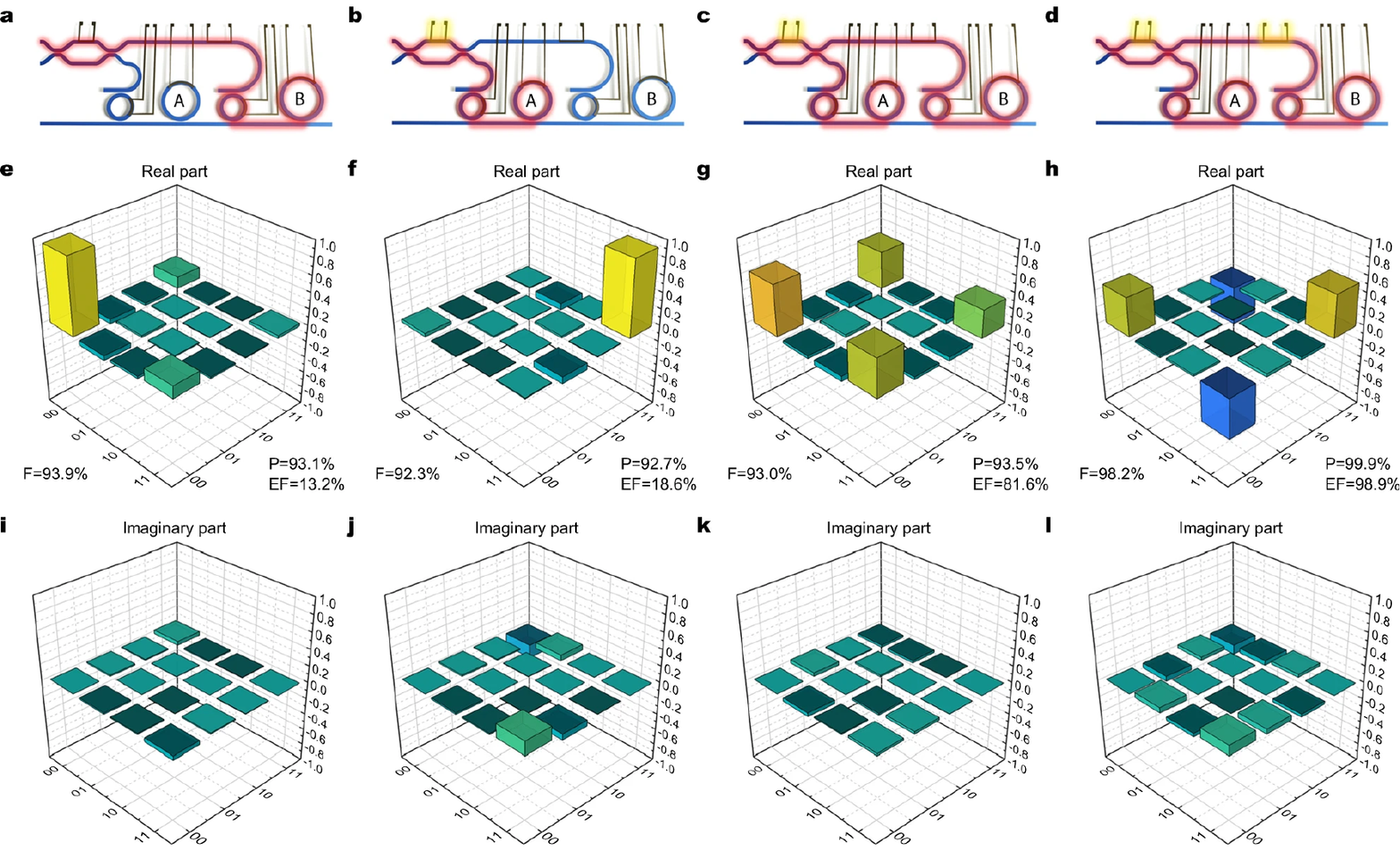}
    \caption{(a) Scheme from Ref. \cite{Clementi2023} for the programmable generation of linear combinations of the $|00\rangle$ and $|11\rangle$ states including the maximally entangled Bell states $|\Phi^+\rangle$ and $|\Phi^-\rangle$. The top row (a--d) shows the integrated circuit geometry along with the excitation patterns for all four states. The source is composed of two microring resonators that can be independently and selectively excited via a Mach-Zehnder interferometer and several integrated thermal phase shifters.   Panels (e--h) show the  real and (g--l) the imaginary part of the reconstructed density matrices for each of the generated states, estimated through the maximum-likelihood method. F, P, and EF indicate, respectively, fidelity, purity, and entanglement of formation of each reconstructed state. Images reproduced with permission from a Creative Commons Attribution 4.0 International License (\url{https://creativecommons.org/licenses/by/4.0/}).}
    \label{fig_fbin_2}
\end{figure*}

The same architecture has also been extended to high-dimensional operations. Lu \textit{et al.} implemented discrete Fourier transform (DFT) gates acting across multiple frequency bins, demonstrating reconfigurable $d$-level transformations that map single-bin states to mutually unbiased superposition bases \cite{Lu2022_DFT}. On the characterization side of frequency-bin states, a work from Simmerman \textit{et al.}  introduced compressive and Bayesian approaches for efficiently reconstructing biphoton frequency spectra \cite{Simmerman2020}. By applying random spectral phase patterns and performing only a limited number of projective measurements, they reconstructed the joint spectral intensity and phase with greatly reduced acquisition time, enabling practical tomography for sources with many frequency modes. Such approaches are particularly relevant in the case of qudits in which Hilbert spaces reach dimensions that quickly become non-manageable for standard tomographic techniques. 

Noise in the measurement of frequency-bin states has been thoroughly studied. At the level of two-photon interference,   the spectral Hong-Ou-Mandel (HOM) effect between a heralded single photon and a thermal field was analyzed in Ref. \cite{Kashi2023}. This study established a quantitative threshold for nonclassical visibility as a function of multiphoton contamination, providing a practical tool for assessing the nonclassicality of spectral interference in realistic conditions.

Recent work has also been devoted to the generation and manipulation of complex entangled states in integrated platforms. Reimer \textit{et al.} demonstrated high-dimensional one-way quantum processing implemented on $d$-level cluster states generated from integrated microresonator frequency combs \cite{Reimer2019}. Their work connected the rich structure of biphoton frequency combs to measurement-based quantum computation in high-dimensional spaces.  The programmability of the emitted state has also been reported using silicon photonics. Programmable frequency-bin quantum states \cite{Liscidini2019} were shown in a nano-engineered silicon device in Ref. \cite{Clementi2023}, incorporating   microring resonators and on-chip spectral filtering elements. This approach dramatically reduces system complexity by embedding generation and manipulation of spectral modes directly on chip (Fig.~\ref{fig_fbin_2} shows examples of the generated states). Similarly, in Ref. \cite{Borghi2023} the same approach was extended to the generation of high-dimensional qudits (with $d$ up to 4), integrating sources and pump control elements to realize complex transformations in a CMOS-compatible platform, with fidelity exceeding 90\% for $4$-level entangled qudits.  Standard continuous-wave spontaneous parametric down-conversion (SPDC) was used in \cite{CabrejoPonce2023} to harness inherent large-scale frequency correlations. The study reported certification of discretized frequency entanglement, successfully verifying a minimum of 33 entangled dimensions using a novel, highly efficient certification technique that requires very few measurements, without requiring additional assumptions about the state.

\subsubsection{Challenges and outlook}
\label{sec:GenAndMeas_FrequencyChallengesAndOutlook}

Despite the rapid progress achieved in recent years, several key challenges remain before frequency-bin quantum technologies can reach the scale, stability, and performance required for large-scale quantum networks or quantum photonic processors. One of the most pressing limitations concerns the outstanding challenge to characterize and verify high-dimensional states as device complexity increases. Although Bayesian and randomized-measurement tomography have substantially reduced the number of required projections \cite{Simmerman2020,Lu2022AUTO}, further advances are needed to enable routine certification of states spanning a large number of frequency modes, particularly in the presence of loss, mode mixing, and non-ideal single photon detectors.

Another major challenge lies in achieving fully integrated and turnkey operation. Recent demonstrations of on-chip biphoton comb sources \cite{Myilswamy2023_TimeResolvedHBT}, programmable silicon devices for frequency-bin state engineering \cite{Clementi2023}, and turnkey entangled-photon generators integrating on-chip pumping \cite{Mahmudlu2023_FullyOnChipSource} highlight a clear trajectory toward compact, stable, and scalable platforms. However, these systems often rely on external modulators, filtering stages (in particular pulse shapers), or pump-stabilization electronics. Integrating active modulation, dispersion engineering, and wavelength-selective filtering on the same chip remains an open problem that will determine the feasibility of a fully integrated frequency-bin quantum photonic platform.

Increasing the dimensionality and complexity of entangled frequency-bin qubits introduces new opportunities for their manipulation. Experiments generating and characterizing broadband polarization-frequency hyperentangled states \cite{Lu2023_UltrabroadPolarFreqHyperentangled} as well as generating time-frequency hyperentangled states on chip \cite{Congia2025} have recently been reported. The use of multiple DoFs allows for deterministic quantum logic operations such as controlled-NOT gates \cite{Lu2024_PolFreqCNOT}, demonstrating the growing importance of generating and controlling quantum states of light over multiple DoFs simultaneously.

\begin{figure}[t!]
    \centering
    \includegraphics[width=0.7\textwidth]{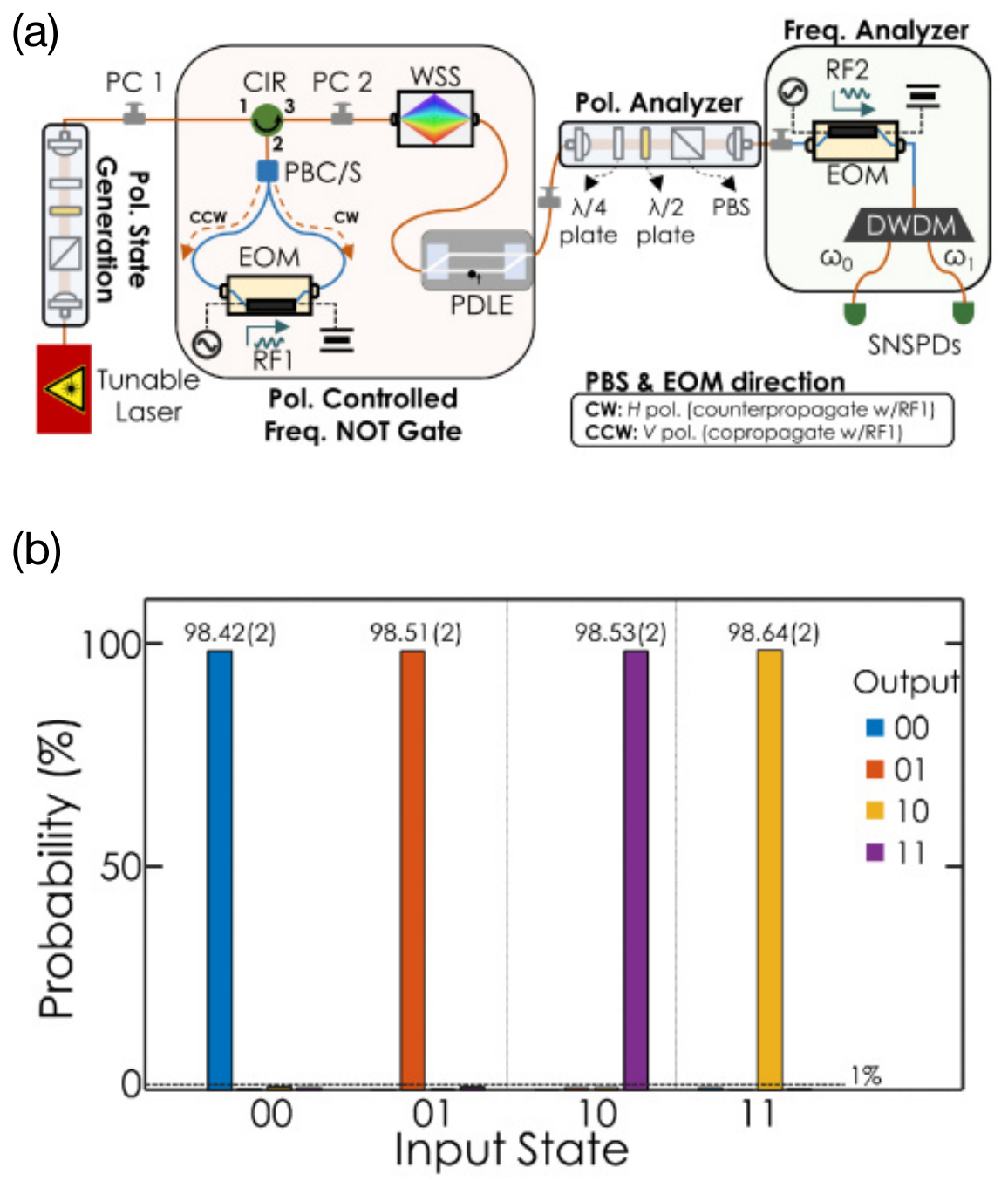}
    \caption{(a) Scheme for the demonstration of deterministic CNOT operation between the polarization and the frequency degree of freedom taken from Ref. \cite{Lu2024_PolFreqCNOT}. (a) Experimental setup for implementation and characterization of polarization-frequency CNOT gate. Acronyms: CIR, fiber-optic circulator; DWDM, 25 GHz dense wavelength-division multiplexer; EOM, electro-optic phase modulator; PC, polarization controller. PDLE: polarization-dependent loss emulator; PBC/S, fiber-based polarization beam combiner/splitter; PBS, polarizing beamsplitter cube; WSS, wavelength-selective switch; SNSPD, superconducting nanowire single-photon detector. (b)  Measured output state probabilities for each computational-basis input state, showing the correct operation of the CNOT gate. Images used with permission from an Optica Publishing Group Open Access License.}
    \label{fig_fbin_3}
\end{figure}

One of the most promising applications of frequency bins lies in quantum communications. Two recent demonstrations of the entanglement-based BBM92 QKD protocol \cite{Bennett1992} showed the possibility of using the natural multiplexing properties of frequency bins to operate in reconfigurable, frequency-multiplexed architectures and over distances up to 26 km \cite{Kashi2025_FreqBinQKD, Tagliavacche2025_QKD}. Nevertheless, realizing wide-area quantum networks based on frequency-bin entanglement will require low-loss, telecom-compatible interfaces, efficient frequency-resolved detectors, and robust methods for compensating spectral phase fluctuations over long fiber links. As shown by \cite{Tagliavacche2025_QKD}, the natural precession frequency bin qubits acquire during propagation in fiber networks can become chaotic in the presence of thermal fluctuations, entailing the need for active compensation mechanisms.

Finally, the application of frequency-bin photonics to boson sampling and photonic quantum simulation presents the opportunity of scaling to a large number of channels but also significant problems due to optical losses. Experiments using squeezed-light microresonators for bipartite Gaussian boson sampling in the time-frequency domain \cite{Borghi2025_BosonSamplingTimeFreqBin} show that integrated frequency combs can naturally generate large mode numbers with strong correlations. However, scaling such demonstrations to classically intractable regimes will require improvements in source brightness, spectral purity, detector efficiency, and programmable multimode transformations. Similarly, fully on-chip pulse shaping of entangled photons \cite{Wu2025_OnChipPulseShapingEntangled} opens the door to chip-scale simulation platforms, but overcoming accumulated fabrication errors and maintaining coherence across widely spaced spectral channels remains a formidable challenge.

In summary, frequency-bin based quantum information protocols offer the way to large-scale transmission and manipulation of quantum states of light, but significant effort needs to be poured into reducing optical losses and ensuring stable coherence of the generated states.

\begin{backmatter}
\bmsection{Acknowledgments}
R.M. acknowledges support from the Canada Research Chair program and NSERC through the following projects: AQUA ALLRP 587602-23, QuEnSi ALLRP 578468-22, Consortium on Integrated Quantum Photonics with Ferroelectric Materials ALLRP 587352-23, HyperSpace ALLRP 569583-21, and from FRQNT, through the project AdéQuATS FRQNT 328872. D.B. acknowledges support by the European Union's Horizon Europe research and innovation programme under grant agreement number 101194170 STARlight and grant number 101070168 HyperSpace; D.B. also acknowledges support from MUR through the PNRR project PE0000023-NQSTI. DJM acknowledges support from the Australian Research Council Centre of Excellence in Optical Microcombs for Breakthrough Science (Grant No. CE230100006).
\end{backmatter}

\subsection{Transverse-spatial modes} \label{Sec:GenAndMeas_Spatial_Modes}
\author{Andrew Forbes\authormark{7}, 
Stephen Walborn\authormark{8,9},
Ebrahim Karimi\authormark{10,11}}
\address{\authormark{7}School of Physics, University of the Witwatersrand, Johannesburg, South Africa\\
\authormark{8}Departamento de Física, Universidad de Concepción, 160-C Concepción, Chile\\
\authormark{9}Millennium Institute for Research in Optics, Universidad de Concepción, 160-C Concepción, Chile \\
\authormark{10}Institute for Quantum Studies, Chapman University, Orange, California 92866, USA \\
\authormark{11}Nexus for Quantum Technologies, University of Ottawa, K1N 5N6 Ottawa, Ontario, Canada}
\subsubsection{Early Work}
Transverse spatial modes as solutions to the wave equation have been known for a long time and can be generated at will since the inception of the laser. More recently, the extensive classical toolkit has been applied to the problem of generating and measuring quantum light carrying transverse-spatial modes. 

\subsubsection*{Back to basics: Transverse modes as quantum states}
Spatial correlations in quantum experiments have traditionally been exploited in the position basis, for example, at the level of pixels, but transverse-space can equally be partitioned into a basis of transverse modes. These modes define vector spaces that serve as powerful carriers of information for classical and quantum information processing. The choice of mode basis depends on the generation technique, the underlying symmetry of the system, and practical laboratory constraints, particularly when physically realisable states are required.

\begin{figure}[hp!]
\includegraphics[width=\textwidth]{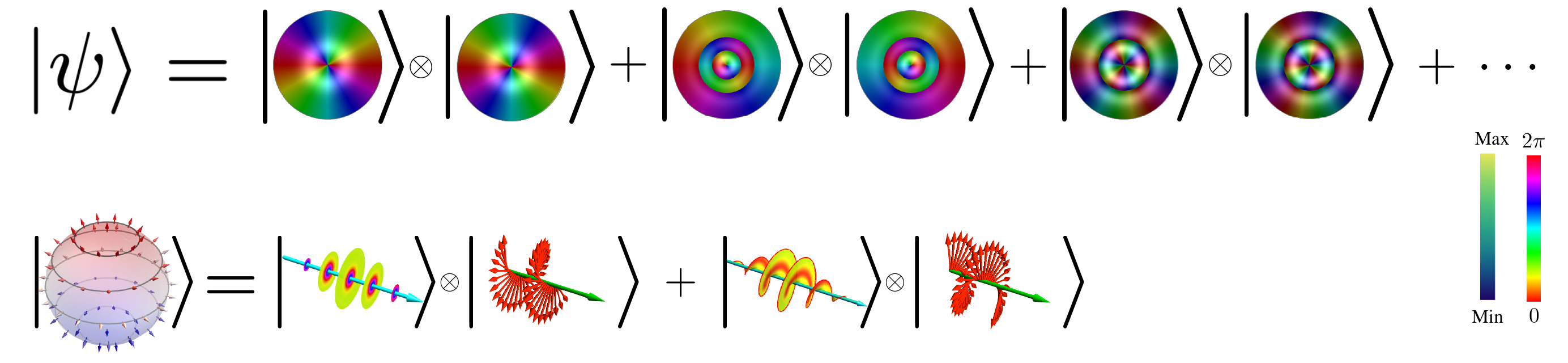}
\caption{Top panel: An example of a two-photon high-dimensional state expressed in the Laguerre-Gaussian basis, with the colour map representing the phase of the light.  Bottom panel: Different degrees of freedom can be mixed to create multi-dimensional states, with this example showing a topological state made by mixing polarization with OAM.}
\label{transmodes1}
\end{figure}

Popular mode families include Laguerre–Gaussian (LG) and Hermite–Gaussian (HG) modes, as well as Bessel–Gaussian, Ince-Gaussian, Hypergeometric–Gaussian (HyGG) and Airy beams, each with characteristic propagation and diffraction properties. Among these, LG and HG modes appear naturally as the eigenmodes of cylindrical and Cartesian symmetries, respectively, with mode number $M^2 = 2p + |\ell| + 1 = m + n + 1$. In LG modes, the indices $p$ and $\ell$ define the radial and azimuthal orders, with $p \ge 0$ and $\ell$ integer-valued, while in HG modes the modal indices $(m,n)$ correspond to orthogonal oscillations along the $x$ and $y$ axes. These bases are orthogonal and complete, and thus form a genuine Hilbert space for encoding information.  An example of a high dimensional state formed from the LG basis is shown in Fig~\ref{transmodes1}.

Although the Hilbert space dimension is unbounded in principle, experimental conditions limit the number of accessible modes in practice. Using SPDC as an example, the maximum number of modes produced at the source scales approximately as $N_\text{max} \approx w_p^2 k_p/L$, where $w_p$ is the pump radius, $k_p$ the pump wavenumber, and $L$ the crystal length—hence the adage of short crystals and large pump beams for a large entangled Hilbert space. The subsequent optical system further filters the modal content through its finite numerical aperture. Finally, different DoFs can be mixed, leading to hybrid transverse-mode states, as shown in the lower panel of Fig~\ref{transmodes1}. 

\subsubsection*{The toolkit: from classical to quantum}
The toolkit for creating, controlling, and detecting spatial modes of light is now highly sophisticated, driven by the seminal work on singular optics and orbital angular momentum (OAM) in the 1990s~\cite{Bazhenov1990, Allen1992, padgett2017orbital} and, more broadly, by the development of spatially structured light~\cite{forbes2021structured,forbes_progress_2025}, where spatial modes are exploited across all their available degrees-of-freedom (DoFs). The field became broadly accessible with the advent of digital technologies such as spatial light modulators (SLMs) and digital micromirror devices (DMDs), which allow programmable mode selection and thus on-demand creation, transformation, and detection of transverse modes~\cite{yang2023review}. Crucially, most of these tools were translated from classical to quantum experiments with little modification.

A decisive step was the 2001 demonstration of OAM entanglement~\cite{Mair2001}, which opened a new research direction in entanglement using the spatial modes of light~\cite{molina2007twisted, erhard2018twisted} [see Box 1]. In this seminal work, spontaneous parametric down-conversion (SPDC) was used to convert a pump photon into a pair of daughter photons (signal and idler), with correlations set by the crystal’s phase-matching conditions and the pump’s spatial structure~\cite{monken1998transfer, srivastav2022characterizing}. In typical SPDC, spatial entanglement appears as near-field correlations and far-field anti-correlations arising from momentum and energy conservation~\cite{Walborn2010}. When the biphoton field is analysed in a transverse-spatial mode basis, the pump profile directly shapes the quantum correlations: for a Gaussian pump, the photons emerge with equal and opposite OAM values. For a Laguerre--Gaussian (LG) pump, the OAM values obey $\ell_s+\ell_i=\ell_p$~\cite{Franke-Arnold2002, Walborn2004}, confirming OAM as a genuine basis for entanglement. 
\par
An early alternative approach to create HD quantum states was the discretization of the transverse field of entangled photons. In a two-photon Young-type experiment \cite{neves2005generation}, multi-slit apertures were used to map the continuous transverse correlations of twin photons onto discrete spatial paths, thereby generating entangled qudits of dimension $d=4$ and $d=8$. In the related “pixel entanglement” approach \cite{OSullivan:2005vh}, the fields were discretized into transverse detection regions (pixels), enabling the realization of $d=3$ and $d=6$ entangled qudits. Both approaches rely on partitioning the continuous spatial degree of freedom into a finite set of orthogonal modes to access high-dimensional entanglement, differing primarily in their physical implementation: coherent path superposition in the multi-slit case versus spatial binning of correlations in the pixel-based scheme.

For measurements, early experiments relied on diffractive holograms to project onto the desired mode: the conjugate hologram of the target mode is displayed, and the on-axis intensity is measured via coupling into a single-mode fibre. This technique implements a positive operator-valued measure (POVM) that is conceptually simple, but non-unitary, lossy and biased~\cite{qassim2014limitations}. To achieve deterministic mode discrimination, unitary mode sorters were developed. The first devices used interferometric techniques~\cite{Leach2002} and required multiple interferometers to discriminate more than two mode classes. A major advance came with the log-polar conformal mapping approach, which converts OAM states into displaced Gaussian-like spots~\cite{Berkhout10}. Although these devices efficiently measure the OAM spectrum, they are typically restricted to specific modal bases. Subsequent refinements have allowed the detection of more than 50 modes including radial information~\cite{Lavery2013}, as well as the first OAM mutually unbiased basis with reduced overall crosstalk~\cite{Mirhosseini2013}. Interferometric mode sorters that exploit Gouy-phase differences can also separate modes according to both OAM and radial quantum number~\cite{Gu2018, Fu2018}. Multi-plane light conversion (MPLC) techniques demonstrated universal, low-crosstalk transformations between arbitrary mode sets, including pixelated wavefronts~\cite{Labroille2014}, albeit with efficiencies ($\sim 30\%$) still below those of custom refractive implementations~\cite{Morizur2010}.

Digital control and fast reconfigurability also enabled automation of quantum experiments with transverse modes. Early work largely repurposed the polarization-qubit toolkit for OAM qubits ($d = 2$), for example, in OAM quantum state tomography~\cite{Jack2009} and Bell tests~\cite{Leach2009A}, at the expense of accessing only a small part of the available Hilbert space. Dedicated tools for genuinely HD control followed soon after~\cite{Romero2012, Dada2011, Agnew2011, nape2023quantum}. Transverse spatial modes have since been exploited in Bessel states~\cite{Mclaren2014}, LG states~\cite{DErrico:21}, Hermite--Gaussian (HG) states~\cite{Walborn2005, Straupe2011, Walborn02007}, Airy states~\cite{lib2020spatially}, and Ince--Gaussian states~\cite{Krenn2013}. Together, these results established spatial modes as a flexible quantum DoF capable of encoding large amounts of information in both discrete and continuous variables. They have the intrinsic capability of ultra-high dimensionality. For example, including the radial quantum number enabled observation of $100 \times 100$-dimensional entanglement in LG modes~\cite{Krenn2014}.  
\par

\subsubsection*{Controlling the spectrum}
Spatial entanglement can be engineered either directly at the source or by manipulating the photons after generation. At the source, the two-photon spatial state produced in SPDC depends on the transverse structure of the pump beam and the phase-matching conditions, making SPDC a versatile platform for quantum-state engineering. The two-photon state can be expanded in transverse-mode sets $\{u_m\}$ and $\{v_n\}$ as $|\Psi\rangle=\sum_{m,n} C_{mn} |u_m\rangle_s|v_n\rangle_i$, with coefficients
\begin{equation}\label{eq:Cmn_momentum}
C_{mn}\;=\;
\iint_{\mathbb{R}^2} \! d^2q_s\,d^2q_i \;
\widetilde{u}_m^{*}(\mathbf{q}_s)\,\widetilde{v}_n^{*}(\mathbf{q}_i)\,
\widetilde{E}_p(\mathbf{q}_s+\mathbf{q}_i)\,\Phi(\mathbf{q}_s,\mathbf{q}_i),
\end{equation}
where $\widetilde{E}_p(\mathbf{q}_s+\mathbf{q}_i)$ is the angular spectrum of the pump beam and $\Phi(\mathbf{q}_s,\mathbf{q}_i)$ is the phase-matching function. By tailoring either quantity, one can shape the joint mode spectrum—and thus the degree and dimensionality of spatial entanglement~\cite{Torres2003, Kovalov2018, Liu2018}. The overall dimensionality depends not only on the source but also on the measurement system used to project into a given basis~\cite{Miatto2012}. 

State engineering can also be carried out after the source, for example via spin–orbit conversion using geometric-phase elements such as $q$-plates~\cite{Marrucci2006, Marrucci_2011}. These devices couple polarization and OAM, enabling the generation and control of hybrid states that combine OAM and polarization~\cite{forbes2019quantum}. A $q$-plate is a birefringent element with an azimuthally varying optic axis. When a circularly polarized beam passes through a $q$-plate tuned for half-wave retardation, its polarization is flipped and the beam acquires an azimuthal phase factor $e^{\pm i2q\phi}$, yielding a vortex beam with OAM $\pm 2q\hbar$ per photon. The process is reversible and highly efficient, although limited by fixed topological charge and wavelength sensitivity associated with liquid-crystal birefringence. This approach has enabled the preparation and detection of hybrid entangled single photons and entangled states \cite{DAmbrosio2012, DAmbrosio2013}. \par 
A complementary strategy is to use quantum interference to select state symmetry. Only antisymmetric states produce antibunching after a beam splitter \cite{Walborn2003PRL}, so conditioning on coincidences allows one to engineer the symmetry of HD states. This principle has been used to sort OAM states~\cite{zhang2016engineering}, HG states~\cite{zhang2016hong}, and to control the radial quantum number~\cite{Karimi:12} in LG states~\cite{Karimi2014}. 

\subsubsection{Recent developments}
\begin{figure}[htp!]
\centering
    \includegraphics[width=0.9\textwidth]{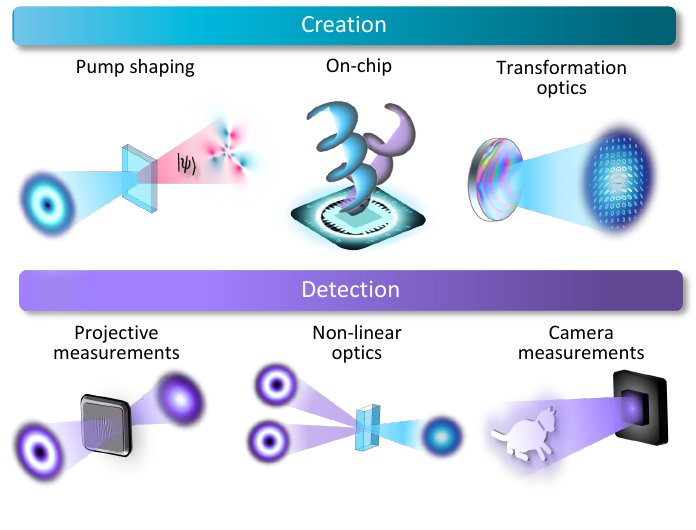}
    \caption{The modern toolkit for the creation of transverse modes as quantum states includes pump shaping in bulk systems, on-chip emitters and the use of in-path transformation optics. Detection includes projective measurements, nonlinear optics and modern cameras.}
    \label{fig_tools}
\end{figure}

Recent developments have pushed the boundaries of generating and measuring quantum states encoded with spatial modes of light, with some examples shown in Figure~\ref{fig_tools}.  These include non-linear operations, bulk to on-chip solutions, the use of digital and metasurface technology for projective measurements and state transformation, and sophisticated cameras for detection. 

\subsubsection*{Developments in mode generation and detection}
The integration of transverse spatial modes with telecommunications and integrated-photonics hardware is both essential and advantageous for advancing classical and quantum photonic communications. A diverse range of devices has emerged for generating, manipulating, and analysing transverse modes. One important example is the photonic lantern---a compact, adiabatic mode converter that provides a low-loss interface between multimode and single-mode optical channels. In such devices, a multimode fibre core gradually tapers into an array of single-mode waveguides, each supporting a supermode of the composite structure~\cite{Leon-Saval2005, Noordegraaf2009, Birks2015}. In the quantum regime, this architecture enables efficient mapping between transverse modes and single Gaussian-mode channels suitable for state generation or measurement. The photonic lantern has been used to demonstrate fibre-compatible mode sorters~\cite{Alarcon2023}. 
\par
An alternative and conceptually distinct approach to generation of HD entangled states is the  \emph{path-identity} method using multiple SPDC sources~\cite{krenn2017entanglement}.  Rather than relying on direct generation of HD entanglement, or post-selection after beam-splitter interference, this technique coherently superposes photon-pair generation processes occurring in multiple nonlinear crystals such that the emission paths of different crystals are rendered indistinguishable.  When the spatial modes associated with each path are structured, the indistinguishability of the emission origins leads to the coherent generation of high-dimensional entanglement across several photons.  This idea was demonstrated by creating high-dimensional OAM entanglement in the spatial degree of freedom using spiral phase plates, showing that the accessible dimension could be increased by simply adding additional crystal/phase plate pairs ~\cite{kysela2020path}.  The path-identity architecture thus offers a modular and intrinsically phase-stable route toward scalable multiphoton generation in structured modes. 
\par
Photonic integrated chips have recently become capable of generating and manipulating spatial modes within compact integrated platforms.  In contrast to those used for generation/manipulation of the path DoF (see Section \ref{sec:GenAndMeas_PathDistribution}), which are typically single-mode, these chips are engineered to exploit transverse spatial structure. They enable the production of photons in predefined spatial modes with tunable structure in some cases with high rates~\cite{cai2012integrated, Sun2014a, Feng2019, Zhao2025, Forbes2025chip}. These advances support scalable on-chip HD quantum states while introducing challenges in mode purity, crosstalk, and reconfigurability. 
\par
On the detection side, carefully engineered measurement schemes are essential for unlocking the full potential of transverse spatial encoding, and recent advances have significantly improved the efficiency and fidelity of HD spatial measurements. In particular, work on optimized spatial mode projections \cite{bouchard2018measuring} demonstrated that carefully designed phase-only elements can dramatically enhance the fidelity of single-outcome projections onto Laguerre–Gaussian modes, enabling >99\% accurate discrimination of both radial and azimuthal components with a simple implementation. Complementing this, a phase-only measurement strategy with specially tailored pixel basis was used to efficiently certify and generate high-quality, large-dimensional pixel entanglement \cite{valencia2020high}, substantially reducing measurement complexity and acquisition time. Machine-learning methods are emerging as powerful tools for recognising spatial modes directly from camera images or intensity profiles, and could be translated across to the quantum realm too. Approaches based on convolutional and diffractive neural networks~\cite{sharifi2020towards, lin2018all, Silva2021} offer real-time, HD discrimination with reduced reliance on interferometric calibration.

\subsubsection*{Photographing entanglement}
Camera technology has evolved to enable direct imaging of single photons and entangled states, offering new experimental capabilities to visualize quantum correlations in transverse modes~\cite{Edgar2012, Ibarra-Borja2019direct, Ndagano2020, Courme2023}. Early demonstrations probed spaces of order $d \approx 50$, while more recent work has pushed this to over $10^6$ spatial modes in complex vectorial structures entangled in polarization and space, detected by quantum cameras with time-tagged pixels~\cite{gao2024full}. Within subspaces of LG or HG modes, complete state reconstruction can be achieved via intensity measurements before and after astigmatic transformations, enabling non-interferometric, informationally complete tomography even in the presence of obstructions~\cite{Gil2025}. Interferometric probes and multiphoton phase retrieval algorithms have also been used to directly measure both the amplitude and phase of the joint spatial wavefunction of biphoton states~\cite{zia2023interferometric, dehghan2024biphoton}, effectively transferring classical full-field approaches into the quantum regime. 
\subsubsection*{Multiphoton entanglement}
The first four-photon OAM entanglement experiment ($\ell=0,\pm 1$) used the direct output of an SPDC source~\cite{Hiesmayr2016}, generating a Dicke-like state via stimulated emission of two photons in the same mode. A key advance for more complex multipartite states was mode-dependent Hong-Ou-Mandel (HOM) interference in the OAM degree of freedom. By combining two independently generated OAM-entangled photon pairs on an interferometric OAM beam splitter incorporating Dove prisms, which impart mode-dependent phase shifts, genuine three-photon HD entanglement was produced, yielding hybrid states with a $(3\times3\times2)$ structure~\cite{Malik2016}. This concept was later extended to a fully symmetric Greenberger–Horne–Zeilinger (GHZ) state involving three photons entangled in three-dimensional OAM subspaces~\cite{Erhard2018}. Four-photon experiments have demonstrated teleportation and entanglement swapping of OAM states~\cite{zhang2017simultaneous}, with high fidelity for qubits and more limited performance for qudits due to the lack of ancillary photons. 

Nonlinear optics offers a possible route to removing the need for ancillary photons \cite{Walborn2007nonlinear}. However, present-day efficiencies make single-photon control challenging. Using nonlinear detectors, quantum states could be remotely transported from a classical beam to a single photon, demonstrated with pixels for images ~\cite{qiu2023remote} and up to $d = 15$ dimensions with transverse modes~\cite{sephton2023quantum}, both using a single entangled pair as a resource. Efficiency limitations have recently been addressed by on-chip light-confinement strategies, achieving faithful teleportation in $d = 3$ with time-bins~\cite{akin2025faithful}, although similar performance in transverse modes remains outstanding. Overall, multiphoton entanglement in transverse spatial modes remains experimentally demanding, constrained by mode purity, loss, and indistinguishability, all of which limit scaling in photon number and dimensionality. 

\subsubsection*{Towards abstract DoFs}
The spatial basis also gives access to more abstract forms of control, such as engineering topological features into the quantum state. Early work demonstrated non-local phase singularities~\cite{gomes2009observation}, entanglement between one photon's polarization with the complex spatial/polarization (skyrmionic) structure of its partner\cite{fickler2014complex}, followed by single-photon (local)~\cite{ma2025nanophotonic} and entangled (non-local) Skyrmions~\cite{ornelas2024non}. In the latter case, the entangled state is a hybrid of polarization and OAM~\cite{karimi2010spin}, with the mapping between one photon’s DoF and the other encoded in a topological wavefunction. Remarkably, each individual photon or DoF may lack topology on its own, while the joint state remains topologically nontrivial. This leads to robustness of topological observables even as entanglement decays~\cite{ornelas2025topological}. 

Another direction is the spatial structuring of quantum interference, where different regions of space are engineered to exhibit distinct HOM signatures~\cite{schiano2024engineering, Ibarra2024imaging}, refining notions of local and global distinguishability. The \qo{common birth zone} of SPDC~\cite{schneeloch2016introduction}---where entangled photon pairs are created at the same time and place in a nonlinear crystal---ensures that space and time can be exploited simultaneously, although this interface remains relatively unexplored. Mixing DoFs in this way is particularly enticing: for example, blending time and space has led to spatiotemporal Airy photons~\cite{wang2024spatiotemporal} with reduced spatial spreading and enhanced robustness to background light, while four-wave mixing in multimode fibers can generate hybrid entangled frequency/transverse mode states \cite{Cruz-Delgado2016}. An even more exotic example is to mix DoFs in time, frequency, polarization, and OAM~\cite{graffitti_hyperentanglement_2020}, where multiple forms of entanglement coexist in a single biphoton state, made possible by combining crystal shaping in SPDC with spin-orbit conversion.

\subsubsection{Challenges and outlook}
\label{sec:GenAndMeas_TransverseChallengesAndOutlook}

Despite remarkable progress in HD quantum information with transverse spatial modes over the past thirty years, several key challenges remain before these systems can be deployed in real-world quantum networks, sensing platforms, or computing architectures. Many studies are still effectively limited to qubits due to incomplete access to the required SU$(d)$ transformations, although this is steadily improving~\cite{brandt2020high,goel2024inverse, dahl2024programable}. On the detection side, deterministic HD measurements that do not rely on sifting through modes are scarce, and the extensive on-chip control of spectral-temporal states does not yet translate straightforwardly to transverse modes. 

\subsubsection*{Challenges}
The main challenges related to all DoFs are efficient state certification and benchmarking methods---such as dimensionality witnesses, entanglement witnesses and compressed-sensing tomography, as discussed in Section~\ref{sec:Theory}. Here, we discuss challenges specific to the generation and detection of transverse-spatial modes.\\
\textit{Loss and mode-dependent fidelity:} HD spatial encodings are intrinsically sensitive to loss, crosstalk, and mode-dependent coupling and detection inefficiencies~\cite{Rojas2021}. For example, mode-dependent insertion losses in MPLC and mode-sorter architectures directly reduce entanglement fidelity and secret key rates. A key goal is to further develop unitary spatial-mode transformers with low insertion loss and low crosstalk. Future work could pursue, among other approaches, (i) low-loss refractive implementations, (ii) mode-selective adiabatic photonic lanterns, and (iii) calibration techniques for compensation of mode-dependent losses.\\
\textit{Scalable, low-loss parallel detection:}
\label{sec:GenTransverseParallelDetection}Although single-photon cameras, SPAD arrays and transverse-mode demultiplexing have advanced, the parallel readout of many spatial modes with high timing resolution and photon-number resolution remains immature, post-processing daunting, and noise control challenging. Integration of spatial-mode demultiplexing with superconducting nanowire detectors or on-chip detector arrays is crucial. Machine learning combined with mode projectors~\cite{ZhaO2025_det} or array-based detection~\cite{Silva2021, wang2024ultrahigh, Gil2025} is emerging as a promising route for scalable state discrimination and certification. Recent advances in superconducting nanowire detectors have enabled demonstrations of arrays with up to 64 independent pixel, timing jitter below 50ps, and system detection efficiencies up to 65\% \cite{fleming2025high, doi:10.1021/acsphotonics.4c00111}.\\
\textit{Integrated photonics for spatial modes:} To scale HD quantum systems and make them robust for field deployment, spatial-mode generation, manipulation, and detection must migrate into integrated photonic platforms. Key challenges include high-fidelity, on-chip generation of spatial modes, on-chip MPLC or mode-sorter demultiplexing networks, and fiber or chip-compatible spatial-mode interconnects that preserve mode purity and relative phase. \\
\par
\subsubsection*{Outlook}
Transverse modes as quantum states offer more than simply a larger Hilbert space. Specific bases provide intrinsic advantages, such as \qo{self-healing} in Bessel beams or the natural Schmidt basis provided by OAM due to momentum conservation in SPDC. Hybrid encodings that mix multiple DoFs can further enhance robustness and experimental convenience, for example, by enabling easier detection and control through proxy DoFs while retaining HD structure in the spatial domain. The design of spatial-mode generators, sorters, demultiplexers and multiplexers increasingly benefits from machine-learning algorithms and inverse-design photonics. Neural-network-guided diffractive optics have already been used for the manipulation of spatial-modes via quantum gates~\cite{wang2024ultrahigh}, illustrating the potential of generative design and real-time adaptive mode control for their generation and detection.

\subsection{Path-encoding} \label{Sec:GenAndMeas_Path}
\author{Yunhong Ding\authormark{12,*}, Stefano Paesani\authormark{13,$\dagger$}, Caterina Vigliar\authormark{12,$\ddagger$}}

\address{\authormark{12}Department of Electrical and Photonics Engineering, Denmark
Technical University, Ørsteds Plads, Lyngby, 2800, Hovedstaden,
Denmark}
\address{\authormark{13}NNF Quantum Computing Programme, Niels Bohr Institute, University of Copenhagen, Blegdamsvej 17, 2100 Copenhagen, Denmark.}
\email{\authormark{*} yudin@dtu.dk}
\email{\authormark{$\dagger$}stefano.paesani@nbi.ku.dk}
\email{\authormark{$\ddagger$}catvi@dtu.dk}

High-dimensional photonic states, i.e. qudits, are powerful as, for a given number of photons, a larger Hilbert space and better noise performance can be provided. This unique feature leads to quantum communications with larger communication capacity and better noise resilience~\cite{Ding2017}, and more efficient quantum computations, compared to qubit systems~\cite{Chi2022}.

However, hardware overheads are introduced to control these larger spaces. 
In fact, one of the main challenges for high-dimensional photonic quantum technologies is developing optical components that can perform complex operations on the qudits at the scale required for useful applications. 
Using the path degree of freedom of single photons comes as a natural way to encode qudits in waveguide-based photonic systems, mapping the different levels of the qudit to different spatially-separated fiber modes or spatially-separated integrated waveguides in a photonic chip.

When encoding qudits using the single photons' path degree of freedom, the latter approach has seen remarkable developments in recent years,  tackling scalability problems enabled by the high-manufacturability of dense devices in commercial chip foundries.

In the following sections, we will give an overview of path-encoded high-dimensional system realizations, with particular attention to integrated photonic systems. The development of technologies capable of supporting these path-encoded systems has been one of the main workhorses for enabling large-scale high-dimensional state generation, control, and detection, spearheading an increased interest in this research topic.  

\subsubsection*{Encoding qudits in spatially-separated waveguides}

A natural way to encode quantum information on quantum photonics is to use the which-path degree of freedom of the photon. In other words, the state of the qubit or qudit is encoded in which path (i.e., which waveguide) it is travelling through. 
\begin{figure*}
    \centering
    \includegraphics[width=0.85\linewidth]{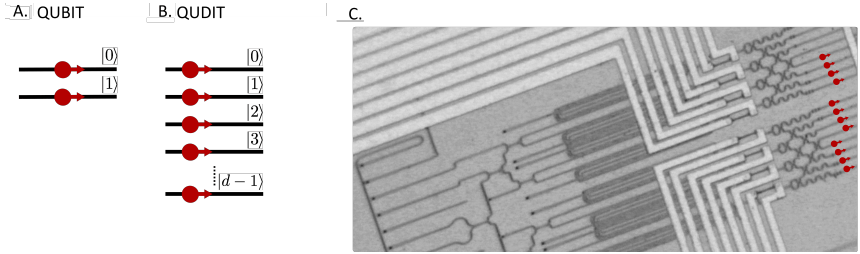}
    \caption{Definition of high-dimensional path encodings. \textbf{A.} A photon propagating in a superposition of two waveguides encodes a qubit. \textbf{B.} A photon propagating in a superposition of $d$ waveguides encodes a $d$-dimensional qudit.\textbf{C.} Microscope picture of an example of a silicon photonic chip that processes multipartite states of ququarts (four-dimensional qudits)~\cite{Vigliar2021}. Spatially separated waveguide modes are highlighted in red.  }
    \label{fig:qudit_encoding}
\end{figure*}
For a qubit, the encoding can be accomplished using a single photon that can travel in two optical waveguides, as illustrated in Fig.~\ref{fig:qudit_encoding}, where the mapping between the logical state of the qubit and the Fock state of the photon is
\begin{align}
    \textbf{Logic}&\textbf{al state} \qquad  \textbf{ Fock state} \nonumber\\ 
    &\ket{0}\ \ \quad\longleftrightarrow  \quad      \ket{1}_{0} \ket{0}_{1}\\
    &\ket{1}\ \ \quad\longleftrightarrow  \quad       \ket{0}_{0} \ket{1}_{1}    
\end{align}
We therefore say that the qubit is in the computational state $\ket{i}$ if the single photon occupies its $i$-th optical waveguide, with $i\in\{0,1\}$. 
This definition can be straightforwardly generalized to encode $d$-dimensional qudits in case the photon can travel through $d$ different optical waveguides. Formally, a photon propagating through $d$ waveguides encodes a $d$-dimensional qudit whose computational state is $\ket{i}$ if the single photon occupies its $i$-th optical waveguide, with $i\in\{0,\ldots,d-1\}$. 
Up to now, path-encoding has been realized on integrated photonic chips for systems with dimensionality up to $d=15$ in the silicon-on-insulator SOI platform~\cite{Wang2018}.
We also note that path-encoding is not limited to photonic integrated chips, but can also be utilized in optical fibers~\cite{Schaeff12:QUNITS}. In particular,  multicore fibers have been a florid platform for investigating path-encoded photonic qudits, where the different cores in the fibers represent the different paths used to encode a photonic qudit~\cite{DaLio21:2KmHDQKD, Ding2017, Marconi24:MCFEntenglement}. 
A technical challenge that has to be considered in path-encoding is maintaining the phase-stability between the different paths of a qudit; the lack of which would introduce decoherence in the qudit state. 
This issue is minimal in integrated photonic circuits, where phase stability between waveguides is intrinsic. 
However, it could be very challenging for fiber-based implementations if separate fibers were used as paths to encode qudits. 
Multicore fibers, where all cores are embedded in the same cladding, offer a good solution for phase stabilizations in fiber-based implementations of path-encoded qudits. Up to now, multi-core fibers with up to 37 cores have been used for encodoing qudits~\cite{bacco2019boosting}, with active phase-stabilization in up to 4 fiber cores~\cite{DaLio21:2KmHDQKD}.

\subsubsection*{Integrated quantum photonic circuits}

Integrated photonics represents a groundbreaking technology where the essential components required to generate, manipulate, and measure light are put together on a common platform.  The field started in the late 1960s, when the first demonstrations of light propagating in waveguides embedded in standard optical materials, such as dielectric organic films and glasses, were reported~\cite{miller1969, marcatili1969}. Thanks to the development of more refined fabrication techniques, integrated photonic circuits can now be fabricated in a wide variety of materials, each providing different properties. These include Complementary Metal-Oxide-Semiconductor (CMOS) compatible silicon-on-insulator (SOI), lithium-niobate (LiNbO$_3$), where fast modulators can be embedded, low-loss materials such as silicon-nitride (SiN), III-V semiconductors (e.g. indium-phospide) allowing laser sources and electro-optics, and silica providing low-loss linear circuits. Since the mid-90s, Silicon has emerged as one of the prominent materials for integrated optics, in particular the SOI technology. In this platform, due to the strong refractive index difference with the substrate, high-confinement waveguides and high-density circuits with intrinsic phase stability are achievable. Moreover, silicon photonics is compatible with the CMOS technology used for silicon microelectronics, meaning not only scalable mass-fabrication of the photonic circuits is possible, but also  hybrid integration of quantum photonics and electronics on the same substrate can be achieved ~\cite{atabaki2018integrating}. Silicon photonic devices are now readily available from commercial foundries, which typically provide the building blocks required for generating and processing photons through standardized process design kits (PDK). We review these building blocks in the following sections. 

\subsubsection*{Integrated waveguides}

\begin{figure*}
    \centering
    \includegraphics[width=0.75\linewidth]{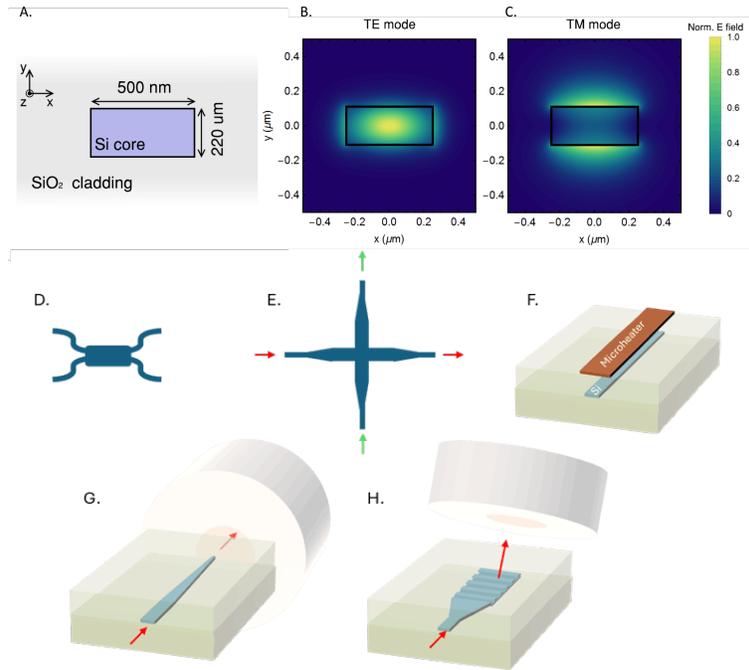}
    \caption{A. Typical geometry of a silicon waveguide with SiO2 cladding, and the corresponding B. TE and C. TM mode profile. Basic linear components of integrated quantum photonics platform, including D. 2x2 multimode interferometer (MMI)-based 3~dB coupler, E. cross-intersection allowing red and green labeled light crossing with each other with low loss and negligible crosstalk, F. microheater-based phase shifter, G. edge coupler, and H. vertical grating coupler.}
    \label{fig:integrated_platform}
\end{figure*}

Waveguides are the optical components that guide photons across a chip. They consist of a core region of a material with refractive index $n_{\text{core}}$ embedded in a cladding of smaller refractive index $n_{\text{clad}}<n_{\text{core}}$. Similarly to optical fibers, light is confined inside the core by total internal reflection. While many types of waveguide geometries exist, the standard is represented by the \textit{strip waveguide} geometry shown in Fig.~\ref{fig:integrated_platform}A, where the core is fully embedded in the cladding and the confinement is maximized. 
Light travels along the waveguide ($z$ direction) according to Maxwell's equations. The transverse field can be described in terms of discrete modes of electric field profiles $A(x,y)$, given by the solutions of the Helmoltz equation~\cite{chang2009}.

For a rectangular strip waveguide, we can define quasi-transverse electric (TE) and quasi-transverse magnetic (TM) modes as those having the electric field principally polarized along the $x$ and $y$ axis, respectively. Single-mode waveguides support only the fundamental modes of transmission, i.e. the TE$_0$ and TM$_0$. In Fig.~\ref{fig:integrated_platform}~B-C examples of these modes for a typical SOI single-mode strip waveguide are shown.

Waveguides support low-loss transmission only in the wavelength range of the material. Small-bandgap semiconductor materials, such as silicon, can have strong absorption in the visible or near infrared region, but become transparent to wavelengths longer than a micron. In such cases it is convenient to operate in the telecom band ($\simeq1550$ nm), where losses of $3$~dB/cm in SOI waveguides are typically recorded. Broader transmission windows and lower losses can be achieved in amorphous materials or large-bandgap semiconductors, for example SiN~\cite{Zhao2025SiN}, LiNbO$_3$~\cite{Zhu2024LN}, and Silica~\cite{Lee2012NCSiO2wg}.

\subsubsection*{On-chip photon sources}
In integrated high-dimensional path-encoded photonic systems, recent demonstrations have seen quantum light directly generated on-chip in the form of the single-photon sources (SPSs) or heralded photon-pair sources (HPPSs). The ideal single photon source is on-demand (close to unity probability), pure, and indistinguishable. The performance of quantum light sources directly determines the error rates, the fidelity, and the scalability of the high-dimensional quantum photonic system. Heralded photon-pair sources based on spontaneous nonlinear process, e.g. spontaneous parametric down-conversion (SPDC) using second-order nonlinearity, and spontaneous four-wave mixing (SFWM) using third-order nonlinearity, have been widely used in exploring large-scale high-dimensional quantum photonic technologies. Specifically, HPPS come with the advantage of working at room temperature, excellent indistinguishability with coherent pumping, and good compatibility with integrated photonics~\cite{AdockJSTQE2021}. The major characteristics of nonlinear HPPSs are pair generation rate (PGR), coincidence-to-accidental ratio (CAR), and spectral purity, which determine the data rate as well as the visibility of quantum interference. SFWM-based HPPSs, consisting of simple silicon straight or spiral waveguides~\cite{Wang2018}, have been widely used to achieve high indistinguishability and purity in combination with narrow spectral filtering. The filtering, however, significantly reduces the brightness. This drawback can be significantly improved by using ring resonators, which enhance the SFWM process and localize the efficient photon pair generation only at the resonance wavelengths, leading to high PGR and spectral purity~\cite{silverstone2015qubit, vernon2017amzimrr, Llewellyn2020}. High coincidence rates up to 23~kcts/s were achieved with 750~$\mu$W pump power. At a spectral purity of 0.92, an enhancement of the pair generation rate of about 230 times can be achieved compared to waveguide HPPS~\cite{Llewellyn2020}. However, due to random phase errors introduced during the fabrication process, ring resonators are typically required to be tunable to achieve high indistinguishability among each HPPSs. Recent proposals are trying to circumvent this issue for ring resonators, e.g., through using multiple passive ring resonators coherently and passively coupled together, which could potentially enable very high purity and indistinguishability without active alignment of the ring cavities~\cite{Alexander2025}.  Furthermore,  improved proposals for waveguide-based SFWM sources which have also been recently demonstrated, such as the inter-modal SFWM-based HPPSs based on TM modes~\cite{paesani2020near}. This approach has been successfully demonstrated to generate photon pairs at the engineered phase-matching window, which can be precisely controlled with high-precision nano-fabrication, making tuning-free HPPSs possible. A more detailed review of silicon non-linear HPPSs is elaborated in~\cite{AdockJSTQE2021}. 
Despite the advantages of non-linear HPPSs, attaining single-photon generation by heralding of one of the photons in the pair requires a low generation probability ($<$5\%) to ensure negligible multiphoton contaminations. An approach that can in principle turn multiple HPPSs into a near-deterministic photon source is multiplexing, which however typically requires significant hardware overheads~\cite{Ekici2025TemporalMultiplexingTFLN}.  The challenge of building low-loss multiplexing systems has so far posed significant limitations to scaling quantum photonic systems based on HPPSs. Possible alternatives to overcome this challenge are approaches based on single-photon sources that use quantum emitters, such as quantum dots (QDs) and color centers, systems that are currently widely investigated~\cite{Sund2023_LNOI_processor, Buzzi2025_StrainTuningSiCenters}.

\subsubsection*{Single-photon processing components}
On-chip quantum information processing is typically realized by elaborated photonic circuits consisting of tens to hundreds of optical components. Basic building blocks include 50:50 beam-splitters (BS, Fig.~\ref{fig:integrated_platform}.D), polarization-managing components, waveguide cross-intersections (Fig.~\ref{fig:integrated_platform}.E), microheater-based phase shifters (Fig.~\ref{fig:integrated_platform}.F), fiber-to-chip couplers (Fig.~\ref{fig:integrated_platform}.G-H), pump rejection filters, as well as high-speed modulators/switches. Scaling-up quantum photonic systems for high-dimensional system processing requires components with utmost performance, in particular with low optical loss and high fidelity. 
Fiber-to-chip couplers are essential for integrated quantum photonics to efficiently couple pump light to the chip, and to couple photons out to high-efficiency single photon detectors (SPDs).
\label{sec:GenPathSinglePhotonProcessing}
Edge couplers typically provide wide coupling bandwidths and high coupling efficiency using efficient spot size converters with smart tapering techniques (typical insertion losses of 1~dB per coupler, even in packaged configurations~\cite{PU20103678Taper, Ben10PTLcoupler, Chen2010Coupler}.) Grating couplers are able to couple light vertically to single-mode fibers, which is very convenient for wafer-level testing and packaging. However, their coupling bandwidth and coupling efficiency can be limited, due to power leakage loss to the substrate. Introducing mirrors, e.g. aluminum~\cite{ding2014fully, Zaoui2014GCmirror, Benedikovic2015GCmirror, HoppeJSTQE20GCmirror} or Bragg gratings~\cite{Nambiar2019GCDBR}, below the grating couplers could greatly enhance the coupling efficiency. Recently, an efficient design based on topological unidirectional guided resonance has demonstrated an ultra-low coupling loss of $0.34$~dB ~\cite{wang2014topo}. Table~\ref{tab:losses} summarizes the state-of-the-art performances for all the key integrated optical components.

\begin{table}[t!]
\begin{adjustbox}{center}
\begin{tabular}{p{4cm}	p{1.5cm}   p{0.7cm}  p{0.7cm} }
Component	& 	Loss (dB)  & (\%)	  &    Ref.			\\
\noalign{\vskip 1mm}    
\hline
\hline

\noalign{\vskip 1mm}
Crosser   &	$0.02$	&   $99.5$\%  & \cite{Zhang2013blochCross}\\
\noalign{\vskip 1mm}
$2\times2$ MMI coupler            &	$0.2$    &  	$96$\%  & \cite{Dumais2016OFCmmi}\\

\noalign{\vskip 1mm}
Edge coupler (SMF)	        &	$0.12$      & 	$97$\%  & \cite{Alexander2025}\\
Edge coupler (tapered fiber)&	$0.3$      & 	$93$\%  & \cite{PU20103678Taper}\\

\noalign{\vskip 1mm}
Grating coupler (1200 nm)    	    &	$0.36$     &  	$92$\%  &\cite{notaros2016ultra}\\
Grating coupler (metal mirror)    	    &	$0.5$     &  	$89$\%  & \cite{HoppeJSTQE20GCmirror}\\
Grating coupler (metal mirror)    	    &	$0.6$     &  	$87$\%  & \cite{ding2014fully}\\
Grating coupler (topological design)    	    &	$0.34$     &  	$92$\%  & \cite{wang2014topo}\\

\noalign{\vskip 1mm}
Waveguide, Si (m$^{-1}$)                  &	$2.7$     & 	$54$\%  & \cite{Biberman2012Siwg}\\
Waveguide, SiN (m$^{-1}$)                &	$0.6$    &  	$54$\%  & \cite{Zhao2025SiN}\\
Waveguide, TFLN (m$^{-1}$)                &	$1.3$    &  	$54$\%  & \cite{Zhu2024LN}\\
Hybrid Si/SiN delay line (m$^{-1}$)      &	$0.12$   &  	$97$\%  & \cite{Puckett2019SiN}\\
Hybrid Si/silica delay line (m$^{-1}$)   &   $0.037$    &  	$99$\%  & \cite{Lee2012NCSiO2wg}\\

\end{tabular}
\end{adjustbox}
\caption{\textnormal{State-of-the-art losses of integrated photonic components. Thermo-optic and Pockels-based modulator losses approach linear waveguide losses, provided that metal elements are sufficiently separated from the waveguide mode. Losses in $2\times2$ directional couplers would also follow linear waveguide loss by weak coupling, which would be much smaller than $2\times2$ MMIs.}}
\label{tab:losses}
\end{table}

\subsubsection*{Single-photon detectors}
\label{sec:GenPathSinglePhotonDetectors}
Single photon detectors are essential components for implementing measurements in quantum photonic systems. Conventionally, InGaAs detectors offer economic solutions~\cite{Namekata09APD, Zhang15SPD} with MHz count rate capability, low detection efficiency of ~10\% and dark-count rate of a few hundred Hz. However, for large-scale high-dimensional quantum photonic system demonstrations, requiring the capability of measuring multiple coincident single-photon events, ultra-high detection efficiency becomes paramount. Superconducting nanowire single-photon detectors (SNSPDs), working at cryogenic temperatures (typically lower than $4$ K) offer excellent performances~\cite{Natarajan2012SNSPD, Rosenberg13SNSPD}. SNSPDs with efficiencies of up to 95.5\%, dark count rates of less than less than 100 Hz, and recovery times which enable more than $10^7$ detections per second are commercially available~\cite{Zhang17SNSPD, QuantumOpus, SingleQuantum, CNPhotec, Pixel, photonspot, IDQuantique}. Such off-chip commercial SNSPD systems have significantly pushed the progress of modern large-scale integrated quantum photonics experiments, but intrinsically rely on efficient fiber-to-chip couplers, leading to inherent sources of single-photon loss. Moving forward, on-chip integration of SNSPD systems is required to further scale-up quantum systems. SNSPDs have been successfully integrated on silicon waveguides in a travelling-wave configuration~\cite{pernice2012high, schuck2013waveguide} with close to unity efficiency \cite{akhlaghi2015waveguide}, demonstrating great potential for integrated quantum photonics systems. SNSPDs have been transfer-printed onto silicon nitride films onto silicon waveguides, successfully performing on-chip $g^{(2)}$ experiments with integrated beamsplitters~\cite{najafi2015chip}. On-chip SNSPDs have also been investigated on a thin-film lithium niobate (TFLN) photonics platform, where a niobium nitride (NbN) nanowire on a TFLN waveguide exhibits 46\% on-chip detection efficiency and 32~ps timing jitter~\cite{sayem2020lithium}. Niobium Titanium Nitride (NbTiN) nanowires have also been integrated on a TFLN waveguide, showing  27\% on-chip detection efficiency and 17~ps jitter~\cite{lomonte2021single, prencipe2023wavelength}. Furthermore, Molybdenum silicide (MoSi) nanowires on TFLN have demonstrated 50\% on-chip detection efficiency and 82~ps jitter at 1550~nm~\cite{colangelo2024molybdenum}.

\subsubsection*{Programmable circuits for processing qudits}
\begin{figure*}[ht!]
    \centering
    \includegraphics[width=0.85\linewidth]{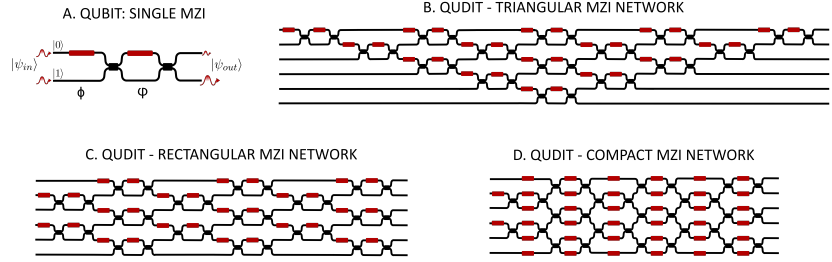}
    \caption{Mach-Zehnder interferometer networks for realizing universal qubit/qudit processing in the path degree of freedom. \textbf{A.} Single-qubit arbitrary state transformations can be realized by a combination of two beam-splitters (black) and two phase shifters (red). \textbf{B.} Example of a six-mode triangular MZI network as proposed by Reck et al.~\cite{Reck_1994} \textbf{C.} Example of a six-mode rectangular MZI network as proposed by Clements et al.~\cite{Clements:16} \textbf{D.} Example of a six-mode simplified rectangular MZI network as proposed by Bell et al.~\cite{Bell21:CompactifyLO}.}
    \label{fig:qudit_processing}
\end{figure*}
To render photonic qudits practical for quantum information processing, it is essential to identify a universal gate set within the which-path encoding framework described in Section I; that is, a collection of gates capable of approximating any arbitrary unitary transformation to an arbitrary degree of precision through a quantum circuit~\cite{WangHuSandersKais2020}. Achieving universality requires the implementation of single-qudit gates in combination with at least one two-qudit entangling operation, such as the controlled-NOT (CX) or controlled-phase (CZ) gate, which may be realized via measurement-induced nonlinearities~\cite{calsamigliaGeneralizedMeasurementsLinear2002, Grice2011, Bacco21:ProposalHDBSM, Bianchi25:NLHDBellMeas}. Control and manipulation of these qudits can be accomplished using appropriate arrangements of linear-optical elements, such as beam-splitters and phase shifters, together with precisely engineered single-photon detection schemes~\cite{Wang2018, Vigliar2021, Chi2022}. Since integrated photonic circuits inherently provide stable phase relationships across multiple optical paths, most of the theoretical studies on the realization of arbitrary unitary transformation on $d$ optical modes have been driven by large-scale implementations of these operations in integrated optical platforms, such as silicon/silica/silicon nitride photonic chips~\cite{Carolan15, Ding2017, Wang2018, Taballione2023, Thomas2025:HDDistr}.

\subsubsection*{Single-qudit processing circuits}
Unitary transformations on a single qudit can be realized by extending the method for realizing path-encoded arbitrary \textit{single-qubit} gates with linear optics. Unitary transformations of a single qubit encoded by a photon placed in a superposition of two optical waveguides can be implemented by decomposing them into three rotations along the $\hat{Z}$, $\hat{Y}$ and again $\hat{Z}$ axes of the qubit space: such rotations can be physically realized by combining two beam-splitters and three phase-shifters to form Mach-Zehnder Interferometers (MZIs). If one is not interested in reading out the final relative phase between the modes, two beam-splitters and two phase-shifters are sufficient, as depicted in Fig.~\ref{fig:qudit_processing}.A~\cite{nielsenchuang}.
Any unitary transformation on a single qudit, encoded by a photon placed in a superposition of $d$ optical waveguides, can be implemented by universal optical interferometric meshes, as firstly proposed by Reck et al.~\cite{Reck_1994, Carolan15}. These take the form of a $d \times d$ multi-port interferometer, built up from a triangular network of two-mode Mach-Zehnder transformations, containing a variable phase shifter in one path, as well as an external phase shifter (see Fig.~\ref{fig:qudit_processing}.B). Any measurement of observables that correspond to discrete Hermitian matrices can also be experimentally realised in this way. To do so, it is required to implement the unitary that relates the eigenbasis of the Hermitian operator to the single-mode occupation basis (or computational basis as defined in section II), by tuning the variable phase shifter elements. The output can be measured with an array of $d$ single-photon detectors, one per optical mode, each corresponding to an orthogonal eigenstate of the Hermitian matrix. The maximum number of MZIs required is $d(d-1)/2$, which is quadratic in the number of modes, i.e. quadratic with the dimension of the encoded qudits. This decomposition has the same number of parameters needed to fully describe a matrix on $d$ modes. An alternative decomposition for achieving universal unitary transformations on $d$ modes with linear optics was subsequently proposed by Clements et al.~\cite{Clements:16}, which consists of a rectangular network of Mach-Zehnder transformations, as shown in Fig.~\ref{fig:qudit_processing}.C. Such a rectangular scheme improves on the triangular scheme in terms of circuit depth, i.e. the number of optical components each photon passes through. Subsequent studies on further compactifying such interferometric meshes have been brought forward using symmetric Mach Zehnder interferometers, requiring only a small number of external phase-shifters that do not contribute to the depth of the circuit~\cite{Bell21:CompactifyLO}. This results in a significant saving in the depth of these schemes, allowing more complex circuits to fit into the scale of photonic chips, and reducing the propagation losses associated with these circuits. Similar savings can be made in alternative schemes which can also provide robustness to fabrication imperfections, such as imbalanced beam-splitters and imbalanced losses~\cite{Saygin2020_RobustUniversalUnitaries, Tsakyridis2022_FidelityRestorableLinearOptics}.

Although not directly linked to high-dimensional realizations, the earliest fully programmable integrated universal interferometers implemented arbitrary $6\times6$ unitaries in silica platforms using triangular MZIs meshes and thermo-optic phase shifters~\cite{Carolan15}, establishing the feasibility of on-chip universal linear optics. Multiple variants of six to twelve-mode processors have since been reported, including devices optimized for specific wavelength ranges and advanced calibration techniques~\cite{Hoch2024, Crespi16, Giordani2023_context, Caruccio2025, deGoede2022_12modeSiN_940nm, perez2017silicon}. Subsequent work in silicon nitride platforms significantly reduced propagation and excess losses, enabling larger meshes: a major milestone was the demonstration of a fully reconfigurable twelve-mode universal photonic processor, capable of implementing arbitrary $12\times12$ unitaries with high fidelity~\cite{Taballione_2021}. 
The largest silicon nitride universal interferometer demonstrated to date operates on twenty spatial modes and employs a Clements-type rectangular architecture comprising roughly 190 MZIs and several hundred thermo-optic phase-shifters~\cite{Taballione2023}. This device achieved multi-percent end-to-end loss with high average fidelity for Haar-random unitaries, representing the state-of-the-art on a mainstream integrated platform.
Looking more specifically at qudit-oriented demonstrations, a sixteen-dimensional triangular network of MZIs and phase-shifters
was implemented in~\cite{Wang2018}, which allowed for performing arbitrary local projective measurements. In this scheme, the measurement outcomes on a specific basis were collected one by one by rotating the qudits' reference frame and using one detector per photon: the collection of the $d^2$ outcomes thus required a total of $d^2$ measurement settings.
Beyond planar silicon-nitride platforms, femtosecond laser-written glass technology has recently enabled even larger mode counts by exploiting three-dimensional routing. Using this approach, a 24-mode universal photonic processor based on an MZI mesh was demonstrated, representing the largest experimentally realized universal linear interferometer to date~\cite{barzaghi2025lowloss24modelaserwrittenuniversal}. Together, these results define the current practical limits of universal MZI networks in integrated photonics.

\subsubsection*{Multi-qudit entangling gates}
For photonic qudits encoded in spatial paths, entanglement is most commonly generated using measurement-induced schemes, as single photons exhibit negligible direct nonlinear interactions. In this approach, independently prepared path-encoded qudits interfere in multiport linear-optical circuits composed of beam splitters and phase shifters, similar to the ones described in the previous section, followed by joint projective measurements that erase which-path information. Conditional on specific detection outcomes, the measurement projects the photons into specific entangled high-dimensional states. This principle is equivalent to qubit-based linear-optical quantum computing, where multi-mode interferometers allow Bell-state measurements, which effectively enable two-qubit gates through post-selection and feed-forward~\cite{Knill2001}. However, in contrast to qubits, as proven by several theoretical studies~\cite{calsamigliaGeneralizedMeasurementsLinear2002, Grice2011, Bacco21:ProposalHDBSM, Bianchi25:NLHDBellMeas}, complete high-dimensional Bell-state measurements are not possible using only linear-optical elements and postselection, but require the introduction of ancillary photons. Moreover, most schemes achieve very low success rates, in the form of a post-selection penalty on the multi-qudit state generation rates, which can vary depending on the number of ancillary photons and modes introduced, and on their entanglement nature. Such schemes naturally integrate with universal linear-optical networks based on MZI meshes, allowing entanglement generation to be embedded within programmable photonic processors.
While such protocols are highly probabilistic, advances in photon-number-resolving detectors, multiplexing, and feed-forward control have made measurement-induced entanglement a cornerstone technique for scalable photonic and hybrid qudit-based quantum information processing~\cite{rudolph2017, ErhardKrennZeilinger2020}.


\subsubsection*{Generation of path-encoded qudits}

\begin{figure*}[t!]
    \centering
    \includegraphics[width=1\linewidth]{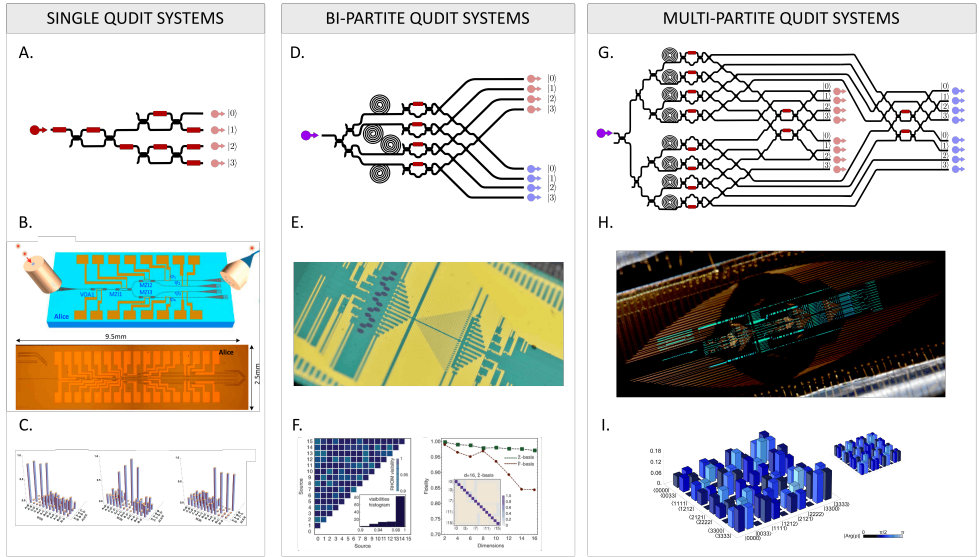}
    \caption{Demonstration of the generation and manipulation of path-encoded qudits from single-qudit systems to multipartite systems. \textbf{A.-C} Four-dimensional weak-coherent state distribution across two silicon photonic chips for Quantum Key Distribution ~\cite{Ding2017}. \textbf{D.-F} Two-photon 15-dimensional GHZ state generation from an array of 15 HPPs on a silicon photonic chip in postselection for a variety of computing demonstrations ~\cite{Wang2018}. \textbf{G.-I} Four-photon four-dimensional entangled state generation from an array of 8 HPPs on a silicon photonic chip with post-selected entangling gates for error-protected computations~\cite{Vigliar2021}.}
    \label{fig:qudit_generation}
\end{figure*}

The simplest high-dimensional path-encoded quantum state, a single qudit, can be realized by placing a single photon or weak-coherent pulses in a superposition of $d$ spatial modes. This is commonly achieved by splitting light through cascaded beam-splitters with variable reflectivity and tunable phase modulators to prepare different basis states~\cite{Ding2017, DaLio21:2KmHDQKD, Genzini24:HDContext}. In practical implementations for such simple systems, weak-coherent states provide an experimentally simple method for encoding information, using fast modulation of an attenuated continuous-wave laser with a very low mean photon number distributed across multiple integrated waveguides~\cite{Ding2017, Genzini24:HDContext} or the distinct cores of a multicore fiber~\cite{DaLio21:2KmHDQKD, Ding2017, Zahidy2024:HDQKDAquila}. Although weak-coherent pulses are not true single-photon sources, operating at sufficiently low photon numbers allows them to approximate single-photon behaviour while remaining compatible with standard fiber-based telecommunications components. As a result, these states are well suited for high-dimensional quantum communication and quantum key distribution, offering increased information capacity and enhanced robustness to noise~\cite{Zahidy2024:HDQKDAquila, DaLio21:2KmHDQKD}. Alternative approaches for the generation of qudits in exact single-photon Fock states have been proposed, for instance, using arrays of non-interacting atoms or quantum emitters~\cite{bell2022protocol}.

The engineering of multipartite entangled qudit systems in the path degree of freedom can be only achieved by precise phase control and stability between spatial modes across many photons~\cite{Schaeff12:QUNITS, Hu20:MultipathHD}. Consequently, experimental realizations in this domain have closely followed advancements in quantum photonic device control (see Fig.~\ref{fig:qudit_generation}). A major milestone was the development of an integrated photonic chip capable of generating, manipulating, and measuring two-photon multidimensional GHZ states, $\sum_{k=0}^{d-1} \ket{k}_1\ket{k}_2/\sqrt{d}$, with demonstrated dimensions up to $d=15$~\cite{Wang2018}. This platform showed that the generation of path-encoded entangled states, implementation of universal operations, and execution of arbitrary multidimensional projective measurements on two path-encoded qudits are feasible using pairwise entanglement produced via spontaneous four-wave mixing (SFWM) and linear optical components. The high fidelity of these operations enabled the verification of multidimensional quantum correlations and the implementation of advanced quantum information protocols. It also offered a clear pathway towards extending such architectures to multipartite systems involving many photons in $d$ dimensions.
A pivotal advancement towards multipartite path-encoded qudit states was achieved in a subsequent study~\cite{Vigliar2021}, which, for the first time, addressed the realization of qudit–qudit entangling gates through post-selection, although not fully reconfigurable in the degree of entanglement. This work successfully verified a four-photon, four-dimensional entangled state, enabling measurement-based protocols equivalent to an eight-qubit Hilbert space.

Scaling up such multipartite states to more than two parties with arbitrary levels of entanglement has been proven to be experimentally demanding, given the large resource overhead needed for entangling separate qudit pairs through Bell-state generalized measurements, as discussed in the previous section~\cite{calsamigliaGeneralizedMeasurementsLinear2002, Grice2011, Bacco21:ProposalHDBSM, Bianchi25:NLHDBellMeas}. Only in very recent years, rapid technological progress in on-chip photonics enabled full initialization, manipulation, and measurement of qudit states and logic gates in highly controllable and programmable architectures using ancillary single photons~\cite{Chi2022}. These developments position integrated photonic quantum technologies as a leading platform for qudit-based quantum computation and communication protocols, offering enhanced capacity, accuracy, and efficiency in both their native $n$-dimensional encoding and in their conversion to binary encodings~\cite{Vigliar2021, Chi2022, Huang2024}.

\subsubsection*{Distribution of path-encoded qudits}
\label{sec:GenAndMeas_PathDistribution}

Distribution of path-encoded photonic qudits is a critical ingredient for scalable and modular quantum photonic architectures for communication or computing applications, enabling the interconnection of photonic processors while preserving access to high-dimensional Hilbert spaces. When using path encoding, the faithful distribution across processors, usually embodied by integrated devices, requires preservation of relative phases, mode orthogonality, and balanced transmission among all paths. These requirements are substantially harder to achieve than in qubit-based schemes, as both loss imbalance and phase noise complexity scale unfavorably with increasing dimensionality and directly reduce the fidelity of high-dimensional quantum states~\cite{ErhardKrennZeilinger2020}.

The most widely adopted method for chip-to-chip path-encoded qudit distribution relies on fiber-array interconnects, where each path mode is coupled to an independent fiber, offering maximal flexibility in routing and straightforward compatibility with standard fiber components and on-chip couplers. This approach is platform-agnostic and compatible with bulk-optics, silicon, silicon nitride, and silica photonic integrated circuits, and has enabled experimental demonstrations of distributed path-encoded states and entanglement for moderate dimensions (up to $d=4$) using bulk or integrated multiport interferometers~\cite{Thomas2025:HDDistr, LanyonEtAl2009}. However, independent fibers exhibit uncorrelated thermal and mechanical fluctuations, leading to a differential phase drift that scales unfavorably with the length of the fibers and the number of paths. These sources of noise typically necessitate advanced active phase stabilization routines (phase-locked loops involving possible additional laser resources) for verifying coherent high-dimensional operation.  Moreover, polarization fluctuations and unequal coupling efficiencies can introduce additional sources of errors and path-dependent losses, necessitating advanced calibration techniques for coherent control~\cite{DaLio21:2KmHDQKD, Zahidy2024:HDQKDAquila}.

A second possible interconnect strategy involves the use of more advanced fiber systems, such as multi-core or multi-mode fibers, in which the spatial modes composing the qudit travel together and experience similar channel losses and phase fluctuations on slower timescales~\cite{Marconi24:MCFEntenglement, Canas2017_HDDistib_MCF}. Notably, a series of experiments established fiber-based distribution of path-encoded qudits using multi-core and few-mode fiber systems, demonstrating robust transmission and manipulation of high-dimensional quantum states over km-scale fiber links, using relaxed phase stabilization techniques~\cite{Canas2017_HDDistib_MCF, Ding2017, DaLio21:2KmHDQKD, Zahidy2024:HDQKDAquila}. The main technical challenges associated with multi-core-based qudit distribution include fabrication-imposed constraints on core spacing, residual inter-core cross-talk, and coupling complexity when interfaced with integrated devices. Nevertheless, multicore fibers provide a particularly attractive route for scalable, phase-stable distribution of path-encoded qudits and constitute a promising interconnect technology for future multi-chip qudit-based photonic quantum networks.

Alternative possible routes for coherent qudit transport rely on the conversion between path-encoding and other degrees of freedom such as polarization and/or time-bin. Qudit distribution via conversion from the path to the polarization degree of freedom, could be be realized generalizing the use of two-dimensional (2D) subwavelength grating couplers, using techniques borrowed from qubit quantum state distribution for chip-to-chip entanglement swapping and teleportation ~\cite{Wang2016_chip2chip, Llewellyn2020, Yu2025ChipToChip}. In these works, coherence between separate silicon photonic chips can be preserved by combining path-encoded spatial modes with polarization and interferometric stabilization, enabling quantum information encoded on one chip to be transferred to and reconstructed on another chip through measurement-induced entanglement and classical feed-forward. Related experiments have exploited hybrid path–polarization encodings to distribute quantum states through free-space or fiber networks, benefiting from polarization as a compact auxiliary degree of freedom ~\cite{Ciampini2016, Ciampini2017nonlocality}. In parallel, other experiments have enabled path to time-bin conversion and vice versa, providing an interface between spatially encoded states on chip and time-bin encodings favorable for low-loss and high-speed integrated platforms such as silicon nitride or thin-film lithium niobate chips~\cite{Liu2025_chip2chip, Ren2025_TimeBinToPath_TFLN}. While all such demonstrations operate at the qubit level, the underlying conversion methods could naturally generalize to higher-dimensional path-encoded qudits.

Overall, although chip-to-chip distribution of path-encoded qudits has so far been convincingly demonstrated only for relatively small dimensionalities and short distances, extension to higher-dimensional systems and/or longer distances appears realistic. This outlook is supported by numerous demonstrations of long-distance, actively-stabilized distribution of two-dimensional photonic quantum states in multiplexed architectures, such as schemes employing multiple cores of a single multicore fiber for parallel qubit transmission, as well as experiments based on path–polarization and path–time-bin conversion~\cite{bacco2019boosting, Zahidy2024:HDQKDAquila, DaLio21:2KmHDQKD}. Collectively, these works indicate substantial progress in ultra-low-loss fiber interconnects, improved passive phase stability, and the integration of active stabilization and control mechanisms, all of which are essential for scaling the distribution of high-dimensional path-encoded quantum states.

\subsubsection*{Acknowledgments}
This work has received funding from the European Union’s HORIZON-CL4-2021-DIGITAL-EMERGING-01 programme under the PROMETHEUS project (Grant Agreement No. 101070195), and from the European Union’s Horizon Europe Research and Innovation Programme under the QPIC 1550 project (Grant Agreement No. 101135785). S.P. acknowledges funding from VILLUM FONDEN (MapQP, No. VIL60743), the European Research Council (ERC StG ASPEQT, No. 101221875), Danmarks Innovationsfond research grant No. 4356-00009B (HyperTenQ), and funding support from the NNF Quantum Computing Programme.

\subsection{Temporal modes}
\label{sec:PulseModes}

\author{Benjamin Brecht,\authormark{14,*} Christine Silberhorn,\authormark{14}}
\address{\authormark{14}Paderborn University, Integrated Quantum Optics, Institute for Photonic Quantum Systems (PhoQS), 33095 Paderborn, Germany}
\email{\authormark{*}benjamin.brecht@uni-paderborn.de} 

\subsubsection{Early work}
Photonic temporal modes (TMs) are broadband wavepacket modes. They are an appealing basis for quantum information science due to their natural compatibility with single mode fiber, their robustness against common distortions such as fiber dispersion, and their high information packing density. Recently, they have generated an increasing interest due to breakthroughs in controlled TM generation, manipulation, and detection \cite{brecht_photon_2015, ansari_tailoring_2018, raymer_temporal_2020, forbes_progress_2025}. While TMs are used in continuous variable quantum optics applications---see, e.g., works by the groups of Nicholas Treps and Valentina Parigi \cite{pinel_generation_2012, roslund_wavelength-multiplexed_2014, walschaers_entanglement_2017, arzani_versatile_2018, walschaers_practical_2020, fabre_modes_2020, ra_non-gaussian_2020, cimini_neural_2020, renault_experimental_2023, kouadou_spectrally_2023, henaff_optical_2024, roman-rodriguez_multimode_2024, tripier-mondancin_optimal_2025}---this chapter of the roadmap will primarily focus on discrete variable quantum optics applications to remain concise. 

\begin{figure}
    \centering
    \includegraphics[width=\linewidth]{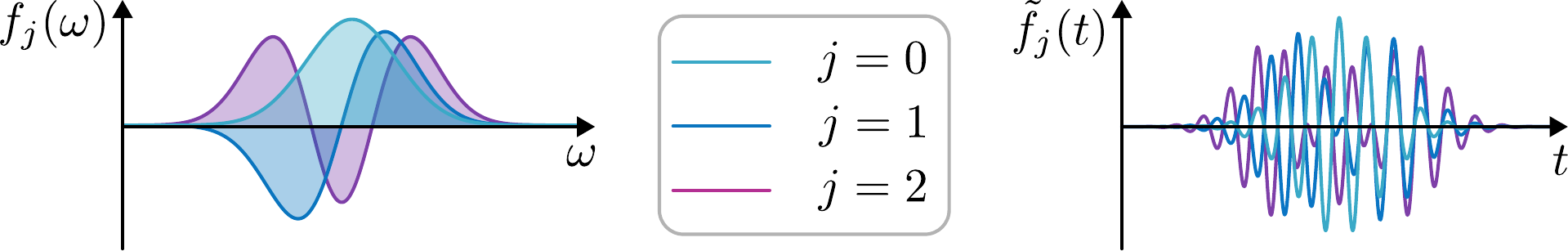}
    \caption{Schematic visualization of the spectral amplitudes $f_j(\omega)$ (left) and temporal profiles $\tilde{f}_j(t)$ (right) of the first three modes of a Hermite-Gauss TM basis. TMs are field-orthogonal although their spectral and temporal intensities overlap.}
    \label{fig:tm_fig_01}
\end{figure}

TMs are characterized by broadband spectral or temporal amplitude functions $ f(\omega) $ and $ \tilde{f}(t) $. A single-photon state occupying a specific temporal mode $ j $ can be expressed as \cite{brecht_photon_2015}
\begin{equation}
    |A_j\rangle = \int \frac{d\omega}{2\pi} f_j(\omega) \hat{a}^\dagger(\omega)|0\rangle = \int dt\, \tilde{f}_j(t) \hat{A}^\dagger(t)|0\rangle \equiv \hat{A}_j^\dagger|0\rangle,
\end{equation}
where $ \hat{A}_j^\dagger $ is the broadband photon creation operator corresponding to the temporal mode $ f_j(\omega) $. This operator creates a single photon in the mode defined by the spectral amplitude $ f_j(\omega) $, or equivalently, the temporal profile $ \tilde{f}_j(t) $, see Fig.~\ref{fig:tm_fig_01}.

Like spatial modes, temporal modes (TMs) are field-orthogonal but exhibit overlapping support in time and frequency. This allows denser packing in the time-frequency domain compared to quasi-orthogonal bases---such as discrete time/frequency-bins---that require guard bands. As a result, TMs are a promising basis for high-dimensional quantum information and multiplexed photonic systems. A key challenge remains the need for specialized hardware for their generation and detection, typically requiring precise dispersion engineering in nonlinear media.

\subsubsection*{Temporal mode generation}
TM states are often generated via the process of parametric down-conversion (PDC). While there are other processes such as four-wave mixing, we will concentrate on PDC to be concise.  

\begin{figure}
    \centering
    \includegraphics[width=\linewidth]{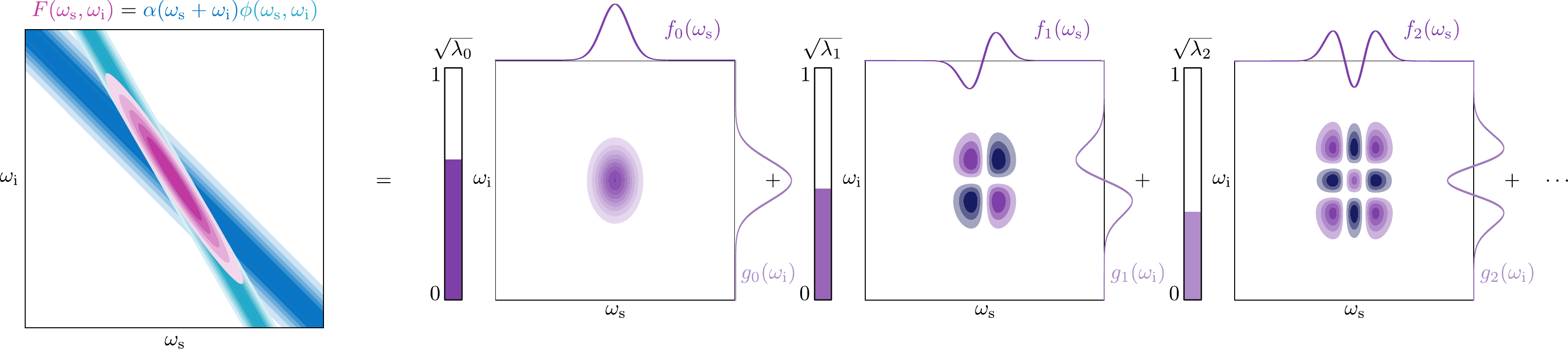}
    \caption{The joint spectral amplitude $F(\omega_\mathrm{s},\omega_\mathrm{i})$ (purple) is the product of the pump envelope $\alpha$ (blue) and phase matching function $\phi$ (teal). It is decomposed into pairs of biorthogonal Schmidt modes $\{f_j(\omega_\mathrm{s})\}$, $\{g_j(\omega_\mathrm{i})\}$ with weights given by the Schmidt coefficients $\{\sqrt{\lambda_j}\}$. The visualization shows the first three terms of the decomposition ($j=0,1,2$). }
    \label{fig:tm_fig_02}
\end{figure}

In 1997, Walmsley and Grice characterized the spectral structure of photon pairs from broadband-pumped PDC, showing that spectral correlations arise from the interplay between energy conservation and phase matching in the nonlinear crystal \cite{grice_spectral_1997}. In 2000, Law, Walmsley, and Eberly applied the Schmidt decomposition to PDC states \cite{law_continuous_2000}, leveraging a formalism introduced by Ekert and Knight \cite{ekert_entangled_1995} for quantifying entanglement. The joint spectral amplitude
\begin{equation}
    F(\omega_\mathrm{s},\omega_\mathrm{i}) = \alpha(\omega_\mathrm{s}+\omega_\mathrm{i}) \phi(\omega_\mathrm{s},\omega_\mathrm{i}),
\end{equation}
which combines the pump envelope $\alpha$ and phase-matching function $\phi$, can be decomposed as
\begin{equation}
    F(\omega_\mathrm{s},\omega_\mathrm{i}) = \sum_{j=0}^{\infty} \sqrt{\lambda_j} f_j(\omega_\mathrm{s}) g_j(\omega_\mathrm{i}),
\end{equation}
yielding biorthogonal TMs---the so-called Schmidt modes---$\{f_j\}$, $\{g_j\}$ and Schmidt coefficients $\{\sqrt{\lambda_j}\}$. This is schematically shown in Fig.~\ref{fig:tm_fig_02}. The full PDC state contains not only photon pairs but higher-order photon-number contributions. In fact, type II PDC, in which pair photons are generated in distinguishable field modes, is nothing but two-mode squeezing. A multimode PDC state can then be written as a tensor product of two-mode squeezed states
\begin{equation}
    |\psi\rangle_\mathrm{PDC} = \bigotimes_{j=0}^{\infty} \exp\left[r_j \hat{A}_j^\dagger \hat{B}_j^\dagger + \mathrm{h.c.}\right] |0\rangle,
\end{equation}
with $r_j = \sqrt{\lambda_j} r$, and $\hat{A}_j^\dagger$, $\hat{B}_j^\dagger$ creating photons in the $j$-th temporal mode. The factor $r$ is the overall squeezing strength of the process. In the photon-pair approximation,
\begin{equation}
    |\psi\rangle_\mathrm{pair} = \sum_{j=0}^{\infty} r_j \hat{A}_j^\dagger \hat{B}_j^\dagger |0\rangle,
\end{equation}
a superposition of mode pairs with probability $r_j^2$. The Schmidt number $K = 1 / \sum_j \lambda_j^2$ quantifies modal entanglement between the Schmidt mode pairs: $K=1$ for decorrelated states, $K\to\infty$ for maximal time-frequency entanglement. Identifying dispersion conditions that yield such states—e.g., asymmetric or symmetric group velocity matching—is known as dispersion engineering \cite{uren_generation_2005}.

In 2008, Mosley et al. demonstrated the first dispersion-engineered, decorrelated PDC source using type II PDC in periodically poled KDP with asymmetric group velocity matching \cite{mosley_heralded_2008}. In 2011, Eckstein et al. realized a similar source in a periodically poled KTP waveguide at $1550\,$nm, enabling integration into photonic quantum networks \cite{eckstein_highly_2011}. In 2011, Christ et al. linked photon statistics to temporal modes \cite{christ_probing_2011}, deriving
\begin{equation}
    g^{(2)} = 1 + \frac{1}{K},
\end{equation}
where $g^{(2)}$ is the pulsed normalized second-order autocorrelation, providing an experimentally accessible measure of time-frequency entanglement given knowledge about the structure of the state.

\subsubsection*{Temporal mode detection}
\label{sec:PulseModesTemporalModeDetection}
TM detection requires coherent temporal filtering on ultrashort timescales \cite{wong_linear_1995, walmsley_characterization_1996}. This is achieved via the quantum pulse gate (QPG), proposed by Eckstein \textit{et al.} in 2011 \cite{eckstein_quantum_2011}. They built on work by Raymer \textit{et al.} which shows that quantum frequency conversion implements a frequency beam splitter operation \cite{raymer_interference_2010}. The QPG uses dispersion-engineered sum-frequency generation in a periodically poled, titanium-indiffused lithium niobate waveguide, where a classical pump pulse co-propagates with the input signal. The pump’s spectral shape $\alpha(\omega_\mathrm{p})$, adjustable via pulse shaping, defines the target TM $f(\omega)$; the filtered mode is converted to sum-frequency light, while the remainder is transmitted. Detecting a photon in the converted output implements a TM-selective projection $|A_j\rangle\langle A_j|$. The spectral-temporal features of the QPG operation are encoded in the transfer function
\begin{equation}
    G(\omega_\mathrm{in},\omega_\mathrm{out}) = \alpha(\omega_\mathrm{out}-\omega_\mathrm{in}) \varphi(\omega_\mathrm{in}, \omega_\mathrm{out})
\end{equation}
that maps input frequencies $\{\omega_\mathrm{in}\}$ to output frequencies $\{\omega_\mathrm{out}\}$. For an ideal QPG, the transfer function is separable and the QPG operation becomes
\begin{equation}
    \hat{U}_\mathrm{QPG} = \exp [\theta\hat{A}\hat{C}^\dagger+\mathrm{h.c.}].
\end{equation}
This is a TM beam splitter with $\theta$ the beam splitter angle and $\hat{A}$ and $\hat{C}$ the broadband operators associated with the input and output TM, respectively. Brecht \textit{et al.} demonstrated the first QPG with attenuated coherent inputs at telecommunication wavelengths in 2014 \cite{brecht_demonstration_2014}. 

There is another approach towards TM-selective operations that is based on broadband quantum memory in either a $\Lambda-$ or ladder-configuration \cite{michelberger_interfacing_2015, saunders_cavity-enhanced_2016, finkelstein_fast_2018, kaczmarek_high-speed_2018}. These memories operate on vastly different timescales compared to the QPG (nanoseconds vs femtoseconds) but are governed by the same underlying physics that is, they are TM beam splitters. Their longer timescales make them interesting for, e.g., applications with quantum dot single photon sources, where they are expected to improve the quality of the emitted photons \cite{gao_optimal_2019}.

\subsubsection{Recent developments}

\subsubsection*{Advanced temporal mode generation}

Two knobs enable TM-engineered PDC: shaping the pump spectrum $\alpha(\omega_\mathrm{s}+\omega_\mathrm{i})$ or modifying the phase-matching function $\phi(\omega_\mathrm{s},\omega_\mathrm{i})$. Pump shaping enables dynamic control of output states; phase-matching engineering supports fixed-mode sources with simple pulsed lasers. Under symmetric group velocity matching, pump and phase-matching functions are orthogonal in the $(\omega_\mathrm{s},\omega_\mathrm{i})$-plane, imprinting the pump spectrum onto both photons. A first-order Hermite-Gauss pump generates TM Bell states \cite{ansari_tailoring_2018}. This does not, however, generalize to higher-order Hermite-Gauss modes; they generate non-maximally entangled states with fixed dimensionality. In contrast, Cosine kernel modes yield maximally entangled states \cite{patera_quantum_2012}, with Serino \textit{et al.} demonstrating dimensionality up to $d=20$ and violating Bell inequalities up to $d=8$ \cite{serino_orchestrating_2024, dekkers_observing_2025}.

Asymmetric group velocity matching aligns the phase matching function with one frequency axis, enabling heralded single-photon generation in programmable TMs: the phase matching defines one photon’s TM, while the pump imprints its shape on the other \cite{ansari_heralded_2018}. A similar effect arises via backward propagation, where one photon co-propagates with the pump and the other counter-propagates, resulting in phase matching parallel to the co-propagating photon’s frequency axis and imprinting the pump spectrum onto it \cite{christ_pure_2009, kuo_photon-pair_2023}.

In high-gain PDC, Lemieux \textit{et al.} achieved spectral engineering via controlled dispersion between two PDC sources \cite{lemieux_engineering_2016}, reducing the spectral Schmidt number from $K \approx 56$ to $K \approx 1.82$, approaching single-TM operation.

The phase-matching function can be engineered via tailored periodic poling in quasi-phase-matched crystals. Branczyk \textit{et al.} introduced an algorithm to customize domain orientation, demonstrating a Gaussian phase match aimed at decorrelated PDC ($K=1$) \cite{branczyk_engineered_2011}. Dixon \textit{et al.} later achieved 97\% spectral purity in heralded single photons without filtering \cite{dixon_spectral_2013}. Subsequent refinements improved shape fidelity \cite{dosseva_shaping_2016, tambasco_domain_2016}, with Chen \textit{et al.} reporting 99\% purity using Gaussian phase matching and symmetric group velocity matching \cite{chen_efficient_2017}. Graffitti \textit{et al.} advanced the method with sub-coherence-length domain engineering \cite{graffitti_pure_2017}, analyzed fabrication and pump effects \cite{graffitti_design_2018}, and demonstrated it experimentally \cite{graffitti_independent_2018}, later realizing first-order Hermite-Gauss phase matching and TM Bell states \cite{graffitti_direct_2020}.

TMs can be combined with time- or frequency-bins to create hyperentangled states—e.g., in another source of Graffitti \textit{et al.} \cite{graffitti_hyperentanglement_2020}, which entangles photons in frequency-bin and TM degrees of freedom.

In broadband quantum spectroscopy, the goal is maximal time-frequency entanglement far from degeneracy. This is achieved by matching group velocities and cancelling group velocity dispersion, enabling ultrashort operation with continuous-wave pumps—eliminating the need for pulsed lasers \cite{vanselow_ultra-broadband_2019, roeder_ultra-broadband_2024}.

Finally, Jin \textit{et al.} introduced quantum optical synthesis: a method using double-pass interference in a nonlinear crystal to Fourier-synthesize target states via coherent superposition of generation amplitudes \cite{jin_quantum_2021}.

\begin{figure}
    \centering
    \includegraphics[width=\linewidth]{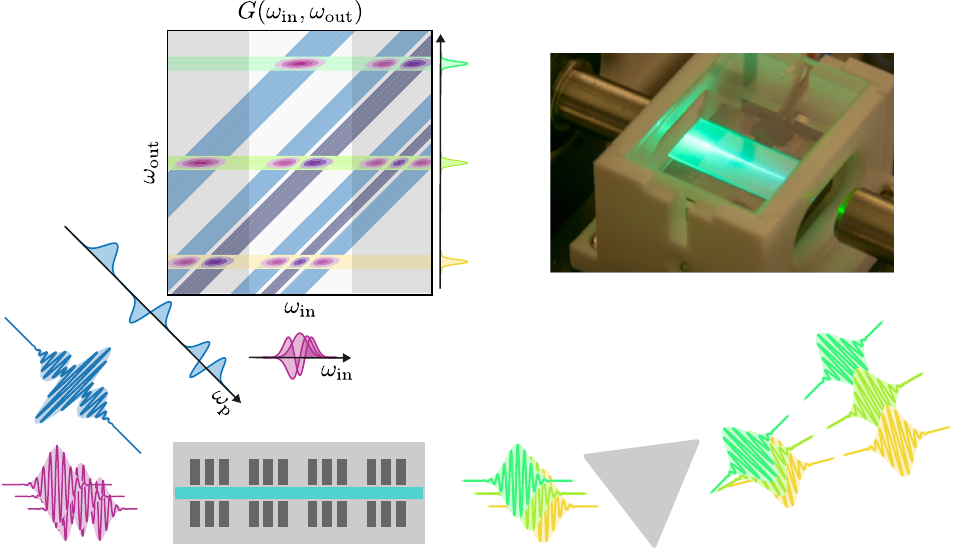}
    \caption{Schematic of an mQPG. A multi-TM signal (purple) is coupled into the mQPG waveguide together with a shaped pump pulse (blue) whose spectrum is comprised of different TMs spaced evenly in frequency space. The waveguide features a superpoling, which yields several phase matching peaks (green to yellow in the plot) whose frequency separation matches that of the pump TMs. The product of pump and phase matching is the transfer function $G(\omega_\mathrm{in},\omega_\mathrm{out})$. Owing to the orthogonality of TMs, each input TM is mapped to a different output frequency. Behind the mQPG, a dispersive element (a prism in the schematic) separates the outputs and frequency resolved photon counting then implements the desired projections. The photograph shows an mQPG sample in one of our optics labs.}
    \label{fig:tm_fig_03}
\end{figure}

\subsubsection*{Advanced temporal mode detection}
\label{sec:GenAndMeas_TemporalModeDetection}

The legacy QPG from \cite{brecht_demonstration_2014} combines telecom-wavelength inputs with 860\,nm pump pulses to generate green output. It enables diverse applications: TM-selective detection supports super-resolution quantum metrology in time and frequency \cite{donohue_quantum-limited_2018, ansari_achieving_2021, de_effects_2021}; TM projections allow QPG detector tomography \cite{ansari_temporal-mode_2017} and are essential for time-frequency quantum state tomography \cite{ansari_tomography_2018, gil-lopez_universal_2021, teo_evidence-based_2024, teo_relative-belief_2024}; time-gating enables temporal sampling of single photons \cite{allgaier_fast_2017, allgaier_streak_2018}; spectral bandwidth compression outperforms filtering \cite{allgaier_highly_2017, shahverdi_quantum_2017}, and when paired with streak cameras, enables ultrafast temporal detection \cite{allgaier_streak_2018, luders_tailored_2023}; single-photon compatible pulse characterization schemes allow for reconstructing the complex spectrum of quantum pulses with arbitrary coherence \cite{bhattacharjee_pulse_2025, bhattacharjee_frequency-bin_2025}. Operating in reverse allows TM shaping \cite{allgaier_pulse_2020}, and telecom compatibility enables integration with dispersion-engineered PDC sources in KTP waveguides. The QPG also enables TM-selective filtering to compress PDC states to $K=1$ \cite{ansari_tomography_2018}, and its coherent filtering acts as an optimal noise filter, with potential for deep-space communications \cite{raymer_time-frequency_2020}.

A key limitation is its single output channel, restricting use in quantum communication. Solutions include cascading multiple QPGs \cite{huang_mode-resolved_2013} or using multi-output QPGs (mQPGs), which employ super-poling structures with alternating poled/unpoled regions to create multiple phase-matching peaks and output channels, c.f. Fig.~\ref{fig:tm_fig_03}. First realized by Serino \textit{et al.} \cite{serino_realization_2023}, the mQPG functions as a high-dimensional temporal mode sorter \cite{serino_programmable_2025}, enabling efficient quantum key distribution, resource-saving state tomography \cite{serino_self-guided_2025}, and tests of foundational quantum physics \cite{serino_complementarity-based_2025}.

\subsubsection*{Quantum photonic systems}

\begin{figure}
    \centering
    \includegraphics[width=\linewidth]{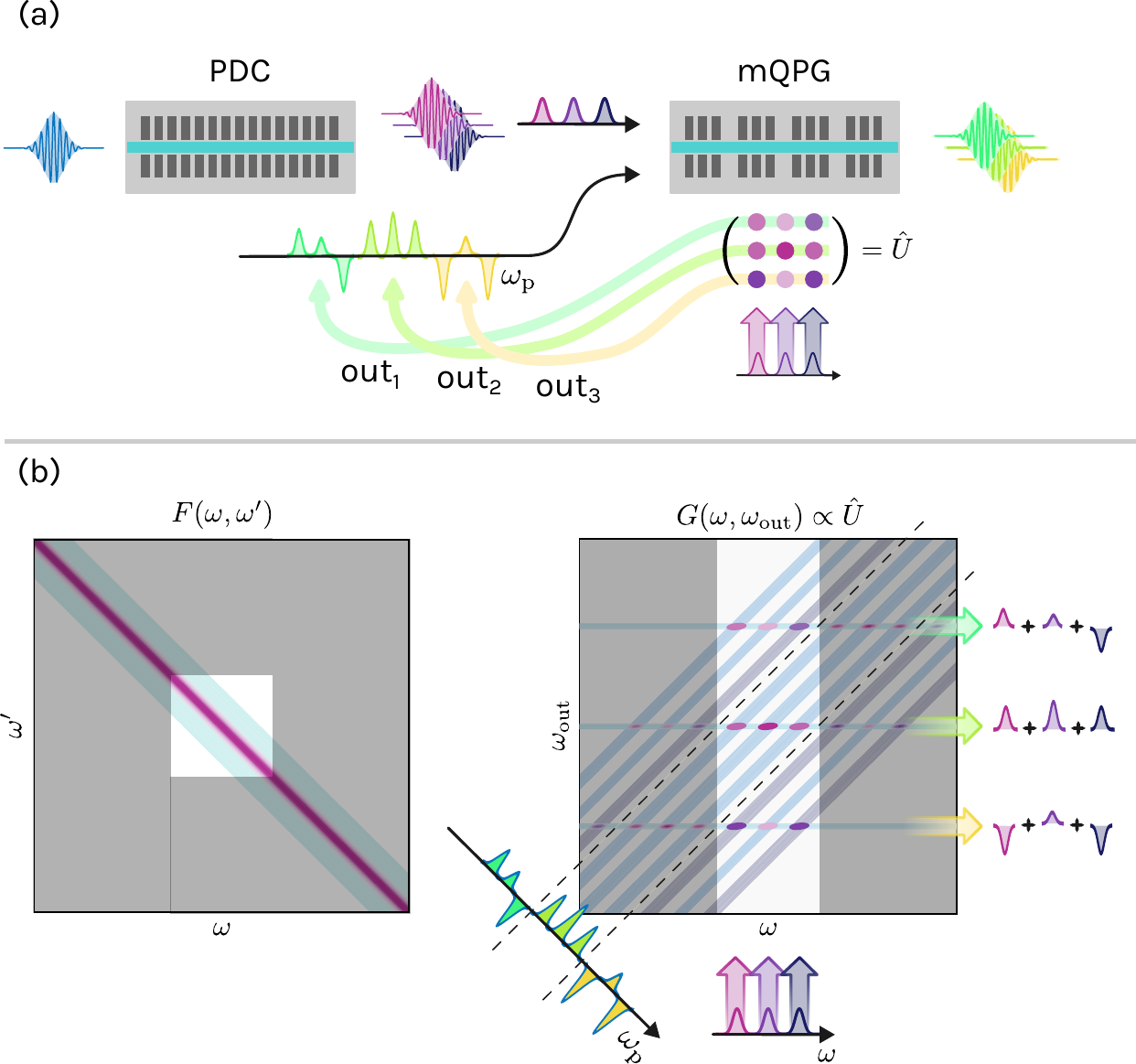}
    \caption{Visualization of a frequency-encoded quantum information network based on TM technology. (a) A type-0 PDC source in a periodically poled lithium niobate waveguide generates broadband quantum light at around $1550\,\mathrm{nm}$. The central part of the spectrum is sent to an mQPG, where it is combined with a pump pulse whose spectrum is comprised of frequency-bins with adapted amplitudes and phase. The exact structure of the pump is defined by the desired quantum network, described by the unitary matrix $\hat{U}$, a $3\times3$ matrix in this sketch. The entries of $\hat{U}$ map to the pump bins. The mQPG then converts the outputs of the quantum network to distinct frequencies, where they can be measured or processed further. (b) The JSA $F(\omega,\omega')$ of the PDC is oriented along $-45^\circ$ in the $(\omega,\omega')$-plane. The central part is selected with spectral filter. The transfer function of the mQPG $G(\omega,\omega_\mathrm{out})$ maps superpositions of input frequencies to different outputs according to $\hat{U}$, which is programmed onto the pump.}
    \label{fig:tm_fig_04}
\end{figure}

The final part of this section focuses on combining a PDC source with an mQPG. De \textit{et al.} demonstrated arbitrary unitary operations on up to 64 modes using QPG-based projections, achieving >99\% similarity to target unitaries \cite{de_realization_2024}. When paired with a broadband, degenerate PDC source, this enables fully programmable frequency-encoded quantum information networks \cite{folge_framework_2024} as sketched in Fig.~\ref{fig:tm_fig_04}.

\subsubsection{Challenges and outlook}

\subsubsection*{Challenges}

Temporal modes (TMs) offer powerful advantages but come with significant challenges. A fundamental issue is time-ordering: the non-commutativity of interaction Hamiltonians at different times leads to strong effects under high nonlinear gain. In PDC, high gain drives the source toward single-TM operation \cite{wasilewski_pulsed_2006, lvovsky_decomposing_2007, christ_theory_2013}, altering TM shapes. Time-ordering effects have been modeled \cite{branczyk_time_2011, quesada_time-ordering_2015, quesada_beyond_2022} and experimentally verified \cite{triginer_understanding_2020}, with recent work providing recipes for gain-agnostic TM structures \cite{houde_perfect_2024}.

For the QPG, high conversion efficiency is counterproductive—it increases multimodedness ($K>1$), degrading selectivity. The maximum selectivity (efficiency × inverse Schmidt number) is limited to 87\% due to intrinsic constraints of TM beam splitter operations, a behavior shared with broadband quantum memories \cite{nunn_mapping_2007}. This can be overcome via multi-stage QPGs with low-efficiency stages and coherent phase control, enabling near-100\% selectivity \cite{reddy_efficient_2014, reddy_sorting_2015, christensen_temporal_2015, reddy_high-selectivity_2018}. In the limit, this converges to a resonant cavity with roundtrip time much shorter than the TM duration—applicable to both QPGs \cite{reddy_photonic_2018} and quantum memories \cite{saunders_cavity-enhanced_2016, nunn_theory_2017}.

Another major challenge is limited conversion efficiency, especially in mQPGs. Splitting the phase matching into multiple peaks via super-poling structures reduces individual peak strength due to reduced poling length per channel. Additionally, shaping pumps for all peaks requires broadband pulse shaping, and if the pump spectrum is not uniformly intense, most energy is discarded in the central region. Simply increasing pump power is impractical. A promising alternative is using a type00 process with nearly degenerate pump and signal, enabling group velocity matching and a 36 times higher nonlinearity in lithium niobate. Operating in the telecom C-band (1535--1565\,nm) leverages mature fiber infrastructure \cite{manurkar_multidimensional_2016, shahverdi_quantum_2017, kumar_spatiotemporal_2021}. However, this risks noise from Raman scattering and cascaded processes. Thin-film lithium niobate offers a compelling alternative through its ultra-high nonlinearity and flexible dispersion engineering \cite{jankowski_dispersion-engineered_2021, ebers_flexible_2022}, enabling efficient, customizable QPGs.

Additional constraints include limited pump bandwidth, resolution limits in pulse shaping (due to finite SLM pixels), and the trade-off between bandwidth and resolution. Finally, TM-based systems rely on bulky, expensive modelocked lasers. Integrated ultrafast lasers or microcombs offer a promising path toward compact, scalable pump sources.

\subsubsection*{Outlook}
TM-based quantum information applications are promising because they enable large, complex systems and are fully compatible with integrated optics. An immediate next step toward practicality is the use of thin-film lithium niobate, as previously highlighted. This material stands out as the most promising platform for integrated nonlinear quantum optics, offering a unique combination of advantages: high second-order nonlinearity, strong electro-optic effects, high integration density, a broad transparency window, compatibility with periodic poling, and advanced dispersion engineering. Together, these properties enable the realization of TM-based applications at novel wavelengths and timescales. A decorrelated PDC source has already been demonstrated \cite{xin_spectrally_2022}. Cascaded devices such as multiple QPGs can be integrated on-chip, with low pump pulse energies compensated by the material’s high nonlinearity \cite{zhu_integrated_2021}. Furthermore, fast electro-optic modulation paves the way for TM applications on timescales of several tens of picoseconds \cite{hu_integrated_2025}, where pulse shaping is achieved directly in the time domain rather than the frequency domain \cite{karpinski_bandwidth_2017, wright_spectral_2017, davis_measuring_2018, sosnicki_interface_2023}. Finally, on-chip resonators can increase the number of available frequency channels, enabling true scalability.

\section{Manipulation, Distribution, and Applications}
\subsection{Time-bin}
\label{sec:appTimeBin}
\author{Fr\'ed\'eric Bouchard\authormark{15}, 
Micha{\l} Karpi{\'n}ski\authormark{16}, Benjamin J. Sussman\authormark{15}}
\address{\authormark{15}National Research Council of Canada, 100 Sussex Drive, Ottawa, Ontario K1A 0R6, Canada.\\
\authormark{16}Faculty of Physics, University of Warsaw, Pasteura 5, 02-093 Warszawa, Poland.}

In both quantum and classical settings, photonic platforms stand out in part because of their extremely large bandwidth, which is directly tied to the temporal degree of freedom. Information carried by optical signals propagates at the speed of light and can be structured into a sequence of well-defined time intervals, or \textit{time-bins}, separated by set delays. Typically the bins are orthogonally defined by the pulse envelope---not carrier phase. By increasing the number of such bins, one naturally obtains high-dimensional time-bin states in which a single photon, a photon pair, or even larger multiphoton entangled states are encoded as coherent superpositions of many temporal modes. Time-bin encoding therefore provides a flexible way to encode, distribute, and process large alphabets in photonic quantum technologies. As introduced in Section \ref{sec:GenAndMeas_TimeBinChallengesAndOutlook}, the scalable generation and measurement of high-dimensional time-bin states remain technically demanding, and are the focus of active experimental and theoretical works. In this section, we review applications of high-dimensional time-bin encoding, and summarize both early work and recent developments in the distribution and manipulation of high-dimensional time-bin states.

\subsubsection{Early work}

\textit{Distribution} -- Once methods for the reliable generation and detection of time-bin, and more generally, high-dimensional time-bin states, were established, the earliest application oriented demonstrations focused on quantum communication. Although time-bin encoding is usually described in terms of discrete temporal modes, it is closely related to time–energy entanglement, an idea introduced by Franson in 1989~\cite{fransonBellInequalityPosition1989}. In particular, time-bin or time-energy entanglement emerged as a robust resource for long-distance distribution in optical fibers~\cite{ribordy2000long}, owing largely to its immunity to polarization fluctuations, and later even in free-space~\cite{steinlechner2017distribution, jin2019genuine}. Foundational experiments distributed time-bin entangled photon pairs over tens of kilometers of fiber~\cite{tittel1998violation, marcikic2004distribution}, and soon after over 100~km~\cite{honjo2007long}. These demonstrations set the stage for high-dimensional time-bin encoding~\cite{Ali-Khan2007} in quantum networks and prompted the development of protocols explicitly tailored to the temporal degree of freedom.

\textit{High-dimensional QKD, higher rates}\label{sec:AppTimeHDQKDLargerRates} -- One of the first motivations for exploring high-dimensional time-bin encoding was the possibility of increasing secret-key rates in quantum key distribution (QKD). For short fiber links or metropolitan-scale channels, detector saturation, rather than channel loss, often becomes the dominant bottleneck. High-dimensional protocols allow each detected photon to carry more than one bit of information, enabling secret-key rates that exceed the raw detection rate of the hardware. Early proposals and implementations of high-dimensional BB84 with time-bins illustrated how a strongly biased basis choice---using the arrival-time basis for key generation and a sparse set of interferometric measurements for checking security---offers a practical route towards high-rate QKD with simple experimental requirements~\cite{mowerHighdimensionalQuantumKey2013, brougham2016information, Islam2017_a, islam2017robust, Islam2019, lee2019large}. Complementary to this, wavelength-multiplexed schemes based on the same time–energy entangled sources have demonstrated parallel entanglement-based QKD channels, providing an additional route to increasing secret-key rates~\cite{pseiner2021experimental}.

\textit{High-dimensional QKD, noise resistance} -- Although high-dimensional encodings are often motivated by their potential for higher key rates, they also offer important advantages in terms of noise resilience. In many cases, a multimode photonic state that would traditionally be coarse-grained into a two-dimensional subspace can instead be treated in its full high-dimensional structure, allowing one to retain information that would otherwise be discarded~\cite{EckerHuber2019}. It is important to note, however, that for protocols such as $d$-dimensional BB84, the apparent increase in error tolerance can be misleading: when measurements require $d$ distinct outcomes, the noise typically grows linearly with the dimension, whereas the information carried by each photon increases only logarithmically in $d$, as already indicated in section \ref{sec:GenAndMeas_TimeBinChallengesAndOutlook}. This trade-off can degrade overall noise performance. The true advantage of high-dimensional encodings often emerges when one performs measurements within lower-dimensional subspaces of a larger high-dimensional state, where the underlying structure can suppress the impact of noise~\cite{Kanitschar2025}.

\textit{High-dimensional QKD, more protocols} -- A further motivation for using high-dimensional encoding in quantum key distribution comes from the family of differential phase–shift protocols~\cite{inoue2002differential}, many of which were conceived with time-bin implementations in mind. In these schemes, the transmitter prepares a train of time-bins and encodes information in the relative phases applied to the pulses within that train. A prominent example is the round-robin differential phase–shift (RRDPS) protocol~\cite{sasaki2014practical}, which attracted considerable attention because its security does not rely on monitoring signal disturbance. Instead, the amount of information an eavesdropper can obtain is intrinsically bounded by the structure of the protocol itself, specifically, by the number of time-bins within each block, making this a genuinely high-dimensional security mechanism. Building on RRDPS, several related protocols have been also proposed that similarly exploit the high-dimensional nature of time-bin encoding to achieve enhanced security features and relaxed monitoring requirements~\cite{chau2015quantum, chau2017experimentally, wang2018proof, wang2021round, Stasiuk2023}. In practice, high-dimensional time-bin encoding is particularly well suited to these strategies. Time-bin states naturally provide access to a large underlying Hilbert space, while still enabling measurements in carefully chosen two-dimensional subspaces using only passive linear optics. Simple time-delayed interferometers can implement these projections with high stability and low loss, allowing one to exploit the noise-mitigation benefits of the high-dimensional structure without requiring full $d$-outcome measurements in a superposition basis. This makes time-bin encoding an especially practical platform for realizing the noise-resilience advantages of high-dimensional quantum communication schemes.

\textit{Manipulation} -- While the preparation and detection of time-bin states matured early, the development of versatile and fully programmable manipulation tools progressed more gradually. Initial demonstrations focused on time-stationary approaches based on fixed interferometric structures, analogous to those used for time-bin detection, introduced in \ref{sec:GenAndMeas_TimeBin_EarlyWork}. Cascaded unbalanced Mach–Zehnder interferometers, multi-arm Michelson configurations, and related multi-delay Franson-type analyzers enabled some of the earliest high-dimensional interference experiments in the temporal domain~\cite{Thew2004PRL_Qutrits}. These schemes established that coherent operations on multiple time-bins could be realized using only passive linear optics and well-controlled path-length differences. For example, a series of unbalanced Mach-Zehnder fiber interferometers were used to realize time-multiplexed pseudo-photon-number-resolved photon detection \cite{achilles2003fiber, kruse2017limits}.

Beyond these fixed architectures, more complex unitary transformations were explored using free-space delay lines and fiber-loops. Early examples were the implementations of discrete-time quantum walks using tunable coin operations in a single spatial mode~\cite{schreiber2010photons, schreiber20122d}. Subsequent experiments showed that time-bin encoding can support additional building blocks, such as programmable beam-splitters and phase shifters on time-bins, required for linear-optical quantum computing~\cite{humphreys2013linear}.

These developments motivated extensive theoretical work aimed at generalizing such primitives into full multiport temporal interferometers. Loop-based architectures were proposed as a route to implementing arbitrary linear unitaries over large temporal-mode spaces using only a small number of dynamically controlled delays and fast switching elements~\cite{motes2014scalable}. Such architectures established a framework in which time-bin encoding could, in principle, support universal linear optical transformations within a single spatial mode. Alternatively, Pant and Englund proposed using time-dependent dispersive propagation to realize unitary transformations for multiple input photons in picosecond time bins in a single spatial mode\cite{pant2016high}.

\begin{figure*}[t!]
	\centering
		\includegraphics[width=1\textwidth]{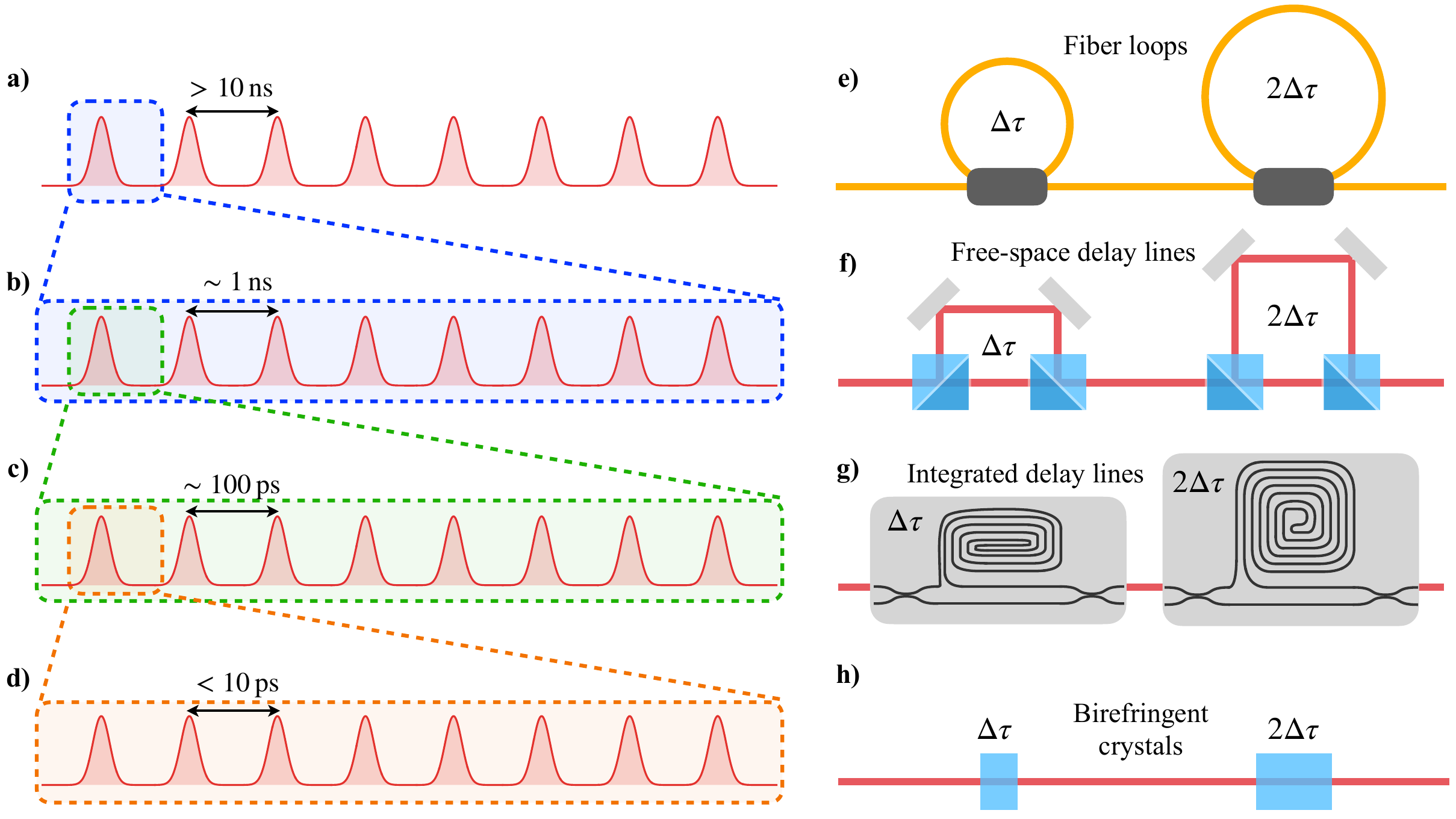}
	\caption{\textbf{Manipulation of time-bin qudits across different timescales.} (a–d) Trains of optical pulses illustrating decreasing time-bin separations from $>10$~ns to $\sim1$~ns, $\sim100$~ps, and $<10$~ps. 
(e–h) Representative architectures for realizing the corresponding time-delayed interferometers, respectively, fiber loops, free-space delay lines, integrated on-chip delay lines, and birefringent-crystal delays with path differences of $\Delta\tau$ and $2\Delta\tau$.}
	\label{fig1}
\end{figure*}

\subsubsection{Recent developments}

\textit{Distribution} -- Recent advances in single-photon detection, particularly the development of next-generation superconducting nanowire single-photon detectors (SNSPDs), have dramatically expanded the accessible temporal resolution in time-bin experiments, such as dispersive-optics QKD. Modern SNSPDs routinely achieve timing jitters in the tens of picoseconds, far beyond the nanosecond regime of avalanche photodiodes (APDs) and even the $\sim$100~ps performance of early SNSPDs. In parallel, progress in integrated photonics has enabled compact, low-loss, time-delayed interferometers, providing stable and scalable temporal-mode analyzers for high-dimensional time-bin measurements. These improvements directly increased the dimensionality of time-bin encoded states that can be resolved~\cite{Liu2019, chang2023large, chang2023experimental}, as well as the distance over which high-dimensional entangled states are distributed, for example over 100~km~\cite{Ikuta2018SciRep_100km_d4}. Combined with new sources such as biphoton frequency combs~\cite{cheng2023high}, these advances are enabling high-dimensional time-frequency entanglement for high-capacity quantum networks and other quantum communication applications~\cite{ogrodnik2025high}.

In addition to sources and detectors, recent progress in high-precision synchronization and clock distribution has also played a key role in pushing high-dimensional time–energy protocols to longer distances. Field demonstrations of high-dimensional QKD over deployed fiber have shown that advanced synchronization schemes can maintain stable interference and low error rates over long distances, for example over 242~km~\cite{liuHighdimensionalQuantumKey2024}. More broadly, these synchronization techniques are becoming increasingly important for quantum networking and quantum internet architectures, where tasks such as entanglement swapping impose stringent requirements on temporal coordination between remote nodes~\cite{davis2025entanglement}.

\textit{Time-multiplexed sources} --
The ability to generate well-defined trains of time-bins, even in cases where the relative phase between bins is not exploited, has also become essential for applications such as time-multiplexed heralded photon sources that approach deterministic operation~\cite{pittman2002single}. In these architectures, high-dimensional time-bin encoding provides the temporal structure that allows multiple, independent heralding opportunities to occur within a single pulse train, while detector timing resolution determines how many such opportunities can be distinguished and utilized. By embedding a probabilistic source inside a fiber or free-space-based delay network, successful heralding events are buffered in distinct time-bins and actively released on demand. This strategy dramatically boosts the single-photon delivery probability while relying solely on linear optics and high-efficiency detectors~\cite{kaneda2015time, hoggarth2017resource, kaneda2019high}. Time-multiplexing can also be extended beyond single-photon generation to construct higher-order nonclassical states by coherently combining multiple heralding events across time-bins~\cite{mccusker2009efficient, engelkemeier2021climbing}. Together, these results demonstrate how temporal multiplexing exploits the high-dimensional structure of time-bin encoding to overcome the intrinsic probabilistic nature of photon generation for scalable photonic quantum technologies.

\textit{Manipulation and applications} -- Recent advances in high-dimensional time-bin manipulation and processing build on the foundational tools introduced in early demonstrations. Modern approaches focus on implementing stable, low-loss temporal-mode interference across a wide range of timescales. Depending on the required time-bin separation, different physical platforms, e.g., bulk optics, fiber-based delays, or integrated photonics, offer complementary advantages in terms of stability, loss, and scalability, see Fig.~\ref{fig1}. These developments enable increasingly sophisticated temporal-mode operations that support large interferometric networks, programmable transformations, and applications in quantum communications and photonic quantum information processing. We now focus on coherent transformations of time-bin modes, whereas projective measurements and readout strategies are discussed in Section~\ref{timebin-recent}

\textit{Fiber loops} -- Fiber loops provide a compact and scalable way to implement large temporal delays, enabling time-bin separations that exceed the timing jitter of modern single-photon detectors. This makes projective measurements in the computational basis straightforward, since different time bins can be resolved directly by their detection time. The drawback is that large delays imply long optical paths, which introduce phase drift and loss that must be carefully managed. Despite these challenges, fiber-loop architectures have enabled several milestone experiments~\cite{he2017time}, including time-division–multiplexed generation of continuous-variable cluster states and demonstrations of measurement-based quantum computation~\cite{yokoyama2013ultra, asavanant2019generation, takeda2019toward, takeda2017universal}, as well as photonic C-NOT gate and multiphoton entanglement generation \cite{meyer2022scalable, pegoraro2024demonstration}. Most prominently, nested fiber-loop interferometers form the basis of platforms such as Xanadu’s Borealis~\cite{Madsen2022}, where time-multiplexed squeezed states are processed through a large temporal network to perform Gaussian boson sampling with quantum computational advantage. These results show that, when engineered with low loss and active stabilization, fiber loops can support complex high-dimensional temporal-mode transformations at scale. Moreover, fiber-loop architectures are broadly compatible with many types of quantum light sources, including quantum dots~\cite{carosini2024programmable}, making them a versatile tool for high-dimensional photonic processing.

\textit{Modulation and dispersion} -- Various modulation approaches, in particular electro-optic phase modulation, enable time-bin manipulation. The most basic manipulation technique involves using electro-optic phase modulation to directly imprint the required phases onto respective time bins. High-speed phase shifts using low-insertion-loss, fiber-compatible, electro-optic phase modulators enable spectral manipulations~\cite{karpinski_bandwidth_2017} and time scale modifications~\cite{sosnicki_interface_2023}. Using highly dispersive elements provides an alternative to time-delayed interferometers for appropriately short time-bin timescales. Here, dispersive broadening of the time-bin pulses combined with time-resolved detection enable sensitivity to the time-bin phases~\cite{sedziak2020tomography, czerwinski2021phase, widomski2024efficient}. Therefore, time-bins and dispersive propagation can, in principle, be used to realize complex interferometers~\cite{pant2016high}, although practical implementations are hampered by the required few-ps timing resolution. In addition, multiphoton interference between photons in distinct time-bins has been demonstrated by erasing the temporal distinguishability using high-resolution, spectrally resolved detection \cite{orre2019interference}. More generally, sequences of time and frequency-domain phase modulation can implement more general unitary transformations on temporal modes~\cite{lukens2018reconfigurable, ashby2020temporal, mazur2019multi}, although propagation loss and imperfect modulation presently pose significant challenges for scaling such schemes in quantum applications. Nonlinear optical processes enable further possibilities for time-bin manipulation. Conversion between time-bin encoding and temporal-mode encoding has recently been demonstrated using the multi-output quantum pulse gate, as discussed in Section \ref{sec:PulseModes}
\cite{serino_programmable_2025}.

\textit{Integrated photonics} -- At shorter temporal separations, on the order of tens to a few hundred picoseconds, fast SNSPDs enable direct resolution of individual time bins while keeping the required path-length differences within the reach of integrated photonic circuits~\cite{montaut2025progress}. In this regime, low-loss waveguide delay lines and compact unbalanced interferometers can implement stable temporal-mode operations with chip-scale footprints~\cite{yuQuantumKeyDistribution2025}. A key element of the manipulation toolbox includes fast phase modulation capabilities. However, mature low-loss integrated photonic platforms (silicon nitride $<0.1$~dB/cm \cite{labonte2024integrated}, silica/glass $<0.1$~dB/cm \cite{tan2022effectively}) are inherently passive, with indirect or hybrid approaches needed for phase modulation at ns or ps time-scales~\cite{hermans2019integrated, lafforgue2025monolithic, Giordani2023}. The silicon photonics platform exhibits moderate losses of $1-2$~dB/cm and enables large-scale manufacturing using CMOS processes. It is a passive platform, yet indirect amplitude and phase modulation approaches have been developed \cite{Silverstone2016, AdockJSTQE2021}. Indium phosphide provides a mature active integrated photonic platform that has enabled integration of lasers and modulators. However, due to its high propagation losses ($1-4$~dB/cm \cite{smit2014introduction}), it is suitable primarily for weak coherent state preparation, with applications in the context of QKD \cite{paraiso2021photonic, widomski2023precise}. Therefore, fast modulation for time-bin manipulation requires novel platforms. Integrated opto-electro-mechanical modulators in aluminum nitride have been shown to enable high-speed phase shifts and spectral manipulations of single-photons~\cite{fan2019spectrotemporal}. Recent fast paced developments in the thin-film lithium niobate platform open new possibilities in high-speed integrated electro-optic manipulation \cite{zhu2022spectral}.

The compatibility between detector timing, waveguide delays, and fast on-chip modulation makes integrated photonics an increasingly attractive platform for scalable manipulation of high-dimensional time-bin states~\cite{monika2025quantum}, although combining these functionalities in a single platform remains a challenge, with a possible route towards further advancements offered by hybrid integration \cite{Elshaari2020AUTO}.

\textit{Ultrafast time-bin} -- Pushing to even smaller temporal separations, at the picosecond or sub-picosecond level, allows time-bin interferometry to be implemented using extremely compact optical elements. For example, at these time-scales, the polarization-dependent group delay in a birefringent crystal naturally provides the picosecond level separation required to realize a time-delayed interferometer~\cite{kupchakTimebintopolarizationConversionUltrafast2017}, as already indicated in Section \ref{sec:GenAndMeas_TimeBin_EarlyWork}. In this ultrafast regime, however, the temporal spacing falls below the timing resolution of even the fastest SNSPDs, making direct time-of-arrival discrimination impossible. Coherent characterization and manipulation therefore rely on nonlinear optical techniques such as sum-frequency generation~\cite{donohue2013coherent, MacLean:2018}, cross-phase modulation~\cite{matsuda2016deterministic, kupchakTerahertzbandwidthSwitchingHeralded2019, england2021perspectives, fenwick2025ultrafast}. Alternatively, electro-optic or nonlinear  temporal imaging can be used \cite{karpinski2021control} (see section \ref{sec:OSTD}). Recent experiments have leveraged this regime to demonstrate coherent measurements of ultrafast time-bin qubits and qudits~\cite{cameron2023ultrafast, bouchardQuantumCommunicationUltrafast2022, Bouchard2023}, quantum walks and multiphoton dynamics in the picosecond domain~\cite{xu2018measuring, wang2018dynamic, fenwick2024photonic, wang2024efficient, fenwick2025multiphoton, white2025robust}, and programmable temporal unitary transformations~\cite{bouchard2024programmable}. These approaches effectively trade detector timing resolution for optical bandwidth and ultrafast optical control, enabling high-dimensional time-bin manipulation without long interferometric paths. As a result, ultrafast time-bin encoding has emerged as a powerful platform for scalable, low-loss, and phase-stable high-dimensional photonic processing.

\subsubsection{Challenges and outlook}

Despite rapid progress, the widespread deployment of high-dimensional time-bin encoding still faces several technological and conceptual challenges. Long-delay interferometers, essential for manipulating large time-bin separations, remain susceptible to phase drift and optical loss, making long-term stability difficult in fiber-based systems. Integrated photonic platforms promise greater robustness, yet achieving low-loss, broadband delay lines and on-chip, reconfigurable temporal interferometers remains an open engineering goal. At the opposite extreme, ultrafast time-bin approaches demand precise nonlinear gating, stringent timing synchronization, and low-jitter pump–probe coordination, all of which introduce their own sources of instability. Bridging these regimes will require continued advances in detectors, integrated photonics, active stabilization, and nonlinear optical control of single-photon-level signals.

Recent improvements in the timing resolution of single-photon detectors as well as in the speed of electro-optic modulators open up possibilities in the intermediate few to tenths of picoseconds timescales. At these timescales, direct detection is becoming possible, whereas the path difference remains at a manageable few-centimeter level. Additionally, the spectral features of such signals become comparable with the resolution of spectral modulators and spectrometers, bringing the prospect of using spectral devices for manipulation and/or detection of time-bin superpositions. This intermediate regime may enable combining the benefits of direct detection and electronic modulation with the stability and flexibility of manipulation offered by the femtosecond regime.  

Looking forward, the convergence of high-efficiency photon sources, low-jitter detectors, and increasingly programmable temporal-mode circuits points toward fully reconfigurable, large-scale time-bin processors operating in a single spatial mode. Electro-optic modulators and nonlinear optical gating offer a promising route, with tunability, low loss, and  fast enough operation speeds to manipulate individual time-bins, suggesting that fast, stable, and compact temporal-mode processors may soon become a practical reality. Taken together, the compatibility of time-bin encoding with fiber networks, integrated photonics, and ultrafast nonlinear optics positions it well for applications in scalable quantum communications and photonic quantum information processing. As these technologies continue to mature, high-dimensional time-bin encoding is poised to become a foundational resource for quantum communication, programmable photonic computing, and large-scale quantum information processing.

\begin{backmatter}
\bmsection{Acknowledgments}
 MK acknowledges in parts the support of the National Science Centre, Poland, under project no. 2023/50/E/ST2/00703, and of the European Union’s Horizon Europe Research and Innovation Programme through grant agreement no.\ 101070700 (MIRAQLS)
 \end{backmatter}

\subsection{Frequency-bin} \label{Sec:App_Freq_Bin}
\author{Micheal Kues\authormark{2},Joseph M. Lukens\authormark{17,18}}
\address{\authormark{2}Institute of Photonics and PhoenixD Cluster of Excellence, Leibniz University Hannover, Hannover, Germany}
\address{\authormark{17}Elmore Family School of Electrical and Computer Engineering and Purdue Quantum Science and Engineering Institute, Purdue University, West Lafayette, Indiana 47907, USA}
\address{\authormark{18}Quantum Information Science Section, Oak Ridge National Laboratory, Oak Ridge, Tennessee 37831, USA}

\subsubsection{Early work}

\begin{figure}    \includegraphics[width=\textwidth]{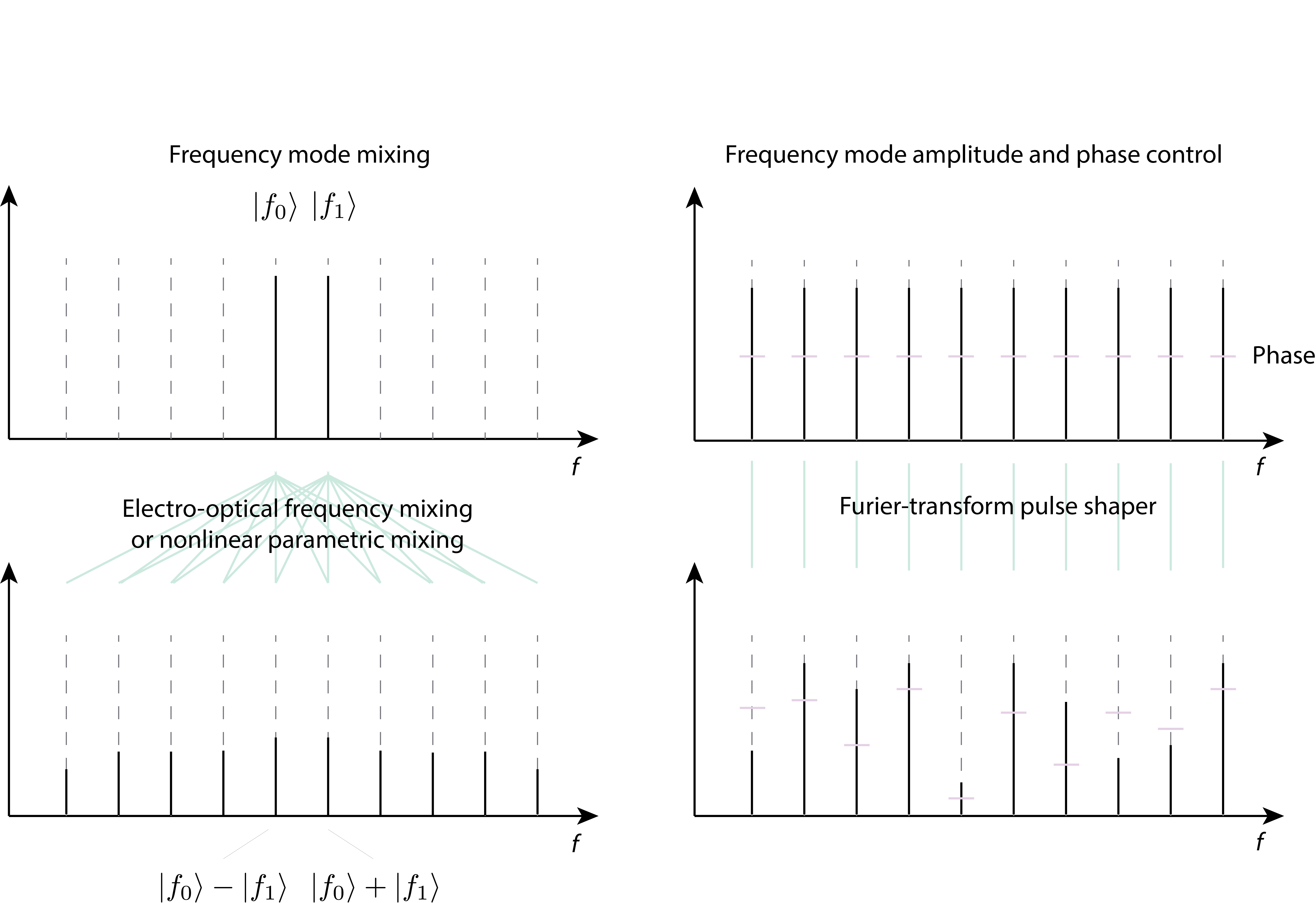}
    \caption{Frequency-bin processing with frequency-mode mixing for projective superpositions and amplitude and phase control of each bin. Cascading these elements allows building frequency networks for quantum applications.}
    \label{fig:modemixing}
\end{figure}

A key requirement in frequency-bin quantum photonics is the ability to mix, interconvert, and distribute discrete optical frequency modes while maintaining quantum coherence. Frequency-bin qubits (or qudits) are typically encoded in well-separated spectral bins---e.g., $\ket{f_0}$ and $\ket{f_1}$ (or more  $\ket{f_n}$)---of a single photon. To perform quantum operations, one must coherently couple these bins, creating and controlling superpositions such as $(\ket{f_0}+e^{i\phi}\ket{f_1})$. Two physical mechanisms are primarily used to achieve this: electro-optic phase modulation (a process in which the phase of the optical field is driven by a microwave field) and nonlinear optical frequency conversion \cite{Kobayashi2016AUTO,Agha2012AUTO} . Both provide frequency-domain analogues of the spatial beam splitter, enabling interference and routing between spectral modes. 
Electro-optic phase modulation (EOM) uses the electro-optic Pockels effect in materials such as lithium niobate to impose a time-dependent phase on an optical field. When an optical mode at frequency $\varpi_0$ passes through a phase modulator driven by a sinusoidal voltage $V(t) = V_{m}\sin{(\Omega t + \phi)}$, its output field is
\begin{equation}
\label{eq:Bessel}
E_\text{out}(t) = E_{0}\ e^{i\varpi_0 t}e^{i\beta\sin{(\Omega t + \phi)}}
\end{equation}
where $\beta = \pi V_{m}/V_{\pi}$ is the modulation index and
$V_{\pi}$ is the half-wave voltage. Expanding the exponential via the
Jacobi--Anger identity gives
\begin{equation}
E_\text{out}(t) = E_{0} e^{i\varpi_{0}t}\left[\sum_{n = 0}^{\infty}{J_{n}(\beta)}e^{in(\Omega t + \phi)} + \sum_{n = 1}^{\infty}{{{( - 1)}^{n}J}_{n}(\beta)}e^{- in(\Omega t + \phi)}\right]
\end{equation}
showing, in the case of the spectral bandwidth of the optical mode being significantly narrower than the modulation frequency $\Omega$, that the optical field acquires sidebands at frequencies \(\varpi_{0} + n\Omega\ \) with complex amplitudes \(J_{n}(\beta)e^{in\phi}\). The Bessel functions \(J_{n}(\beta)\) determine how power is distributed among these bins. For small \(\beta\), only the carrier and first sidebands (\(n = \pm 1)\) have significant amplitude, making the modulator an effective two-mode frequency coupler. For larger \(\beta\), more sidebands are generated, allowing the mixing of several frequency modes and thus high-dimensional processes. This process can be viewed as a transformation between discrete frequency modes---analogous to a spatial beam splitter. By tuning the modulation index \(\beta\) and RF phase \(\phi\), one controls the mixing ratio and relative phase between bins. 
Multi-tone RF drives or cascaded modulators allow arbitrary mode-mixing networks, as shown in Ref. \cite{Lukens2017}. The pure amplitude and phase of individual frequency modes can be accessed independently by the use of programmable spectral filters, operating via the Fourier-domain principle, where the wave function is spatially dispersed and a liquid crystal array controls amplitude and phase.  In practice, combinations of EOMs and Fourier-domain pulse shapers form a universal toolkit for frequency-domain linear optics (Fig.~\ref{fig:modemixing}), realizing Hadamard gates, tritters \cite{Lu2018a}, and interferometers \cite{Kues2017} in the spectral domain.

Beyond electro-optic modulation, frequency-bin states can also be mixed using nonlinear optical interactions that directly couple two spectral modes. Nonlinear $\chi^{(2)}$ processes such as sum- and difference-frequency generation enable partial conversion between frequencies \cite{Takesue2008AUTO}, effectively functioning as tunable frequency-domain beam splitters. Experiments using $\chi^{(2)}$ conversion have shown that operating the nonlinear interaction below full conversion causes a single input frequency to be coherently distributed across two output frequencies---analogous to how a spatial beam splitter divides a photon between two paths. By adjusting the pump power, the relative weighting of the two output frequencies can be precisely controlled \cite{Kobayashi2016AUTO}. In $\chi^{(3)}$-based Bragg-scattering four-wave mixing, two strong pump fields are launched into a nonlinear medium, and their frequency difference defines the separation between two target frequency bins. When the phase-matching condition is satisfied, the nonlinear interaction couples these two target frequencies: part of the energy from each bin is transferred to the other, while the relative phase between them is preserved. In this way, the two excitation fields mediate a frequency-domain beam splitter, enabling coherent superpositions and controlled rotations of the quantum state within the two-frequency subspace \cite{McGuinness2010AUTO,Agha2012AUTO,Joshi2020AUTO,Clemmen2016AUTO}. Together, $\chi^{(2)}$ and $\chi^{(3)}$ processes provide tools for fully coherent manipulation of frequency-bin qubits in quantum photonic systems. While these processes work all-optically, which is advantageous for further advancements in integration and fast control, they have only recently been extended beyond two modes~\cite{Oliver2025}, with additional pump fields required to continue scaling---compared to electro-optic modulators that natively support many-bin couplings via Eq. (\ref{eq:Bessel}).

\subsubsection*{Frequency-bin Hong-Ou-Mandel Interference}
\begin{figure}    \includegraphics[width=\textwidth]{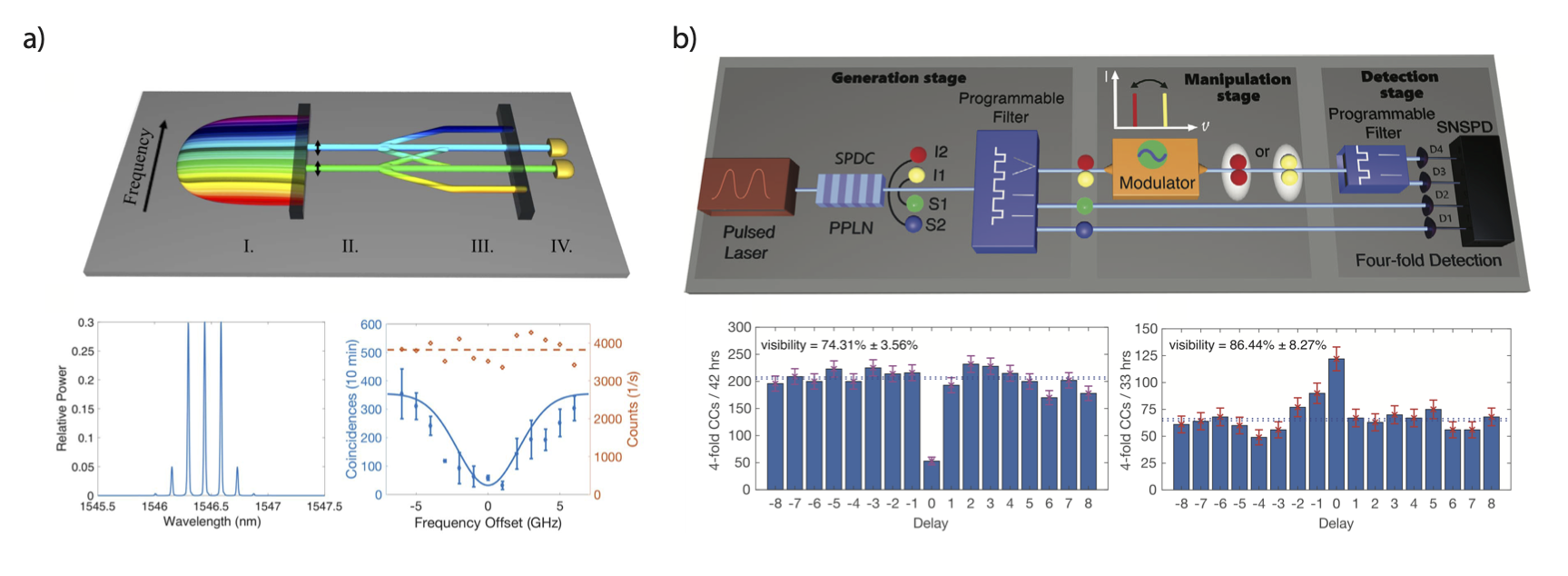}
    \caption{Spectral Hong-Ou-Mandel interference with dependently (a) and independently (b) created photons. Images reproduced with permission from Optica Publishing Group \cite{Imany2018c} and a Creative Commons Attribution 4.0 International License (\url{https://creativecommons.org/licenses/by/4.0/}) }
    \label{fig:hom}
\end{figure}

Hong--Ou--Mandel (HOM) interference \cite{Hong1987}  is a foundational effect in quantum photonics because it enables Bell-state measurements, quantum teleportation, entanglement swapping, and a wide range of multiphoton protocols that rely on photon indistinguishability at a beam splitter. Demonstrating HOM interference in the frequency domain is therefore essential for extending these capabilities to frequency-encoded qubits and qudits. In the frequency-bin setting, two single photons prepared in distinct spectral modes are mixed through a device that implements a two-mode unitary transformation---typically realized using electro-optic phase modulation combined with Fourier-transform pulse shaping, or through nonlinear frequency conversion \cite{Kobayashi2016AUTO}. When the frequency-domain beamsplitter is tuned to 50:50 mode coupling, the photons become indistinguishable in their joint spectral-temporal mode at the output, leading to the characteristic suppression of coincidence counts that signals HOM interference.
The first demonstrations of modulator-based frequency-bin HOM interference used spectrally distinct photons generated from the same nonlinear source and employed electro-optic modulation to construct a frequency-domain beam splitter \cite{Imany2018c}, showing clear photon bunching into the same output colors and thereby erasing their initial spectral distinguishability [Fig.~\ref{fig:hom}(a)]. Subsequent experiments extended this concept to independently generated photons, a critical requirement for scaling frequency-domain quantum circuits \cite{KhodadadKashi2021AUTO}. These experiments also highlighted the reconfigurability of spectral mixing: by adjusting the relative phase within the frequency beam splitter, the same setup could be switched from bosonic bunching to fermionic-like anti-bunching, illustrating full control over two-mode interference in the frequency basis.

\subsubsection*{Quantum gates for frequency-bin states}
Cascading electro-optic modulation with Fourier-transform pulse shaping enables the construction of linear frequency networks and the implementation of quantum gates \cite{Lukens2017}. In particular, adding multiple layers improves the operation fidelity in most cases. A three-dimensional frequency-domain Hadamard gate has been realized by driving modulators with a two-tone microwave and shaping sidebands, coherently mixing multiple frequency modes and achieving near-unity fidelity across the C-band \cite{Lu2018a}. Fully arbitrary single-qubit rotations have also been demonstrated, with single photons mapped to several different points in the high-dimensional Bloch space and average fidelities above 0.98 \cite{Lu2020_ArbitraryControl}. Entangling gates, such as controlled-NOT (CNOT), have been implemented in the frequency basis using modulators combined with line-by-line pulse shaping, achieving two-photon gate fidelities of $\sim$0.91 \cite{Lu2019a}. These results illustrate that frequency-bin quantum information processing supports both single-qudit and multi-qubit operations within a fiber-compatible, highly parallelizable platform. Drawbacks compared to other DoFs are, however, the high insertion losses of the optical components used, which limit scalability toward many layers or large numbers of photons. 

\subsubsection*{Quantum key distribution with frequency-bin states}
\label{sec:AppFreqbinQKD}
Quantum key distribution (QKD) using frequency-bin-encoded photons leverages the discrete spectral modes as the computational basis for quantum states. In such schemes, logical states $\ket{0}$ and $\ket{1}$ correspond to photons occupying distinct frequency bins (e.g., $\omega_0$, $\omega_1$), while conjugate-basis states are realized as coherent superpositions across these bins. This approach is compatible with standard telecom-band infrastructure and dense wavelength-division multiplexing (WDM), making it attractive for practical fiber-based quantum communication. The frequency bins provide an intrinsically phase-stable encoding because all modes travel in a single spatial and polarization mode of the fiber. Basis switching---essential for QKD---is then achieved via previously discussed frequency shifting, creating projection states within the required mutually unbiased bases. Compared to polarization or time-bin encodings, frequency-bin methods offer high environmental robustness and natural compatibility with classical multiplexed systems, while enabling straightforward parallelization across many channels.

The first proposals of frequency-coded quantum key distribution using prepare and measure schemes use approaches where information was encoded in the optical frequency of weak coherent pulses. A laboratory-scale BB84 system using electro-optic modulation to generate four frequency states and optical filtering for basis measurement was demonstrated in Ref. \cite{Bloch2007AUTO}. Subsequent work extended this concept using more compact and stable architectures: for example, the use of phase-modulated continuous-wave lasers to define frequency-shift keying bases. Moreover, studies have combined frequency-bin and time-bin encodings into joint time-frequency protocols (TF-BB84), analyzed theoretically and numerically \cite{Rodiger2017AUTO}. These schemes exploit the mutual conjugacy of time and frequency (or time-bin superpositions) to realize BB84 bases \cite{Rödiger2021AUTO} and have achieved low quantum bit error rates (QBER < 4\%) over several kilometers of fiber \cite{islam2017robust, Islam2017_a}. Frequency-bin BB84 systems benefit from low polarization sensitivity and the ability to reuse mature WDM components. However, they face technical constraints including spectral resolution and filter loss that degrade interference visibility. Superposition-basis stability requires active phase tracking or modulation synchronization. Analyses adapting decoy-state BB84 models to frequency-mode imperfections reveal the potential that system-level secret key rates comparable to polarization-based QKD are achievable with current technology.

\subsubsection*{Generation of high-dimensional cluster states }
To provide a crucial resource for measurement-based quantum computing, frequency-bin encoding has been used to create photonic cluster states. On-chip quantum frequency combs based on microresonators have been used to produce multi-frequency entangled states \cite{Kues2019AUTO}. A time-bin excitation has been added to create hyperentangled states. In particular, the time-bandwidth product being much larger than unity allows the consideration of time and frequency as independent, allowing the mapping to four parties within two photons, i.e. each photon has a time and frequency DoF. These structures are then transformed into cluster states via coherent control. Specifically, by addressing individual frequency bins through frequency-time mapping and electro-optic modulation, an experiment has demonstrated one-way cluster states in the time-frequency domain, marking the first realization of high-dimensional cluster-state entanglement \cite{Reimer2019}. It was demonstrated that such high-dimensional cluster states are more tolerant to noise, with increasing dimensionality. These results show that hyperentangled time-frequency states allow scalable entanglement across many modes while remaining fiber-compatible, opening the door to large-scale, high-dimensional measurement-based quantum architectures.

\begin{figure}    \includegraphics[width=\textwidth]{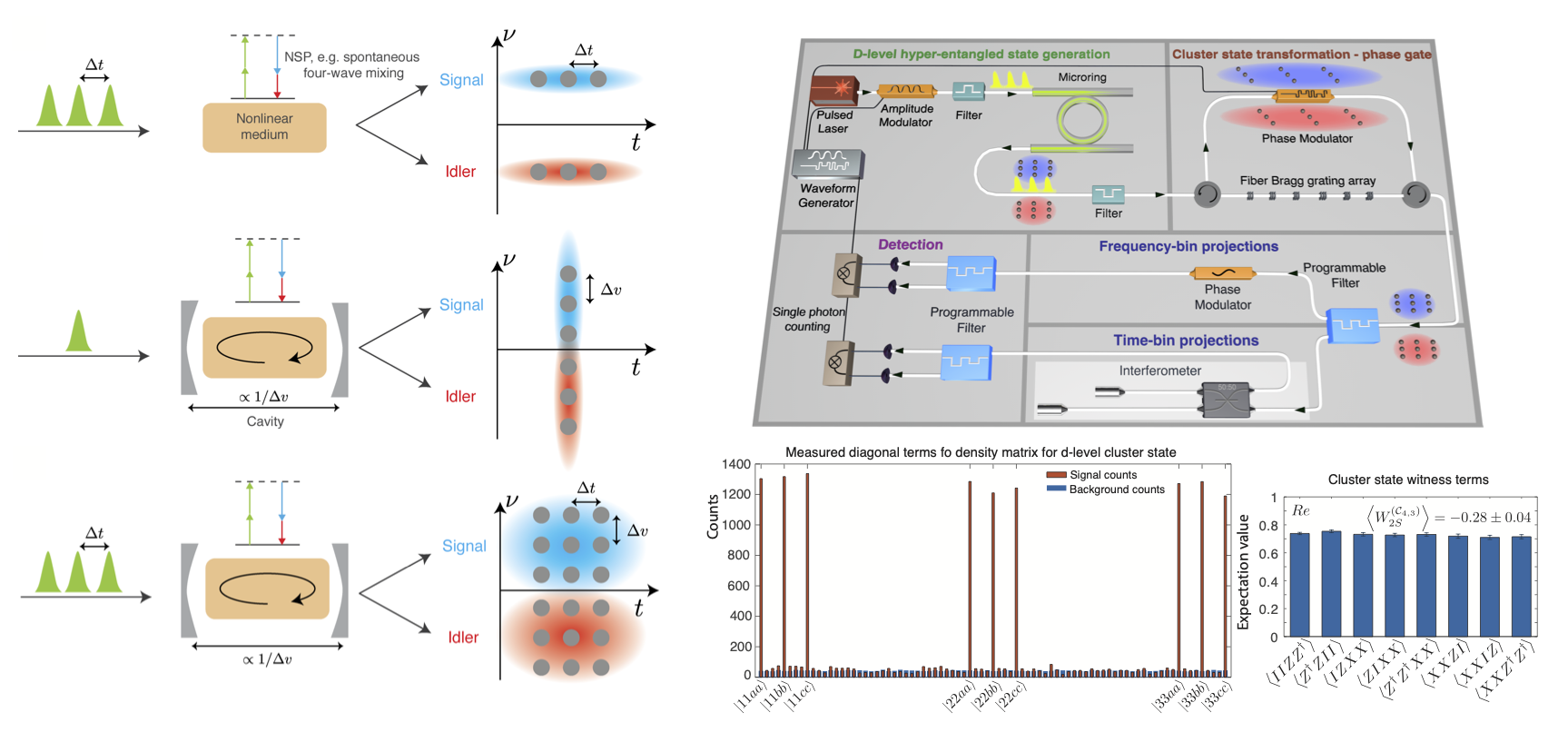}
    \caption{Hyperentanglement scheme for frequency and time, allowing with coherent manipulation the creation of a high-dimensional cluster state. Images reproduced with permission from Springer Nature \cite{Reimer2019}.}
    \label{fig:Hyper}
\end{figure}

\subsubsection{Recent developments}

\subsubsection*{Integrated frequency-bin photonics}
One of the initial inspirations for frequency-bin quantum information processing---particularly  in the electro-optic-based quantum frequency processor paradigm \cite{Lukens2017}---stemmed from its intriguing synergies with integrated photonics. As described in more detail in Sec. \ref{sec:GenAndMeasFreqBins}, frequency-bin qudits are naturally generated by integrated quantum microcombs \cite{Kues2019AUTO}, and ubiquitous microring-resonator-based circuits can facilitate not only state generation, but also multiplexing, pulse shaping, and modulation of frequency-bin resources \cite{Myilswamy2025AUTO}. Historically, the experimental surge of frequency-bin processing tools coincided with the first full tomographies of quantum microcombs \cite{Kues2017, Imany2018b}, and in the following years a variety of subsequent insights into on-chip states have been enabled by frequency-bin processing techniques \cite{Reimer2019, Imany2019, Lu2022AUTO, AndreaSabattoli2022AUTO, Myilswamy2023_TimeResolvedHBT, Clementi2023, Borghi2023}.  Yet despite the confluence of on-chip frequency-bin sources, all the examples above leveraged off-chip technology (e.g., discrete modulators and pulse shapers) for state control, leaving the original dream of fully integrated frequency-bin photonics unfulfilled.

\begin{figure}
    \centering
    \includegraphics[width=0.9\linewidth]{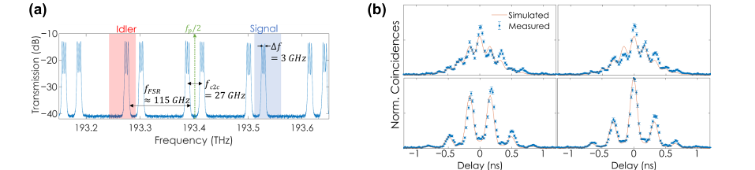}
    \caption{Temporal shaping of frequency-bin-entangled qutrits with an integrated pulse shaper. (a) Classical power spectrum highlighting the locations of the signal-idler frequency bins. (b) Coincidences measured for multiple phase settings, showing coherent control of biphoton correlations. Images reproduced from Ref.~\cite{Wu2025_OnChipPulseShapingEntangled} with permission under a Creative Commons Attribution 4.0 International License (https://creativecommons.org/licenses/by/4.0/)}
    \label{fig:chip}
\end{figure}

In the quest toward realizing this vision, we see the greatest challenges on the electro-optic modulation front. The most efficient electro-optic materials are not CMOS-compatible, suggesting three distinct directions moving forward: (i) alternative modulation modalities like aluminum nitride acousto-optics \cite{Zhou2024AUTO} for fully CMOS solutions, (ii) monolithic circuits in less mature but potentially more flexible non-CMOS platforms like thin-film lithium niobate \cite{Boes2023AUTO}, or (iii) heterogeneous integration combining devices from multiple platforms together \cite{Elshaari2020AUTO}. All approaches offer their own advantages and disadvantages, revealing an open-ended future with no obvious frontrunner.
In the meantime, advances in specific on-chip capabilities for frequency-bin encoding are emerging at a rapid pace. In one recent example \cite{Wu2025_OnChipPulseShapingEntangled}, a programmable six-channel microring filter bank enabled the first control of frequency-bin-entangled photons with an on-chip pulse shaper. Leveraging a recently demonstrated multiheterodyne and dual-comb spectroscopy technique for efficient heater tuning \cite{Cohen2024AUTO}, signal and idler frequency bins could be placed precisely on an ultra-narrow 3 GHz grid [Fig.~\ref{fig:chip}(a)], after which the temporal correlations of frequency-bin photon pairs (produced off chip) were shaped in excellent agreement with theoretical expectations Fig.~\ref{fig:chip}(b).

Beyond the implications for future all-in-one frequency-processing circuits, this pulse shaper unlocks exciting opportunities for spectrally efficient wavelength control beyond what is currently possible with bulk devices. Because the ultimate resolution of a microring-based shaper depends on the ring linewidth and not free-spectral range, ultra-narrow frequency spacings can be resolved in a line-by-line fashion. Indeed, the 3 GHz resolution realized in \cite{Wu2025_OnChipPulseShapingEntangled} is already significantly tighter than the $\gtrsim 10$ GHz typically attainable from commercial devices \cite{Ma2021AUTO}; and with straightforward incorporation of higher-quality-factor rings---e.g., $Q>10^6$ is fairly routine in many platforms \cite{Xie2020AUTO, Zhu2024LN, Puckett2021AUTO}---resolutions down to the hundred MHz-level should be readily attainable with existing technology.

\subsubsection*{Entanglement-based spectral quantum key distribution}

\begin{figure}    \includegraphics[width=\textwidth]{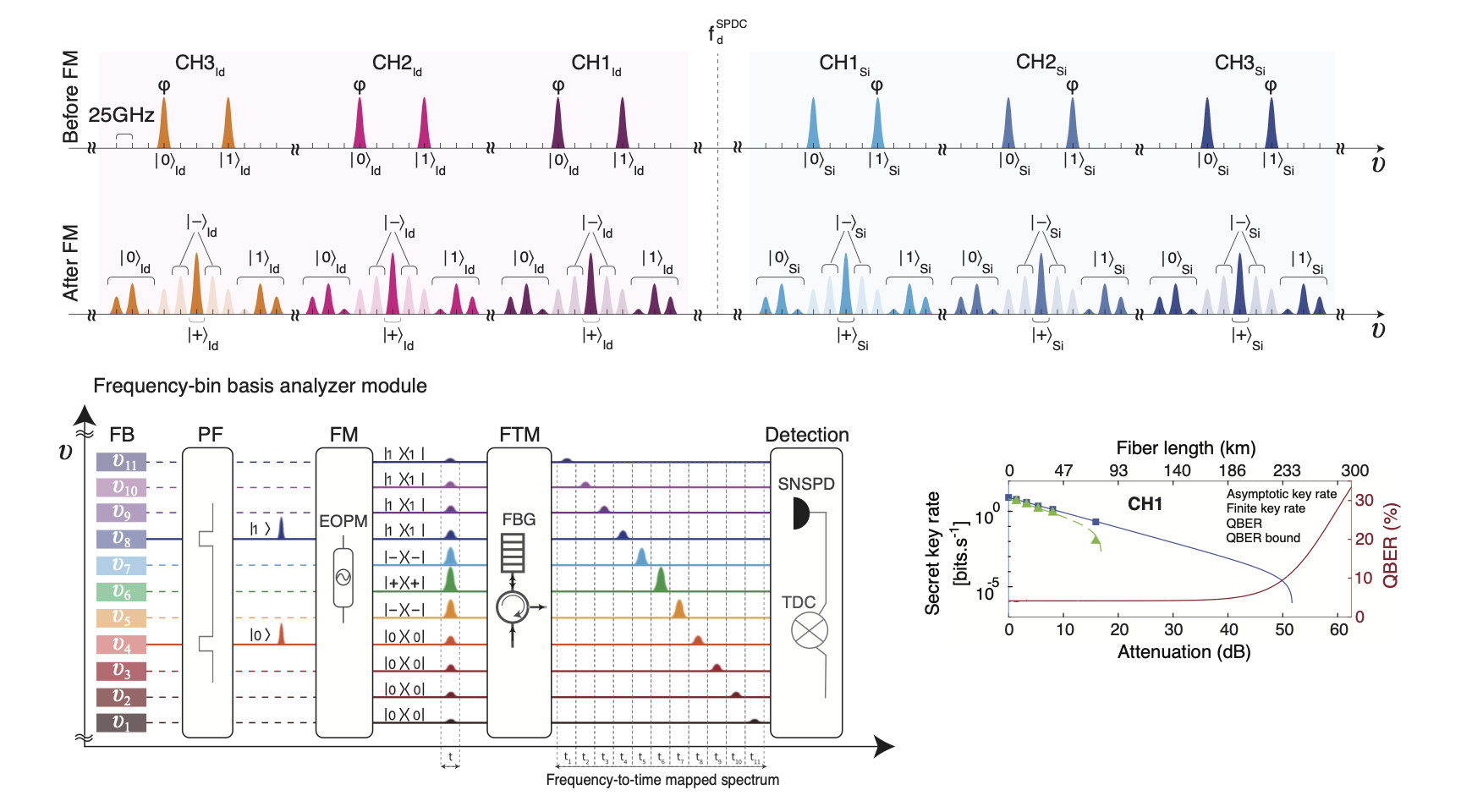}
    \caption{Frequency-bin entanglement-based quantum key distribution. A basis state analyzer allows the simultaneous measurement of four projections out of two mutually unbiased bases. A frequency-to-time mapping approach allows the measurement of these frequency projections with one detector.  }
    \label{fig:entanglement}
\end{figure}

Entanglement-based quantum key distribution uses---opposed to encode-and-measure schemes---nonlocal correlations to create secure keys, with the benefit of being more secure due to source independence as well as allowing entanglement distribution for future entanglement-based networks. In several works, entanglement-based quantum key distribution has been demonstrated where different DoFs such as polarization or time-bin have been used for encoding and the spectral domain has been exploited for channel multiplexing \cite{Appas2021AUTO}. More recently, frequency bins have been used for encoding in entanglement-based QKD [Fig.~\ref{fig:entanglement}]. Spectral encoding can provide new forms of reconfigurability and scalability that are difficult to achieve with conventional encodings.

The first implementation of frequency-bin-encoded entanglement-based QKD was demonstrated in a fiber-based system \cite{Kashi2025_FreqBinQKD}. It allowed introducing a dynamically reconfigurable frequency-bin measurement module capable of passive random basis selection while requiring only a single detector per user. \label{sec:AppFreqFreqToTime} This was enabled through frequency-to-time mapping, which permitted projective measurements in multiple frequency-bin bases without additional hardware. This architecture substantially reduces hardware overhead and dark-count contributions, mitigating detector-related side-channel vulnerabilities. Moreover, the scheme enabled adaptive frequency multiplexing without modifying the physical setup. The results furthermore highlight how frequency-bin encoding can support multi-user operation and resource-minimized distribution of entanglement, making it attractive for scalable quantum networks.

A further study in parallel demonstrated that frequency-bin encoding can also be realized with integrated sources \cite{Tagliavacche2025_QKD}. Using two independent high-finesse ring resonators on silicon, an entanglement-based BBM92 protocol with one modulator per party for frequency-bin projective measurements and passive basis choice has been demonstrated. To ensure stable operation, they introduced a real-time adaptive phase-tracking scheme that compensates thermal phase noise and enabled reliable key generation over a 26-km fiber link, achieving secure rates above several bits per second. Together, these demonstrations establish frequency-bin entanglement as a practical and versatile resource for QKD, compatible with existing telecom infrastructure and suitable for future high-capacity, networked quantum communication systems. 

\subsubsection{Challenges and outlook}
Despite recent progress, several challenges remain for frequency-bin-based quantum processing and entanglement distribution, which define key directions for future research. Ultrafast manipulation of frequency-bin states requires high-bandwidth electro-optic or nonlinear modulators capable of generating and controlling coherent superpositions across many frequency bins without adding excess phase noise; achieving low-loss, low-drive, and sub-nanosecond operation remains a central technological bottleneck. Particularly, all-optical nonlinear modulations can be explored, which need to extend to multi-level mixing. Chip-integration and scalability demand compact, thermally stable, and low-crosstalk photonic components---such as high-finesse micro-resonators \cite{Cohen2024AUTO} and programmable frequency processors---whose performance must remain uniform across large spectral bandwidths. This requires advances in hybrid integration, where passive wavelength splitting components and fast modulators need to be combined. At the same time, network-level synchronization becomes increasingly critical as frequency-bin systems scale: stable frequency references, active phase-tracking, and compensation of thermally induced drifts are necessary to preserve entanglement over long distances and across many simultaneous channels. Finally, hybrid architectures that combine frequency-bin encoding with other DoFs (e.g., time-bin, polarization, spatial modes) offer pathways to multidimensional and multiplexed quantum networks, but require precise inter-domain interfacing to maintain coherence. Addressing these challenges will enable robust, high-capacity, and interoperable networks that fully exploit the spectral domain for large-scale quantum information processing.

\begin{backmatter}
\bmsection{Acknowledgments}
M.K. acknowledges funding from the European Research Council (ERC) under the European Union’s Horizon 2020 research and innovation programme under grant agreement No. 947603 (QFreC project), and the German Research Foundation (Deutsche Forschungsgemeinschaft; DFG) within the cluster of excellence PhoenixD (EXC 2122, Project ID 390833453).

\end{backmatter}

\subsection{Transverse-spatial modes} \label{Sec:App_Spatial_Modes}
\author{Mehul Malik\authormark{1}, Yaron Bromberg\authormark{19}, Robert Fickler\authormark{20}}
\address{\authormark{1}Institute of Photonics and Quantum Sciences (IPAQS), Heriot-Watt University, Edinburgh, EH14 4AS, United Kingdom}
\address{\authormark{19}Racah Institute of Physics, The Hebrew University of Jerusalem, Jerusalem, 91904, Israel}
\address{\authormark{20}Photonics Laboratory, Physics Unit, Tampere University, Tampere, FI-33720, Finland}

\subsubsection{Early work}

The challenge of manipulating high-dimensional quantum states encoded in the transverse-spatial photonic DoF has echoes back to the very first classical imaging system---the camera. While the optical lenses used in such systems perform fundamental transformations in the continuous transverse-spatial domain, their ability to perform generalized operations on discrete transverse-spatial modes is quite limited. The high-dimensional equivalent of waveplates and beam-splitters for spatial modes simply do not exist, and significant effort has been invested over the past 25 years in the development of devices that can perform such operations. The potential of encoding high-dimensional quantum states, or qudits, on the transverse-spatial structure of single and entangled photons was recognized for free-space quantum communication \cite{Molina2001, Gibson:2004fw} as well as for improving the security of quantum key distribution \cite{Bourennane:2001cp, MolinaTerriza:2004vk}. 

Early work in this direction focused heavily on holographic methods for preparing and measuring single-photons carrying orbital angular momentum (OAM), which is reviewed in Section \ref{sec:GenAndMeas_TransverseChallengesAndOutlook}. However, the first device to manipulate photonic spatial modes was an interferometer designed to efficiently sort single-photons based on their OAM-mode parity \cite{Leach:2004kp}. This interferometer used rotated dove prisms placed in its paths to introduce an OAM-mode-dependent phase such that constructive or destructive interference was obtained depending on the parity of the OAM-mode carried by the photon. While designed for sorting single-photons \cite{Erhard:2016wk}, this device was instrumental in the first demonstration of high-dimensional multipartite entanglement, as it also allowed the OAM amplitudes of two independent, input photons to be mixed \cite{Malik2016, Erhard2018}. 

The interferometric OAM-parity sorter was a key building block for the first experimental implementation of HD quantum gates \cite{Schlederer:2015uf, Babazadeh:2017js}---high-dimensional generalizations of the qubit Pauli X and Z gates, which were first discussed by Daniel Gottesman in 1999 \cite{Gottesman1999ey}. While initially proposed in the context of quantum computation, these gates were first used to generate the complete set of four-dimensional Bell states \cite{Wang:2017}. Interestingly, the experimental setups for these gates were found through the use of a computational algorithm called MELVIN that searched through combinations of bulk optical elements such as beam splitters and dove prisms to find useful solutions \cite{Krenn2016ds}. MELVIN was also used to find bulk optical setups for generating a large class of asymmetric, multipartite HD entangled states where the local dimension of each particle is different \cite{HuberDeVicente2013}.

A related bulk-optics device that emerged a few years after the OAM-parity sorter was the Q-plate \cite{Marrucci2006}. This device used patterned birefringence to couple the spin and OAM of a single-photon via the Pancharatnam-Berry phase \cite{Pancharatnam:1956}. By construction, Q-plates operate in a four-dimensional polarization-spatial modal space and have been used in a variety of different quantum information applications such as HD quantum key distribution and quantum random walks \cite{Rubano:19}. Moving away from bulk optics, the emergence of the field of complex light scattering heralded a new approach for manipulating transverse-spatial modes. Scattering media are equivalent to very large, complex optical networks that support a large number of photonic spatial modes. However, gaining control over the scattering process is a significant challenge. Early work studied two-photon speckle produced by the passage of spatially entangled photons through a ground glass diffuser \cite{Peeters:2010kx}. The ability to measure the transmission matrix of a complex scattering medium \cite{Popoff:2010} led to the first experiments demonstrating two-photon interference inside a multi-mode fiber \cite{Defienne:2016dk} and a layer of Teflon \cite{Wolterink:2016bc}.

The first demonstrations of high-dimensional QKD using transverse-spatial modes were performed in 2006 and implemented both the BB84 \cite{Walborn2006} and Ekert (entanglement-based) protocols \cite{Groeblacher2006}. The BB84 experiment used an attenuated laser, a movable pinhole, and Fourier/imaging lenses to prepare and measure photonic states in up to $d=37$, while the Ekert experiment used spatially entangled photons and OAM-holograms to generate a key with OAM-modes in $d=3$. A key challenge for QKD was the ability to sort single-photons carrying HD spatial modes. The four-element refractive OAM sorter \cite{Mirhosseini2013} discussed in Chapter X enabled the first multi-outcome demonstration of HD-QKD with a prepare-and-measure experiment in $d=7$ \cite{Mirhosseini_2015}.

\subsubsection{Recent developments}
\label{sec:Applications_TransverseRecentDevelopments}

\begin{figure*}[t!]
	\centering
		\includegraphics[width=1\textwidth]{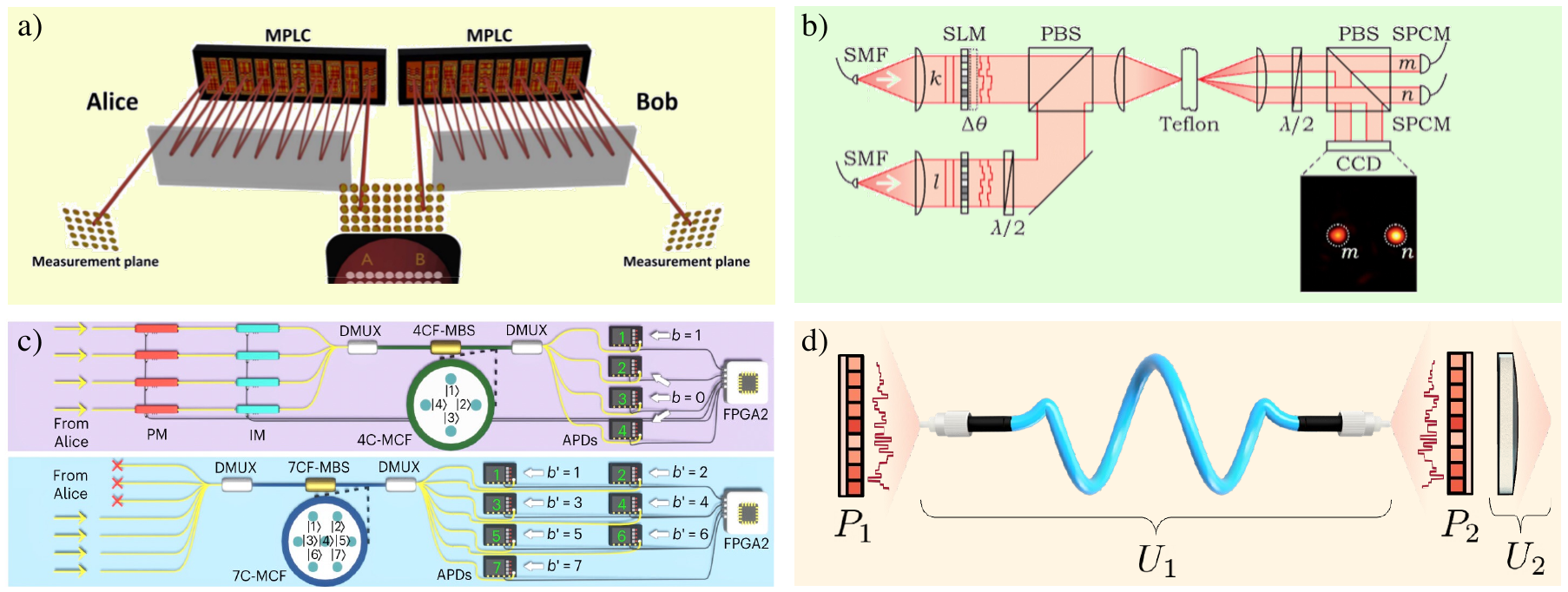}
	\caption{Current approaches for spatial-mode qudit manipulation include a) multi-plane light conversion, b) complex scattering media, c) multi-core fibers, and d) multi-mode fibers. Figures adapted from references \cite{lib2025high,Wolterink:2016bc,martinez2023certification, goel2024inverse}.}
	\label{XPfig1}
\end{figure*}

To go beyond proof-of-principle demonstrations of quantum information processing with high-dimensional encoding, efficient tools for manipulating qudits had to be developed. The number of elements required to realize a general $d \times d$ transformation of a qudit scales as $d^2$, making it impractical to scale using bulk optical components such as $q$-plates, Dove prisms, or spiral phase plates. Spatial light modulators (SLMs) offer millions of degrees of control over the transverse modes of photons, yet their operation is fundamentally limited. Expressing the $d$ transverse modes in the pixel basis at the SLM plane reveals that the SLM can only implement diagonal transformations in this basis, covering just a $d$-dimensional subspace of the full $d^2$ space of possible operations.

A practical solution is provided by multi-plane light conversion (MPLC), which uses multiple SLM planes separated by free-space propagation \cite{Morizur2010}. Diffraction between consecutive planes allows light illuminating one pixel on a given plane to overlap with multiple pixels on the following plane. With a sufficient number of planes, this architecture can implement arbitrary linear transformations. Originally developed for spatial-division multiplexing of classical light, MPLC has greatly expanded the degree of control achievable over qudits. It has been used to demonstrate Hadamard gates on heralded single photons encoded in the orbital angular momentum (OAM) basis \cite{brandt2020high} and in the pixel basis \cite{li2020programmable}, as well as an implementation of unambiguous state discrimination \cite{Goel2023USD}. By passing photon pairs through an MPLC device, Hong–Ou–Mandel interference between two $d=4$ qudits encoded in OAM modes was observed \cite{hiekkamaki2021high}, and entanglement between two $d=3$ qudits encoded in the pixel basis was certified \cite{lib2022processing, LibLiuShekelHeHuberBrombergVitagliano2025}. The reconfigurability of the MPLC, which allows smooth transitions between measurement bases, was utilized to advance measurement-based quantum computation with photons entangled in transverse modes \cite{lib2024resource}.  This approach was recently extended to coherently exciting and collecting light from up to five quantum dots on a single chip \cite{Goel:26}, showcasing the potential of spatial-mode control for matter-based quantum systems. Typically, the SLM phase patterns required for a given MPLC transformation are found using iterative optimization algorithms such as wavefront matching \cite{fontaine2019laguerre}. More recently, methods based on diffractive neural networks designed for qudit gates have shown improved fidelities \cite{wang2024ultrahigh}. 

An alternative approach for realizing non-diagonal transformations in the pixel mode basis is to couple the pixel modes into a multimode fiber. When the qudit dimension $d$ is much smaller than the number of guided modes in the fiber, the fiber couples enough SLM pixels to realize nearly arbitrary transformations. Using this approach, reconfigurable gates in up to $2 \times 22$ dimensions (two input modes and 22 output modes) were implemented to control pairs of photons coupled into the fiber, demonstrating a wide range of transformations, including non-unitary ones, for efficient manipulation of high-dimensional two-photon interference \cite{leedumrongwatthanakun2020programmable, makowski2024large}. By adding another SLM after the fiber, $7 \times 7$ gates were realized and used to manipulate and certify high-dimensional entanglement \cite{goel2024inverse}. Combining this configuration with multiple input photons from entangled pairs further enabled the routing and swapping of entanglement between 8 users across a network~\cite{valencia2026large}. Interestingly, the SLM–MMF–SLM configuration acts as a two-plane multi-mode fiber converter (MMFC), where the mode mixing induced by the MMF plays a key role. Numerical studies have shown that repeating this structure, similarly to the MPLC, offers a promising route for scaling up the dimension $d$ \cite{goel2024inverse}. Hence, in combination with the ease and efficiency of generating and measuring multiple mutually unbiased bases by simple phase modulations, encoding and manipulating qudits using the pixel mode basis has become a vital player in quantum photonics with transverse spatial modes \cite{osullivanhale2005pixel,valencia2020high}. 

A different fiber-based approach for implementing high-dimensional gates utilizes multi-port splitters based on $d$-core fibers. These are realized by tapering a multi-core fiber, followed by a phase modulator for each core, forming a sequence that again resembles the diagonal–mode-mixing–diagonal transformation architecture of MPLC. This approach enabled the realization of $d=4$ and $d=7$ gates with remarkably high fidelities (above 0.99), implemented for weak coherent states \cite{carine2020multi} and entangled photon pairs \cite{gomez2021multidimensional}, and was further used to demonstrate high-quality generalized measurements in dimension $d=4$ \cite{martinez2023certification}. Being fully compatible with standard telecommunication fibers, such devices offer an attractive solution for fiber-based applications.

Based on these advances in control and manipulation of transverse modes, recent years have also seen substantial progress in developing applications, in particular for quantum communication. The enhanced resilience of high-dimensional quantum key distribution (QKD) to noise has attracted significant attention since the early proposal by Cerf et al. \cite{cerf2002security}. A modern understanding of the noise-robustness of high-dimensional entanglement, together with experimental demonstrations showing that such states tolerate noise far better than two-dimensional entanglement \cite{EckerHuber2019}, has stimulated the development of QKD protocols that exploit this immunity \cite{Hu2021}. These insights have also motivated both theoretical and experimental studies of genuine high-dimensional quantum steering in the presence of noise \cite{Designolle2021, qu2022retrieving} and loss \cite{Srivastav2022}, paving the way towards high-dimensional device-independent protocols.

In parallel, research on HD-QKD with transverse-mode encoding has progressed along two main directions. At the laboratory level, significant work has focused on reducing the bit error rate for moderate dimensions using polarization–spatial hybrid encoding ($d=4$) \cite{Mafu2013, ndagano2017deterministic, Wang2019}  and improving mode sorting ($d=7$) \cite{larocque2017generalized}. At the same time, substantial effort has been devoted to increasing the dimensionality accessible in proof-of-principle demonstrations. Examples include entanglement-based QKD with projective measurements in dimensions up to $d=21$ \cite{otte2020high}, and more recently $d=25$ using an MPLC-based mode sorter \cite{lib2025high}, as well as $d=360$ using near-field and far-field projections to realize mutually unbiased measurements \cite{scarfe2025spatial}. HD-QKD has also been employed to implement new protocols developed over the past decade. A prominent example is the round-robin protocol \cite{sasaki2014practical}, which relies on higher dimensionality to bound information leakage without the need for active monitoring, even though its logical encoding remains effectively two-dimensional. Implementing such protocols with transverse modes can significantly simplify the required optical setup \cite{bouchard2018round, bouchard2018experimental}.

The second direction has shifted toward long-distance demonstrations. Both entanglement-based protocols, where Alice retains one photon in the laboratory and transmits its entangled partner to Bob, and prepare-and-measure protocols, where Alice sends weak coherent states prepared in the desired transverse mode, face similar challenges. Loss and optical aberrations reduce the raw key rate and increase the bit error rate in both scenarios. For free-space links, early demonstrations of classical transverse mode distribution over 143 km \cite{krenn2016twisted}, together with binary ($d=2$) OAM-entanglement over 3 km \cite{krenn2015twisted}, motivated the transition to high-dimensional implementations outside the laboratory. A free-space implementation of HD-QKD was carried out by Sit et al., who realized $d=4$ HD-QKD using polarization–spatial hybrid modes distributed over a 300 m free-space link \cite{Sit:17}.

\begin{figure*}[t!]
\centering
\resizebox{0.7\linewidth}{!}{
  \begingroup
\newcommand{\labelshiftx}{0mm}
\newcommand{\labelshifty}{0mm}
\newcommand{\labxmax}{10}

\begin{tikzpicture}

\begin{axis}[
    name=top,
    width=\linewidth,
    height=3.65cm,
    xmin=2, xmax=50000,
    xmode=log,
    ymin=330, ymax=420,
    xlabel=\empty,
    ylabel=\empty,
    xtick={10,100,1000,10000,100000},
    xticklabels=\empty,
    x tick style={draw=none},
    minor x tick num=8,
    axis x line=bottom,
    x axis line style={draw=none},
    axis y line=left,
    axis line style={very thick},
    tick style={very thick},
    tick label style={font=\Large},
    clip=true,
    grid=both,
    major grid style={line width=0.4pt, draw=gray!40},
    minor grid style={line width=0.2pt, draw=gray!20},
     ytick={360},
    yticklabels={360},
]
\addplot[draw=none, fill=gray!12, forget plot] coordinates {
  (2.1,340) (\labxmax,340) (\labxmax,419) (2.1,419)
} \closedcycle;
\node[
    font=\Large\bfseries,
    anchor=north west,
    align=left
]
at (axis cs:2.1,422)
{Lab\\demos};
\addplot[only marks, mark=*, mark size=2.8pt]
coordinates {(3,360)};
\node[font=\Large, anchor=west]
  at ($(axis cs:2.8,360)+(\labelshiftx,\labelshifty)$) {\cite{scarfe2025spatial}};
\end{axis}
\begin{axis}[
    name=bottom,
    at={(top.south west)},
    anchor=north west,
    yshift=-0.25cm,
    width=\linewidth,
    height=7.cm,
    xmin=2, xmax=50000,
    xmode=log,
    ymin=0.5, ymax=26,
    xlabel={\Large Distance},
    ylabel=\empty,
    xtick={10,100,1000,10000,100000},
    xticklabels={10 m, 100 m, 1 km, 10 km, 100 km},
    ytick={5,10,15,20,25},
    axis x line=bottom,
    axis y line=left,
    axis line style={very thick, -},
    tick style={very thick},
    tick label style={font=\Large},
    label style={font=\Large},
    clip=true,
    grid=both,
    major grid style={line width=0.4pt, draw=gray!40},
    minor grid style={line width=0.2pt, draw=gray!20},
    minor x tick num=8,
    legend style={
      draw=black,
      fill=white,
      font=\Large,
      at={(0.62,1.4)},
      anchor=north west,
      legend cell align=left,
      /tikz/column sep=8pt
    },
]
\addplot[draw=none, fill=gray!12, forget plot] coordinates {
  (2.1,0.6)
  (10,0.6)
  (10,26)
  (2.1,26)
} \closedcycle;
\addplot[FreeSpace]
coordinates {(3,25) (3,21) (3,7) (300,3.6)};
\addlegendentry{Free-space}
\node[font=\Large, anchor=west]
  at ($(axis cs:2.8,25)+(\labelshiftx,\labelshifty)$) {\cite{lib2025high}};
\node[font=\Large, anchor=west]
  at ($(axis cs:2.8,21)+(\labelshiftx,\labelshifty)$) {\cite{otte2020high}};
\node[font=\Large, anchor=west]
  at ($(axis cs:2.8,8.0)+(\labelshiftx,\labelshifty)$) {\cite{larocque2017generalized}};
\node[font=\Large, anchor=west]
  at ($(axis cs:80,3.3)+(\labelshiftx,\labelshifty)$) {\cite{Sit:17}};
  
\addplot[MultiMode]
coordinates {(3,6) (1000,3) (1000,4) (25000,4)};
\addlegendentry{Multimode fiber}
\node[font=\Large, anchor=west]
  at ($(axis cs:2.8,6)+(\labelshiftx,\labelshifty)$) {\cite{valencia2020unscrambling}};
\node[font=\Large, anchor=west]
  at ($(axis cs:750,1.6)+(\labelshiftx,\labelshifty)$) {\cite{cao2020distribution}};
\node[font=\Large, anchor=west]
  at ($(axis cs:750,5.6)+(\labelshiftx,\labelshifty)$) {\cite{cozzolino2019orbital}};
\node[font=\Large, anchor=west]
  at ($(axis cs:13000,5.6)+(\labelshiftx,\labelshifty)$) {\cite{Wang2021high}};
\addplot[MultiCore]
coordinates {(3,4) (300,4.4) (400,4)};
\addlegendentry{Multicore fiber}
\node[font=\Large, anchor=west]
  at ($(axis cs:2.8,4)+(\labelshiftx,-1mm)$) {\cite{Ding2017}};
\node[font=\Large, anchor=west]
  at ($(axis cs:80,5.2)+(\labelshiftx,\labelshifty)$) {\cite{Canas2017_HDDistib_MCF}};
\node[font=\Large, anchor=west]
  at ($(axis cs:280,5.6)+(\labelshiftx,\labelshifty)$) {\cite{achatz2023simultaneous}};
\end{axis}
\coordinate (ybreak) at
  ($(top.south west)!0.50!(bottom.north west)+(0,-2.2mm)$);
\draw[very thick]
  ($(ybreak)+(-0.15cm, 0.15cm)$) -- ++(0.45cm,0.45cm);
\draw[very thick]
  ($(ybreak)+(-0.15cm,-0.05cm)$) -- ++(0.45cm,0.45cm);
\path let \p1=(bottom.west), \p2=(top.west) in
  node[rotate=90, font=\Large, anchor=center]
  at (\x1-14mm, {(\y1+\y2)}) {Dimension, $d$};
\end{tikzpicture}

\endgroup}
  \caption{Reported distances and dimensions $d$ for the distribution of transverse spatial modes. Free-space laboratory demonstrations have achieved the highest dimensions to date, while fiber-based systems enable long-distance transmission but are currently limited to $d=4$. The distribution of higher-dimensional states over long distances remains an outstanding challenge.}
  \label{XPfig2}
\end{figure*}
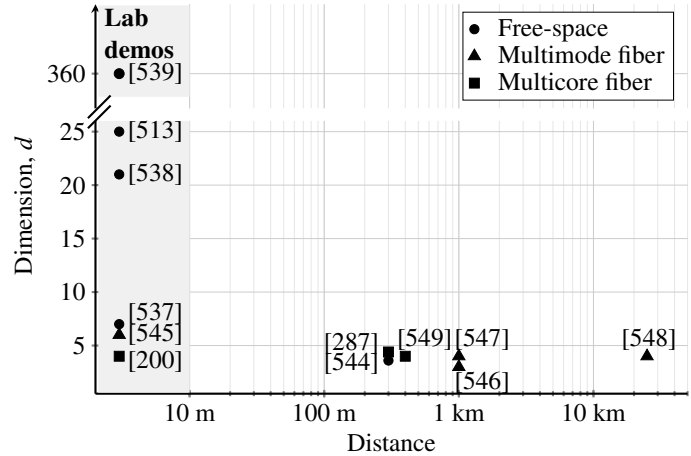

The main advances in distribution over long distances, however, have come from transmitting transverse modes through fibers. Distributing such modes in multimode fibers is challenging because of mode-mixing and modal interference. As a result, most demonstrations have focused on specially designed fibers with reduced mode-mixing. Entanglement has been distributed over $d=3$ OAM modes through a 1 km, specially designed step-index fiber \cite{cao2020distribution}, and HD-QKD has been demonstrated with $d=4$ OAM modes using an air-core fiber \cite{cozzolino2019orbital}, as well as $d=4$ hybrid polarization–transverse modes over a 25 km ring-core fiber \cite{Wang2021high}. Multi-core fibers offer another approach, where each core acts as an individual mode, with negligible inter-core coupling. HD-QKD with $d=4$ was realized using a 3 m multi-core fiber and silicon-based photonic integrated circuits for multiplexing and demultiplexing \cite{Ding2017}, as well as over 300 m of fiber \cite{Canas2017_HDDistib_MCF}, albeit with inefficient demultiplexing. More recently, hyperentanglement was distributed over 400 m of multi-core fiber, using polarization or temporal degrees of freedom to achieve $d=4$ with only two cores per photon \cite{achatz2023simultaneous}.

Using standard multimode fibers would be far more desirable than these specialized fibers, but strong mode-mixing and modal interference in commercially available fibers present a significant challenge. Nevertheless, inspired by recent advances in spatial-division multiplexing for classical optical communication, it was shown that shaping the two-photon wavefront with an SLM can compensate for mode-mixing, enabling the distribution and certification of $d=6$ entanglement across a standard 2 m multimode graded-index fiber \cite{valencia2020unscrambling}. Interestingly, entanglement was used here as a means of characterizing the fiber transmission matrix via channel-state duality, which allowed entanglement to be unscrambled by manipulating the photon that did not go through the fiber.

Finally, recent advances in the implementation of HD-QKD with transverse modes have also been reported at the source level. For prepare-and-measure protocols, a laser that emits the desired hybrid spatial–polarization modes directly, without the need for external modulation, was recently demonstrated \cite{Zhang2025}. In addition, for prepare-and-measure protocols that are immune to photon-number-splitting attacks, transverse-mode modulation has been implemented for photons emitted by a single quantum dot \cite{suprano2023orbital, halevi2024high}.

\subsubsection{Challenges and outlook}
While the progress made in controlling and distributing qudits encoded in transverse-spatial modes of light has been remarkable, several key challenges must be addressed to unlock the full potential of this platform.
One of the major challenges lies in scaling up the dimensionality of the accessible Hilbert space. Although demonstrations of tens or hundreds of spatial modes are now routine at least for certain operations \cite{fontaine2019laguerre}, moving toward thousands or maybe even reaching the regime of one million or more fully controlled dimensions poses significant experimental and theoretical difficulties. These difficulties are manifold: the physical complexity of constructing and stabilizing optical setups grows rapidly with dimensionality, rendering standard bulk optics approaches impractical. In addition, even seemingly small losses and errors inevitably become a problem when accumulated across vast mode spaces, thereby severely reducing the quality of the high-dimensional quantum operation.

Given such immediate challenges with increasing Hilbert space sizes, heuristic or manual approaches to system design and manipulation schemes might no longer suffice. As such, a promising approach will be the development of smart functionality-driven architectures, where hardware implementations are optimized for specific tasks rather than universal control. Here, concepts from compressed sensing, inverse-design, and machine learning will become important tools to allow scalable and efficient system design for a given task, with the first steps already being demonstrated \cite{wang2024ultrahigh}. Moreover, it will also be essential to optimize the design for the best operation under implementation-specific error and imperfection constraints \cite{lib2025building}. Such approaches will result in the need for a deeper theoretical understanding of how noise, loss, and imperfections scale with the state-dimensionality and how their effects might be minimized, thereby guiding the experimental realizations.

Along with these challenges, it is also essential to push innovations forward at the device level. Current spatial light modulators and bulk optical components are insufficient due to a lack of efficiency, speed, and scalability. The future will likely require advanced bulk-optic modulation approaches including programmable metasurfaces \cite{yang2022active} or high-speed MPLC systems \cite{rocha2025self}. Similarly, it will be promising to work towards integrated solutions that are programmable and operate with low loss. Here, a possibility will be to sort spatial modes into waveguide arrays \cite{butow2024generating, sharma2025universal} for which the scheme of arbitrary manipulations is known, or convert spatial modes to multimode waveguide solutions \cite{zhang2025fully}.
\label{sec:AppTransverseToWaveguides} The latter can be seen as a promising route \cite{tzang2019wavefront}, albeit a route that also poses significant hurdles. Mastering modal dispersion, suppressing modal crosstalk, and developing integrated multimode phase modulators are some of the key challenges here \cite{li2019multimode}.
Additionally, complex spatial mode transformations might also be realized in an integrated manner, for which initial steps have already emerged in the form of integrated laser-written but static modulation schemes \cite{wang2025ultracompact}.
However, overcoming these challenges will not only enable the miniaturization of current table-top experiments, but also enable robust deployment in real-world quantum networks.

\label{sec:AppTransverseDetectorChat}In addition, efficiently detecting high-dimensional quantum systems will demand equally sophisticated measurement capabilities. Current standard single-pixel photon detectors must evolve into detector arrays \cite{fleming2025high} with millions of pixels, operating at high speed, low noise, and with minimal jitter \cite{venza2025research}\label{sec:AppTransverseMillionsOfPixels}. 
Along with these hardware improvements, the challenge of handling the enormous amount of generated data will become another essential aspect that will need to be addressed. New strategies for real-time processing and information compression of the data sets, potentially on the order of terabytes per second, will be required, such that this intersection of quantum optics and ``big data'' science is likely to become a defining challenge of the field.
One of the most promising near-term applications is high-dimensional QKD, benefiting from increased information capacity and noise tolerance. 
However, the progress in practical deployment will depend on overcoming key challenges in the transmission and interconnection in a long-distance quantum network.  To increase the distance, reach and channel capacity one can draw on concepts from space-division multiplexing (SDM)~\cite{Richardson2013, Xavier2020}. Structured fibers designed to guide structured modes can provide low-loss, phase-stable, and mode-preserving channels for HD entangled photons and single-photon qudits encoded in transverse modes \cite{Wang2021high, Villalba2023}. Co-design of SDM hardware with quantum requirements--minimising dispersion, mode coupling, birefringence, and inter-core crosstalk \cite{cao2023controlling, ma2023scaling}--will be central to the practical deployment of spatial-mode quantum communication systems and their integration into future quantum networks.  In terms of free-space links,  atmospheric turbulence is a considerable obstacle \cite{wang2022orbital}. 
Adaptive optics, wavefront shaping, and multiplexing strategies may provide partial solutions but will require significant improvement over current implementations.  
Moreover, on the theoretical side, security proofs must be extended and adapted to realistic, high-dimensional implementations, for example through accounting for imperfect devices, crosstalk, realistic noise sources, and finite-key effects. 
Ultimately, robust and efficient QKD protocols along with a user-friendly, real-world implementation tailored to high-dimensional photonic systems will be essential for transforming laboratory advances into field deployments.

Looking ahead into the far future, the grand vision for photonic high-dimensional quantum information in general, and for spatial modes qudits in particular,  is to achieve a robust fully reconfigurable spatial-mode control at massive scale. 
In addition, distortion-free transmission channels with low loss over large distances will be another important milestone to reach, which would not only revolutionize quantum communication (and computation), but also provide a powerful platform for fundamental studies of complex quantum systems. 
Achieving this will be a truly interdisciplinary effort requiring advances in materials science, integrated photonics, control theory, and quantum optics.

\begin{backmatter}
\bmsection{Acknowledgments.}
We thank our lab members throughout the years and our colleagues for their valuable contributions to this field. YB acknowledges support from the Zuckerman STEM Leadership Program and the Israel Science Foundation (Grant No. 2497/21).
RF acknowledges support through the Research Council of Finland (PREIN - decision 346511) and the European Research Council under project TWISTION (Grant Agreement 101042368). MM acknowledges support from the UK Engineering and Physical Sciences Research Council (EPSRC) (EP/Z533208/1, EP/Z533166/1), European Research Council (ERC) Starting Grant PIQUaNT (950402), and the Royal Academy of Engineering Chair in Emerging Technologies programme (CiET-2223-112).
 \end{backmatter}

\subsection{Path-encoding} \label{Sec:App_Path}
\author{Taira Giordani\authormark{21} 
Fabio Sciarrino\authormark{21}, 
Yun Zheng\authormark{22}, 
Jianwei Wang\authormark{22}}
\address{\authormark{21}Dipartimento di Fisica, Sapienza Universit\`{a} di Roma,
Piazzale Aldo Moro 5, I-00185 Roma, Italy}
\address{\authormark{22} State Key Laboratory for Mesoscopic Physics, School of Physics, Peking University, Beijing, 100871, China}

\subsubsection{Early work}

The transition from bulk optical setups to integrated waveguide architectures represents a crucial evolution toward realizing high-dimensional quantum photonic processors \cite{Wang2020_review, Pelucchi2022, Moody_2022, Giordani2023}. The encoding strategy most compatible with photonic integrated circuits (PICs) is path encoding, wherein quantum information is mapped onto the discrete spatial modes of single photons \cite{OBrien2009Photonic, Flamini_rev}. Early efforts focused on miniaturizing multi-port linear interferometers—the primary optical devices for manipulating path-encoded states—into monolithic chips to enhance phase stability and scalability. In seminal investigations, fundamental building blocks such as beam-splitters and phase retarders were implemented as integrated directional couplers and phase shifters (see Sec. 2.4). These early works demonstrated the feasibility of fabricating such components in silica-on-silicon waveguides \cite{Politi08} and in glass \cite{Sansoni2010} via femtosecond laser writing (FLW) \cite{Gattass2008}. Reported proof-of-principle demonstrations include static logic gates, such as the first integrated probabilistic CNOT gates \cite{Politi08, Crespi2011}, and small-scale algorithms. These range from the quantum simulation of transport with bosonic/fermionic statistics \cite{Peruzzo2010, sansoni2012quantum, crespi2013anderson, Crespi16, peruzzo2011} and proof-of-concept implementations of Shor’s and Grover’s algorithms \cite{Politi2009, Ciampini2016}, to a variety of Boson Sampling experiments \cite{Tillmann13boson, Crespi2013, Broome13boson, Spring2013, Carolan14, Spagnolo14, Bentivegna2015, Giordani18}.

Subsequent technological advances sought materials capable of higher component density and lower optical loss, such as silicon (Si) and Silicon-Nitride (SiN) waveguides \cite{Silverstone2016}. A pivotal development was the achievement of active phase control using thermo-optic phase shifters, where metallic heaters deposited on the waveguides modify the refractive index via the thermo-optic effect \cite{Silverstone2013, Flamini2015}. These achievements paved the way for reconfigurable PICs, enabling the investigation of complex interferometers and the development of integrated parametric single-photon sources \cite{Silverstone2013, Atzeni:18}. Most representative experiments conducted on this second generation of PICs include the realization of Variational Quantum Eigensolvers (VQEs) \cite{peruzzo2014variational, Santagati2019}, simulations of quantum transport phenomena \cite{pitsios2016photonic, Harris2017}, and quantum machine learning algorithms, such as Hamiltonian learning \cite{wangpaesani}, quantum reinforcement learning \cite{Saggio2021}, and quantum optical memristors \cite{Spagnolo2022}. Progress has also been made in phase estimation \cite{Paesani2017}, quantum metrology algorithms \cite{Polino:19, cimini2024variational} and chip-to-chip interconnections \cite{Wang2016_chip2chip, Llewellyn2020, Zheng2023, ChipToChip2025, Liu2025_chip2chip}.

\begin{figure}
    \centering
    \includegraphics[width=0.8\linewidth]{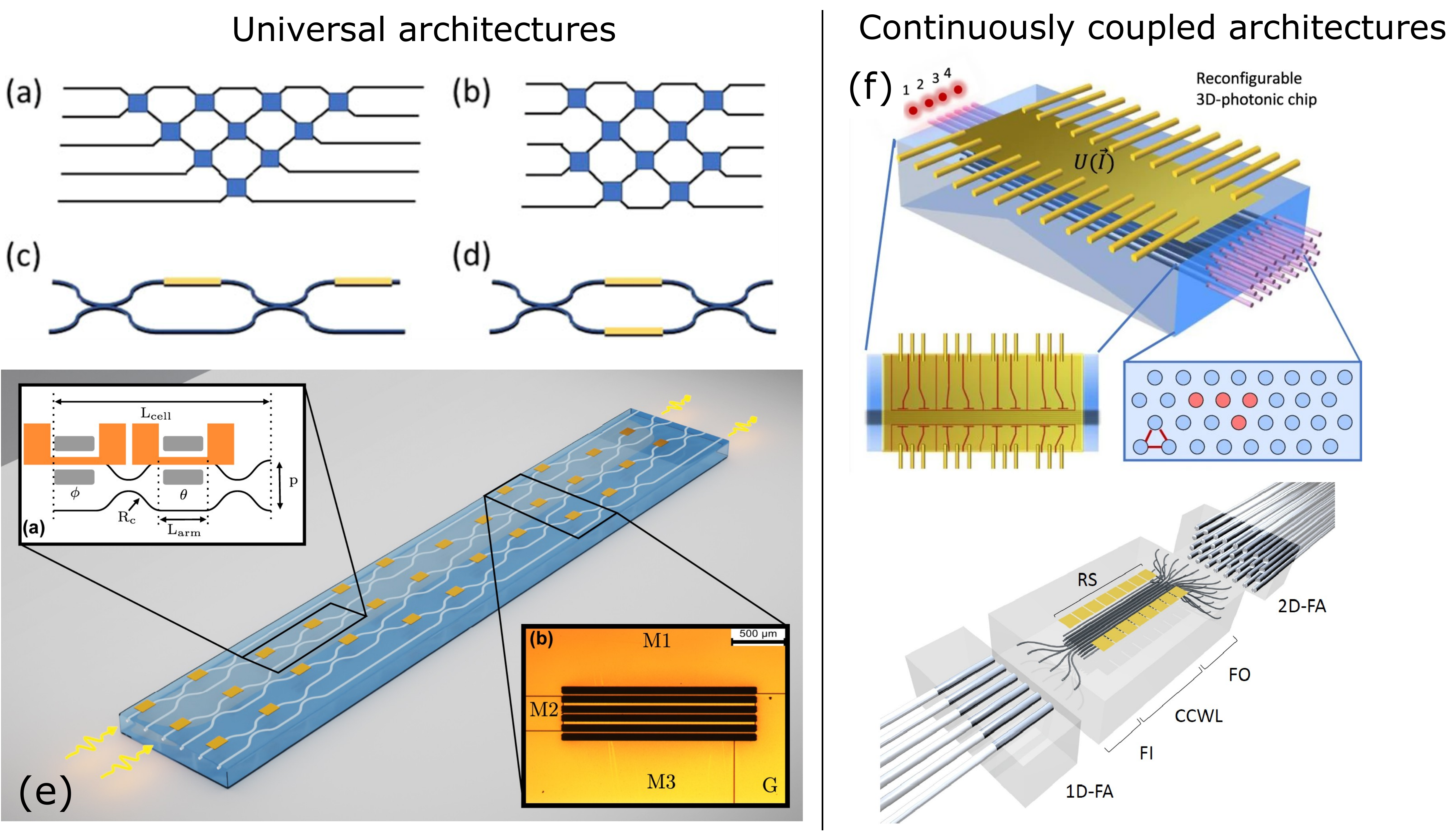}
    \caption{\textbf{Path-encoding architectures.} (a) Illustration of the Reck decomposition \cite{Reck_1994}, in which the blue squares represent tunable beam-splitters and phase-shifters. (b) The Clements decomposition \cite{Clements:16}. (c) Tunable beam-splitters constructed from MZIs featuring a single internal phase-shifter and an external phase-shifter to constitute a full unit cell. (d) A symmetric MZI variant that uses two internal phase shifts to reduce the footprint of the unit cell \cite{Bell21:CompactifyLO}. Panels a-d) from Ref. \cite{Bell21:CompactifyLO} reproduced without modifications under CC-BY 4.0 license. (e) Example of a 6-mode reconfigurable universal integrated circuit according to the Clements layout. Figure from Ref. \cite{Pentangelo2024} reproduced without modifications under CC-BY 4.0 license. (f) Example of a reconfigurable continuously-coupled device with 32 modes arranged in a triangular lattice. In the inset, the resistors (RS), the fan-in (FI) and fan-out (F0) sections, and the continuously-coupled waveguide lattice (CCWL). Figure from Ref. \cite{Hoch2022} reproduced without modifications under CC-BY 4.0 license.}
    \label{fig:architecture}
\end{figure}

In these demonstrations of path-encoded quantum algorithms, the Mach-Zehnder Interferometer (MZI)—comprising two unbiased directional couplers and phase shifters—serves as the fundamental computational unit (Sec. 2.4). These interferometers enable arbitrary single-qubit rotations \cite{OBrien2009Photonic, Flamini_rev} and function as building blocks for larger linear optical networks, facilitating the design of universal layouts. These decomposition schemes allow for the implementation of any unitary transformation over a set of $m$ spatial modes, opening the path toward programmable, multi-purpose high-dimensional quantum processors. The first identified universal layout was the triangular scheme \cite{Reck_1994}, which was demonstrated in the first programmable and universal 6-mode PIC for elementary gates, Boson Sampling \cite{Carolan15}, and the simulation of molecular vibronic spectra \cite{Sparrow2018}. Recent studies have shown that this triangular layout can be condensed into a rectangular mesh of $m(m-1)/2$ MZIs, as first proposed in Ref. \cite{Clements:16} and refined in Ref. \cite{Bell21:CompactifyLO} (see Fig.~\ref{fig:architecture}a-e). These approaches have been employed for the more recent demonstrations of reconfigurable universal PICs with 6- \cite{Giordani2023_context, Hoch2024, Yin2025}, 8- \cite{Rodari2025, hoch2025quantum}, 12- \cite{Monbroussou2025}, and 24-mode \cite{barzaghi2025lowloss24modelaserwrittenuniversal} via the FLW fabrication method \cite{Ceccarelli2020, Pentangelo2024}. Universal devices with 12- \cite{Taballione_2021, maring2024versatile, CorreaAnguita2025} and 20-modes \cite{Taballione2023} have also been reported in SiN chips.

The MZI networks are not the exclusive architecture for manipulating path-encoded quantum states. Advanced waveguide fabrication technologies enable the realization of optical lattices with diverse geometries, facilitating the creation of arrays comprising multiple parallel waveguides that interact via continuous evanescent coupling. This architecture represents an alternative strategy for realizing large-scale circuits, offering favorable scaling properties with respect to optical losses and fabrication imperfections. In this framework, the evolution implemented by the circuit is entirely determined by the lattice geometry, which defines the system's Hamiltonian, $H$. Consequently, continuously coupled devices are particularly well-suited for quantum simulation, as they allow for the direct mapping of a physical system's Hamiltonian onto the circuit \cite{Pulios, Preiss_waveguide_array, caruso2016fast, Jiao2021, Tang2021}, as well as for large-scale demonstrations of Boson Sampling \cite{Paesani2019}. While earlier works were limited to static circuits—representing a single, fixed random unitary transformation—recent advancements have introduced the dynamic modulation of circuit parameters (see Fig.~\ref{fig:architecture}f). This capability enables reconfigurable devices capable of generating multiple unitary evolutions, as reported in a 32-mode PIC with thermal phase shifters realized via femto-second laser writing \cite{Hoch2022} and in Lithium Niobate chips featuring electro-optical modulators \cite{Youssry2024, Yang2024, Yang2025}.

\subsubsection{Current developments}
\label{sec:Applications_Path_CurrentDevelopments}

Recent progress in path-encoded single-photon processing has been defined by the synergy between high-performance material platforms and the implementation of photon-native algorithms via adaptive, variational quantum protocols designed for near-term integrated photonic quantum technologies. A key driver of this progress has been the refinement of Boson Sampling (BS) \cite{AA} and its variants—Gaussian Boson Sampling (GBS) \cite{Lund2014, Hamilton2017}, Nonlinear Boson Sampling (NLBS) \cite{spagnolo2023nonlinear}, and Adaptive Boson Sampling (ABS) \cite{chabaud2021quantum, hoch2025quantum, Monbroussou2024}—which exploit the complexity of multiphoton interference to address computational problems intractable for classical systems. The field is evolving from demonstrating quantum computational advantage via the non-universal Boson Sampling problem \cite{AA}, solvable naturally on photonic platforms through high-dimensional linear optical circuits \cite{Zhong_GBS_supremacy, zhong2021phase, Madsen2022, Deng2023, liu2025robustq}, toward the ambitious goal of universal quantum computation, which necessitates nonlinear operations and active feed-forward control \cite{Knill2001, Briegel2009, Bartolucci2023Fusion}.

\paragraph{Sampling algorithms} Initial efforts have focused on leveraging the intrinsic properties of the Boson Sampling distribution under fully linear evolution, governed by matrix permanents and hafnians, for practical applications. This approach relies on the mathematical connection between the matrix permanent and the counting of perfect matchings in bipartite graphs, while the more general hafnian relates to perfect matchings in generic undirected graphs \cite{VALIANT1979189, Caianiello}. By encoding adjacency matrices into the sub-matrices of an integrated interferometer, researchers have formulated GBS-based algorithms to solve complex graph-related problems, such as identifying dense subgraphs, finding maximum cliques, and estimating graph similarity \cite{Shuld_GBS_graphsimilarity, GBSGraphTheory1, GBSGraphTheory3, Bromley_2020}. Successful tests on small-scale integrated photonic devices have validated the platform's potential for combinatorial optimization \cite{Arrazola2021, Qiang2021}. Simultaneously, exploiting the similarity between the photon’s Hamiltonian in path-encoded circuits and the vibronic transitions of molecules \cite{Banchi_vibronic, Huh2015_vibronic}, programmable integrated chips have been deployed to simulate molecular vibronic spectra \cite{Sparrow2018, Arrazola2021}, serving as specialized hardware for chemical dynamics. Furthermore, multiphoton interference, exemplified by the Hong-Ou-Mandel (HOM) effect \cite{HOM, Bouchard2020}, enables the direct measurement of quantum state overlaps. This parallelism with the SWAP-test \cite{Garcia-Escartin} has been investigated in reconfigurable PICs for estimating multiphoton indistinguishability \cite{Pont2022, Rodari2025, CorreaAnguita2025} and in Quantum Machine Learning (QML) for computing kernels of data encoded in path-based states \cite{Yin2025}. Finally, integrated Boson Samplers are being explored for randomness manipulation, including modular photonic Quantum Bernoulli Factories \cite{Hoch2024} for accelerating the generation of Bernoulli variables \cite{Dale2015, Patel2019} and manipulating qubit amplitudes independently from the input bias \cite{Jiang2018, Liu2021}, as well as accelerating Monte Carlo simulations \cite{anguita2025_boson_sampler}.

\paragraph{Variational photonic integrated circuits.} Building upon linear evolution, a central pillar of recent algorithmic development is the adoption of the variational quantum algorithm framework, which introduces classical feedback loops to train circuit parameters. These algorithms operate via a hybrid quantum-classical loop where a parameterized quantum circuit prepares a state $|\psi(\theta)\rangle$ dependent on tunable parameters $\theta$, such as the phase shifts in PICs \cite{peruzzo2014variational, Santagati2019, wangpaesani, Zhang2021_neural, maring2024versatile, Hoch2025_cloning, Baldazzi2025}. A classical optimizer minimizes a cost function by updating these parameters, a task particularly suited for universal and programmable PICs given their high component density. A critical challenge lies in efficient gradient estimation since standard finite-difference methods are noise-susceptible on hardware. To address this, the field has developed a "photonic parameter shift rule" tailored for integrated processors \cite{cimini2024variational, pappalardo2025photonic, Hoch2025, facelli2024}. This technique exploits the trigonometric dependence of measurement probabilities on phase parameters \cite{Schuld2019, Wierichs2022} to calculate exact gradients by evaluating the cost function at macroscopically shifted values (e.g., $\pm \pi/2$). This noise-robust approach has been successfully applied in VQE demonstrations \cite{Hoch2025} and variational quantum metrology \cite{cimini2024variational}. Beyond gradient-based methods, significant progress using gradient-free optimization has been made in tasks ranging from Hamiltonian eigenvalue estimation \cite{Santagati2019, wangpaesani, Baldazzi2025, maring2024versatile} to variational searches for U-NOT gates \cite{Hoch2025} and quantum cloning \cite{Hoch2025_cloning}.

\paragraph{Adaptive platforms towards quantum neural networks} To transcend the limits of linear optics and enable advanced QML protocols, researchers are introducing nonlinearities via adaptive operations. Two primary paradigms have emerged. The first, Adaptive Boson Sampling (ABS), involves dynamically reconfiguring the linear optical circuit based on intermediate measurement outcomes \cite{chabaud2021quantum, hoch2025quantum}. This adaptive mechanism updates the unitary transformation for the remaining photons—a technique emulated via post-selection in recent experiments with 6- and 8-mode universal PICs for kernel estimation \cite{hoch2025quantum}. The second paradigm, adaptive state injection \cite{Monbroussou2024}, shifts adaptivity to the generation stage, where measurement patterns dictate photon re-injection. This architecture was realized in a multi-processor platform (8- and 12-mode chips) functioning as a Photonic Quantum Convolutional Neural Network for image classification \cite{Monbroussou2025}, thanks to the parallelism of the architecture with Hamming-weight preserving networks \cite{monbroussou2025subspace}. Also demonstrated via post-selection, this work provides a proof-of-concept that adaptive interconnects can effectively act as non-linear activation functions in quantum neural architectures \cite{Monbroussou2025}.

\begin{figure}[htbp!]
    \centering
    \includegraphics[width=0.8\linewidth]{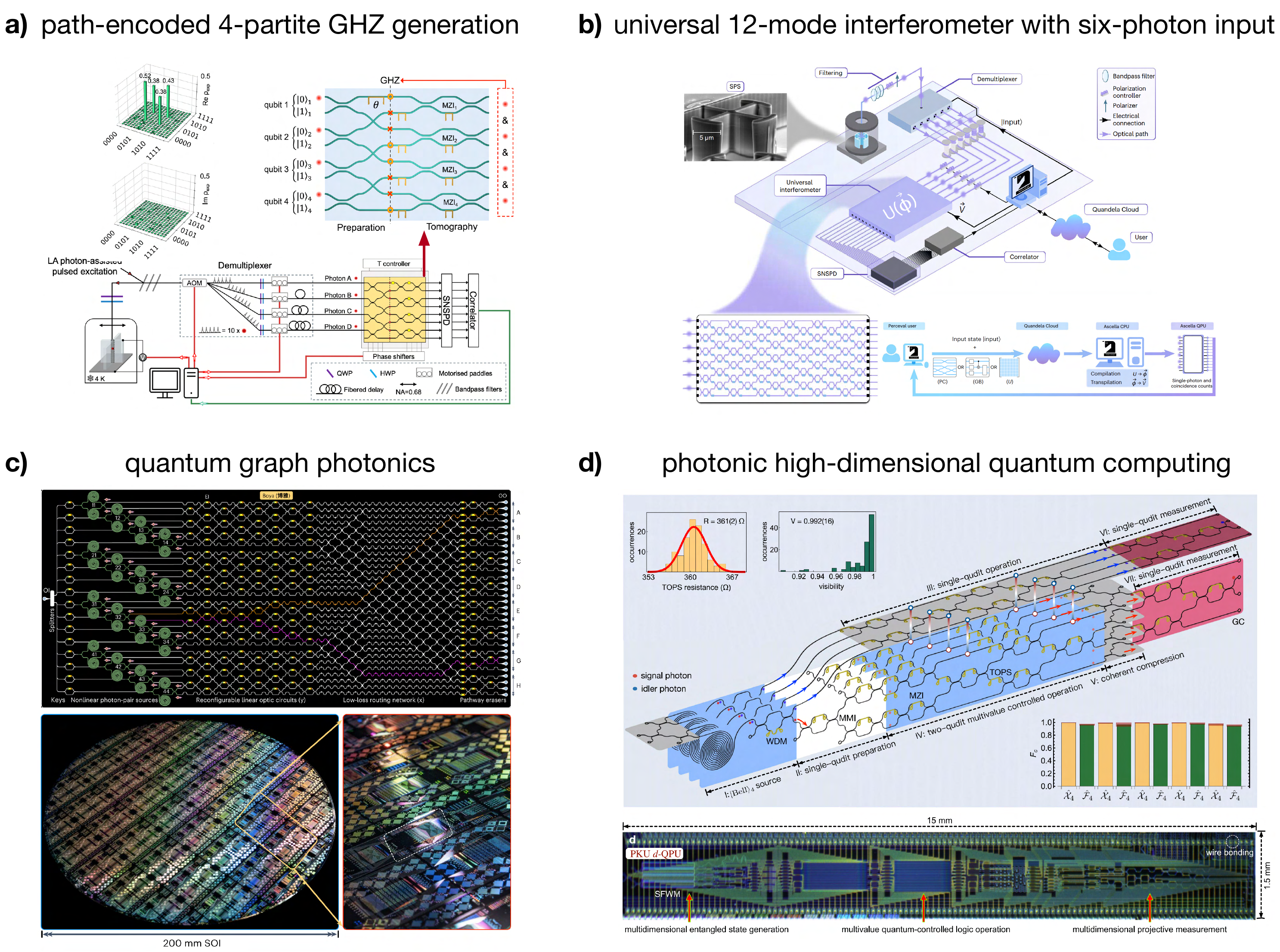}
    \caption{\textbf{Complex states generation and quantum computing.} 
    a) and b): on-chip multi-mode interferometers enable the generation of complex multiphoton entangled states and QIP, driven by off-chip quantum-dot single-photon sources.
    c) very-large-scale integrated quantum graph device. Multiphoton high-dimensional genuine entanglement could be generated and processed with on-chip parametric sources and MZI arrays.
    d) two-partite 4-dimensional photonic quantum computing platform.
    Figures from Ref. \cite{Pont2024High}, Ref. \cite{maring2024versatile}, Ref. \cite{Bao2023} and Ref. \cite{Chi2022} reproduced without modifications under CC-BY 4.0 license.}
    \label{fig:computing}
\end{figure}

\paragraph{Resources states generation and quantum computing.} Parallel to these algorithmic implementations, significant strides have been made in the on-chip generation and manipulation of complex entangled resources, which are foundational for measurement-based quantum computing and quantum error correction \cite{Briegel2009, Bartolucci2023Fusion}. By leveraging post-selection and heralding mechanisms within programmable linear optical networks, researchers have successfully engineered high-fidelity entangled states, such as four-photon Greenberger-Horne-Zeilinger (GHZ) states \cite{Pont2024High, Caruccio2025} in post-selection (Fig.~\ref{fig:computing}a) and in a heralded six-photon scheme (Fig.~\ref{fig:computing}b) in Ref. \cite{maring2024versatile} by interfacing high-brightness deterministic single photon sources based on quantum dot \cite{Somaschi2016, Ding_quantum_dot, Margaria2025} with PICs devices.  Versatile fully-on chip graph state generation has been previously reported in silicon platforms \cite{Adcock2019}. These architectures enable the integration of parametric sources on the same chip and the implementation of two-qubit processing through reconfigurable fusion gates \cite{Qiang2018, Adcock2019}, paving the way for exploring error-protected logical qubits encoded in photonic chips \cite{Vigliar2021}. Furthermore, the dimensionality of the generated states has been expanded beyond standard qubits to high-dimensional qudits \cite{Wang2018, Chi2022,Huang2024} (Fig.~\ref{fig:computing}c-d).  
A landmark demonstration showcased the monolithic integration of more than 550 photonic components on a single chip, enabling the on-chip generation, manipulation, and measurement of bipartite entangled states with dimensions up to $15\times15$\cite{Wang2018}.
A typical framework could be implemented to effectively process high-dimensional quantum information with linear combination of unitary operations\cite{Zhou2011}, which consists of the entanglement generation, space expansion, local unitary operation, and coherent compression. This approach has a broad range of applications\cite{Qiang2018, Santagati2019, wangpaesani} and has recently been used to realise on-chip high-dimensional quantum computation\cite{Chi2022} (Fig.~\ref{fig:computing}d) and to generate hypergraph states\cite{Huang2024}. By exploiting high-dimensional encoding, multiple qubits can be embedded within a single photon, thereby enhancing information capacity\cite{Li2025} and enabling multi-qubit gate operations to be implemented deterministically via equivalent local transformations in a higher-dimensional Hilbert space—operations that would otherwise be intrinsically probabilistic in linear-optical systems due to the lack of photon–photon interactions\cite{Knill2001}. With the very-large-scale integration of silicon PIC (Fig.~\ref{fig:computing}c), a graph-theoretical programmable quantum photonic device was developed, advancing both the number of controllable photons and the accessible dimensionality\cite{Bao2023}. The device features a general architecture that can be reprogrammed to generate genuine multipartite HD entanglement and to measure the probability distributions of the modulus-squared permanent and hafnian matrix functions for more complex graphs, according to the graph theory. About 2,500 components were monolithically integrated, representing the largest quantum PIC up to date.
These advancements rely on the interference of photons at reconfigurable MZI arrays, enabling the construction of proof-of-concept cluster states, via measurements, required for fault-tolerant optical quantum computing.

\begin{figure}[htbp!]
    \centering
    \includegraphics[width=0.8\linewidth]{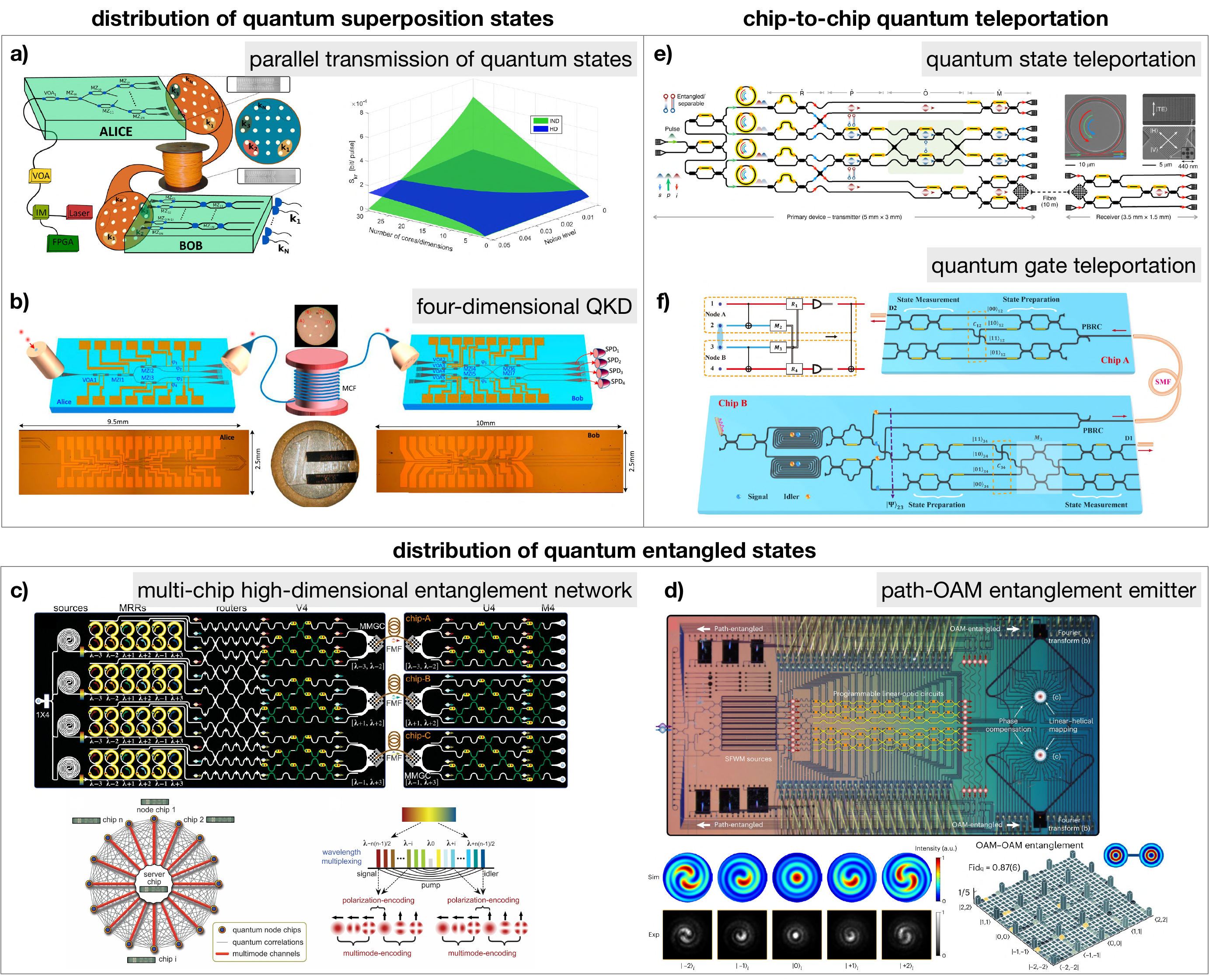}
    \caption{\textbf{Distribution of quantum states between chips.} 
    a) and b): Superposition states transmission via one-to-one mapping from on-chip paths to MCF spatial modes.
    c) and d): Entangled states distribution by using coherent DoF conversion techniques. 
    e) and f): chip-to-chip quantum state and gate teleportation.
    Figures from Ref. \cite{Bacco2017}, Ref. \cite{Ding2017}, Ref. \cite{Zheng2023}, Ref. \cite{HuangOAM25}, Ref. \cite{Llewellyn2020} and Ref. \cite{ChipToChip2025} reproduced without modifications under CC-BY 4.0 license.}
    \label{fig:distribution}
\end{figure}

\paragraph{Distribution of quantum states} As the scale of monolithically integrated qPICs continues to grow, to meet the demands of quantum networks\cite{QuantumnetworkReview}, interconnection between multiple chips has become an interesting focus. A primary challenge is to achieve high-fidelity coherent transmission of quantum states or entangled states, particularly those encoded in high dimension, across different chips. For on-chip path encoding, a natural approach is to map each path directly onto a corresponding spatial mode in fibre or free space, e.g. multicore fibres (MCF). 
Using grating-coupler-based MCF fan-in/fan-out techniques, multiple qubits (Fig.~\ref{fig:distribution}a) and four-dimensional quantum states (Fig.~\ref{fig:distribution}b) could be distributed between silicon PICs, enabling the demonstration of parallel communication \cite{Bacco2017} and high-dimensional quantum key distribution\cite{Ding2017}. Insertion loss requires further optimisation, and active phase-stabilisation techniques must be incorporated to ensure stable coherent transmission over long fibres\cite{DaLio21:2KmHDQKD}.
Another way is to exploit coherent conversions between different photonic DoFs. Coherent path-to-polarization conversion provides an effective solution, allowing entanglement distribution between chips through single-mode fibre channels\cite{Wang2016_chip2chip, Llewellyn2020, Hui2022, ChipToChip2025, Liu2025_chip2chip}. However, this approach is inherently incompatible with high-dimensional encoding. By employing hybrid multiplexing techniques, multiple photonic DoFs can be coherently controlled. Wavelength-multiplexed entangled photon pairs can establish correlations across many quantum nodes, and coherent conversion from on-chip path-modes to spatial-mode and polarization DoFs in few-mode fibres enables the distribution of high-dimensional entangled states between chips, forming a fully connected multichip entanglement network\cite{Zheng2023} (Fig.~\ref{fig:distribution}c). An integrated optical entangled quantum vortex emitter was developed to generate and control vortex entanglement in free space, coherently transitioning from on-chip high-dimensional path entanglement\cite{HuangOAM25} (Fig.~\ref{fig:distribution}d). Such chip-to-free-space interfaces offers the possibilities of processing quantum information on a chip and regulating the quantum state transmission beyond the chip, which will be practically important in quantum communication and networks\cite{Zheng2023}.
The ability to control multi-qubit on-chip and distribute entangled states across different quantum nodes opens the possibility for advanced quantum applications, such as quantum teleportation, which plays a vital role in QIP\cite{Gottesman1999, teleportchip, polacchi2024teleportation}. Four-qubit GHZ states were generated on chip with one photon transmitted to another chip via path-polarization DoF conversion\cite{Llewellyn2020} (Fig.~\ref{fig:distribution}e). Bell measurement was carried out to realise the quantum teleportation of the heralded single-qubit state between chips. Although the demonstration relied on multi-qubit states, the combination of complex on-chip quantum state manipulation with coherent inter-chip connectivity provides important insights for the development of quantum networks. Later, chip-to-chip teleportation of a CNOT gate was demonstrated facilitated by high-dimensional path-encoded quantum states\cite{ChipToChip2025} (Fig.~\ref{fig:distribution}f).
Integrating high-dimensional entanglement distribution between chips\cite{Zheng2023} with high-dimensional Bell-state measurements\cite{Luo2019} promises to enable far more complex QIP tasks across multiple photonic chips.

\subsubsection{Challenges and outlook}
\label{sec:Applications_PathChallengesAndOutlook}

\begin{figure}
    \centering
    \includegraphics[width=\linewidth]{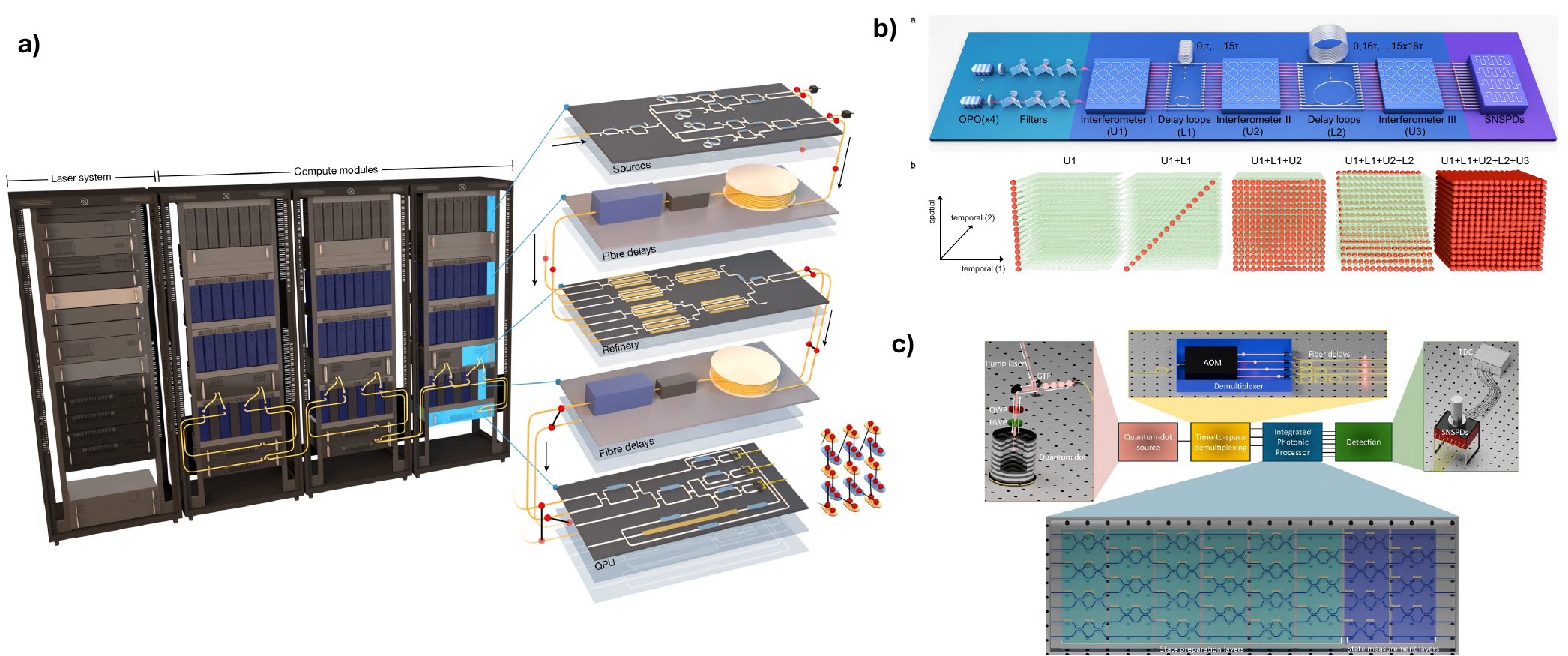}
    \caption{\textbf{Toward modular and hybrid path-encoded integrated photonic platforms.} a) The modular photonic quantum processor \emph{Aurora} envisages off-chip fast modulation and optical delays to enable feed-forward operations. Figure from Ref. \cite{AghaeeRad2025} reproduced without modifications under CC BY-NC-ND 4.0 license. b) The \emph{Jiuzhang 4.0} processor combines path-encoding and temporal-encoding to scale the size of Boson Samplers. Figure from Ref. \cite{liu2025robustq} reproduced without modifications under arXiv.org perpetual, non-exclusive license. c) Hybrid high-dimensional platform featuring interfaces of universal PICs and quantum dot sources. Figure from Ref. \cite{Caruccio2025} reproduced without modifications under CC-BY 4.0 license}
    \label{fig:vision}
\end{figure}

Despite the remarkable progress in scaling integrated photonic processors, the roadmap toward fault-tolerant high-dimensional processing faces critical engineering hurdles. The most pervasive challenge remains optical loss, which scales detrimentally with circuit depth and complexity. Beyond passive loss, the active control of these devices presents a fundamental bottleneck. Current large-scale programmable circuits rely heavily on thermo-optic phase shifters, which suffer from slow modulation speeds (kHz range) and significant power dissipation. This latency precludes the implementation of rapid adaptive operations and active feed-forward, which are strict requirements for scalable universal quantum computing schemes \cite{Briegel2009, Bartolucci2023Fusion} and adaptive protocols \cite{chabaud2021quantum, hoch2025quantum}. Consequently, the inability to reconfigure the circuit within the coherence time of the photonic qubit currently limits the transition from static sampling tasks to deterministic logic.

To overcome these limitations, the future outlook envisions a paradigm shift toward monolothic approaches with all the components, single-photon sources, evolution and detection integrated in the same PIC \cite{Alexander2025} or to modular and distributed architectures where computational tasks are distributed across networked photonic chips \cite{Wang2016_chip2chip, Llewellyn2020, Alexander2025, AghaeeRad2025}. Realizing this vision requires the development of ultra-low-loss chip-to-chip interconnections allowing high-dimensional states to flow coherently between modules\cite{Zheng2023}. To bypass the bandwidth limitations of on-chip thermal tuning, architectures are increasingly incorporating off-chip fast modulation or integrating high-speed electro-optic materials via hybrid bonding, enabling the nanosecond-scale switching necessary for feed-forward logic. Such a vision has been recently validated in the modular photonic processor \emph{Aurora} \cite{AghaeeRad2025} that demonstrated feed-forward operations among PICs modules (Fig.~\ref{fig:vision}a).
Also, scalable fault-tolerant quantum computing requires hardware that combines millions of qubits with equally scalable classical control, a major unresolved challenge. Silicon photonics offers a path toward such integration by co-integrating electronics and quantum photonics\cite{Kramnik2025}, with future scalability likely depending on cryogenic operation\cite{Alexander2025} and heterogeneous integration\cite{Xiang2023}.

Moreover, the long-term evolution of the platform points toward hybrid encoding schemes. To maximize information capacity, future processors will likely exploit hybrid encoding across DoFs, for example combining spatial path-encoding with time-bin multiplexing. This strategy has already enabled robust quantum computational advantage in bulk high-dimensional GBS experiments with over 3000 photons and 8000 spatial-temporal modes, as reported by the \emph{Jiuzhang 4.0} machine \cite{liu2025robustq} (Fig.~\ref{fig:vision}b). Other hybrid encoding are possible through the interface of deterministic single-photon sources with low-loss passive routing and evolution in integrated optical circuits, and high-efficiency detection. A particularly promising direction involves interfacing path-encoded photons with deterministic solid-state emitters such as quantum dots \cite{Pont2022, Pont2024High, Wang2023_SiN, maring2024versatile, Rodari2025, hoch2025quantum, Monbroussou2025, Caruccio2025}, like in \emph{Ascella}\cite{maring2024versatile} (Fig.~\ref{fig:computing}b) and in \emph{Qolossus} machines \cite{Rodari2025, hoch2025quantum, Monbroussou2025} (Fig.~\ref{fig:vision}c). These systems can also serve as hybrid spin-photon interfaces in which the spin of the artificial atom provides a stationary memory that enables deterministic generation of spin-photon graph states \cite{Huet2025Deterministic}. In the long-term vision, a hybrid platform working with quantum emitters of spin-photon states and processing in path-polarization encoded integrated circuits would minimize the resource overhead for fusion-based computing \cite{Wein2024minimizing}.

\begin{backmatter}
\bmsection{Acknowledgments}
 This work is supported by the PNRR MUR project PE0000023-NQSTI (Spoke 4) and by the European Union’s Horizon Europe research and innovation program under EPIQUE Project (Grant Agreement No. 101135288). Y.Z. and J.W. acknowledge support from the National Natural Science Foundation of China (Grant Nos. 12325410 and 62505005) and the Quantum Science and Technology–National Science and Technology Major Project (Grant Nos. 2021ZD0301500 and 2024ZD0302401).

\bmsection{Disclosures}
The authors declare they have no competing interests.

\bmsection{Data Availability Statement}
There are no data associated with the paper.

\end{backmatter}

\section{Theory}
\label{sec:Theory}

\author{Marcus Huber\authormark{23,24},
Armin Tavakoli\authormark{25},
Roope Uola\authormark{26,27},
Nicolas Brunner\authormark{28},
Nicolai Friis\authormark{23}}
\address{\authormark{23} Atominstitut, Technische Universit\"at Wien, Stadionallee 2, 1020 Vienna, Austria}
\address{\authormark{24} Institute for Quantum Optics and Quantum Information (IQOQI), Austrian Academy of Sciences, Boltzmanngasse 3, 1090 Vienna, Austria}
\address{\authormark{25}Physics Department and NanoLund, Lund University, Box 118, 22100 Lund, Sweden.}
\address{\authormark{26}Department of Physics and Astronomy, Uppsala University, Box 516, 751 20 Uppsala, Sweden}
\address{\authormark{27}Nordita, KTH Royal Institute of Technology and Stockholm University, 10691 Stockholm, Sweden}
\address{\authormark{28}Department of Applied Physics University of Geneva, 1211 Geneva, Switzerland}

\subsection{Early Work}

{\noindent}With the advance of quantum technologies that enable access to and control over higher-dimensional Hilbert spaces comes the question of how high-dimensional quantum systems and the entanglement therein can be compactly described and quantified, as well as how this degree-of-freedom can be made use of and how one can assess this usefulness. In this section, we briefly describe the essential theoretical tools available for addressing these questions, along with recent advances and open challenges in the theory of high-dimensional quantum information processing. 
For more detailed reviews on high-dimensional entanglement theory, we refer to~\cite{GuehneToth2009, FriisVitaglianoMalikHuber2019, ErhardKrennZeilinger2020}, while for applications of HD systems we refer to reviews on quantum communication \cite{Cozzolino2019} and quantum computing \cite{WangHuSandersKais2020, KiktenkoNikolaevaFedorov2025}.


\subsubsection{HD entanglement and distillabilty}

{\noindent}One of the earliest focus of work in quantum information theory has been the characterisation of entanglement in HD systems. For bipartite pure states, the problem is solved via the Schmidt decomposition. For mixed states, the situation is much more complicated. On the one hand, all entangled states can be detected in low dimensions ($2\times 2$ and $2\times 3$) via the the positive partial transpose (PPT) criterion~\cite{Peres1996, HorodeckiMPR1996}. On the other hand, there exist entangled states in higher dimensions that satisfy the PPT criterion. Such PPT-entangled states are termed \emph{bound entangled}. Notably, such states cannot be distilled, meaning that it is impossible to extract from them a pure and maximally entangled state (the unit of bipartite entanglement) via local operations and classical communication (LOCC). This highlights a fundamental difference between HD entanglement compared to the case of qubits, since all qubit entangled states are distillable. The detection and characterization of bound-entangled states has sparked considerable activity early on~\cite{Horodecki1997, BennettDiVincenzoMorShorSmolinTerhal1999, BrussPeres2000, DuerCiracLewensteinBruss2000, DiVincenzoMorShorSmolinTerhal2003, BertlmannKrammer2008b, BertlmannKrammer2008c}.


\subsubsection{Generalized Bloch decomposition}

{\noindent}At the same time, the study of bound entanglement necessitated the development of tools suitable for the description of high-dimensional quantum systems in the first place. For this purpose, the generalization of the Bloch sphere~\cite{Bloch1946} from two-dimensional qubits to $d$-dimensional qudits proved to be useful and was widely considered~\cite{Kimura2003, KimuraKossakowski2005, KryszewskiZachzial2006, JakobczykSiennicki2006, Mendas2006, BertlmannKrammer2008}. Specifically, the \emph{generalized Bloch decomposition} entails the choice of a matrix basis $\{ \Gamma_i\}_{i}$ that includes the identity $\mathds{1}_{d}$ along with $d^2 \,-\, 1\,$ traceless and orthogonal operators $\Gamma_{i}$, i.e., which satisfy $\Tr(\Gamma_i) \,=\, 0$ and $\Tr\bigl(\Gamma^\dag_i \Gamma_j \bigr) \,=\, 
d\delta_{ij}$, 
see, e.g.,~\cite{KrammerPhD2009, AsadianErkerHuberKlockl2016} or~\cite[Chapter~17]{BertlmannFriis2023} for compact overviews (but note the variety in conventions). Then, any qudit density  matrix $\rho$ in a $d$-dimensional Hilbert{\textendash}Schmidt space can be written as 
\begin{equation} \label{bvgenbasis}
    \rho \,=\, \tfrac{1}{d}\,\Bigl(\mathds{1} \,+\, 
    \sum\limits_{i=1}^{d^{2}-1}b_{i} \,\Gamma_{i}\Bigr) \;,
\end{equation}
where the real values $b_{i}=
\Tr\bigl(\rho\,\Gamma_i\bigr)$ form the components of a generalized Bloch vector $\vec{b} \in \mathbb{R}^{d^2-1}$ with $|\vec{b}\nr|^{2}\leq d-1$. This expression reduces to the usual Bloch decomposition for $d=2$ where the $\Gamma_{i}$ are the usual Pauli matrices. For arbitrary $d$ and $\Gamma_{i}$, any valid density operator admits such a decomposition in a generalised sphere, but not every choice of $|\vec{b}\nr|^2\leq d-1$ yields a positive semidefinite operator $\rho\geq0$.


\subsubsection{HD entanglement and Bell nonlocality}

Performing local measurements on a shared entangled state can result in nonlocal quantum correlations, as witnessed via a Bell inequality violation. This phenomenon has also been discussed for HD entanglement. Tailored Bell inequalities have been developed \cite{Collins2002} and the relation between Bell nonlocality and entanglement has been shown to be non-monotonous \cite{Acin2002}. It has also been shown that arbitrarily large Bell inequality violations are possible for systems of increasing dimension \cite{Buhrman2012}. On the other hand, classes of entangled states have been shown to admit local hidden variable models \cite{Werner1989, Almeida2007, Wiseman2007}, implying they cannot lead to Bell inequality violation, whereas bound entangled states can violate Bell inequalities \cite{Vertesi2014}. 

\subsubsection{Quantifying entanglement dimensionality}

{\noindent}A central question is to quantify the dimensionality of entanglement in HD systems. A prevalent approach consists in quantifying how many entangled qubit pairs can be coherently encoded in a given high-dimensional system. 
Two important figures of merit for high-dimensional entanglement in a bipartite state $\rho$ are the \emph{Schmidt number} $d_{\rm{ent}}$ or \emph{entanglement dimensionality} (ED)~\cite{Terhal2000},
\begin{align}
    d_{\rm{ent}}:=\inf\limits_{\mathcal{D}(\rho)} \left( \max\limits_{\ket{\psi_{i}}} \Bigl\{ \operatorname{rank}\bigl[\,\Tr\subtiny{0}{0}{B}\bigl(\ket{\psi_{i}}\!\!\bra{\psi_{i}}\bigr)\bigr]\Bigr\}\right),
    \label{eq:Schmidt number}
\end{align}
and the \emph{entanglement of formation} (EOF, in units of `ebits')~\cite{BennettDiVincenzoSmolinWootters1996},
\begin{align}
    E_{\rm{F}}(\rho) \,=\, \inf_{\mathcal{D}(\rho)}\,\sum_i \,p_i\,S\bigl[\Tr\subtiny{0}{0}{B}\bigl(\ket{\psi_{i}}\!\!\bra{\psi_{i}}\bigr)\bigr], 
    \label{eq:EOF}
\end{align}

\noindent where $S(\sigma)=-\Tr\bigl(\sigma\log(\sigma)\bigr)$ is the von~Neumann entropy, and the infima in both expressions are taken over all pure-state decompositions of bipartite quantum states $\rho_{AB}$. 
That is, $\mathcal{D}(\rho)$ is the set of all sets $\{p_{i},\ket{\psi_{i}}\}_{i}$ with $0\leq p_{i}\leq1$ and $\sum_{i}p_{i}=1$ for which $\rho=\sum_{i}p_{i}\ket{\psi_{i}}\!\!\bra{\psi_{i}}$. 
For each given decomposition the maximization in $d_{\rm{ent}}$ is carried out over all pure states in the decomposition. For pedagogical introductions to these quantities we refer to, e.g.,~\cite{FriisVitaglianoMalikHuber2019} or~\cite[Chapter~17]{BertlmannFriis2023}. 
The ED and EOF represent two complementary quantities, but while the Schmidt number is bounded from below by the exponential of the EOF, the EOF can be arbitrarily small (but nonzero) even for diverging Schmidt number. 
At the same time, the ED is believed to be strongly limited for certain weakly entangled states such as PPT bound-entangled states~\cite{SanperaBrussLewenstein2001}. Nevertheless, the ED was shown to typically scale linearly with the local dimension for PPT states~\cite{HuberLamiLancienMuellerHermes2018}. 

In this context it is also important to note that some works use the term ``Schmidt number" to refer to the inverse purity rather than the canonical entanglement dimensionality above. 
For pure states $\ket{\psi}=\sum_i\lambda_i \ket{i,i}$, the inverse purity is related to the Schmidt coefficients $\lambda_i$ via $1/(\sum_i\lambda_i^4)$, and roughly quantifies the number of local dimensions that contribute to the observable coincidences. Introduced in~\cite{LawEberly2004}, this quantity was used to describe pure continuous-variable systems, where the Schmidt number of pure two-mode squeezed states $[1/\cosh(r)]\sum_{n=0}^{\infty}[-\tanh(r)]^{n}\ket{n,n}$ is infinite for $r\neq 0$ while proper entanglement measures like the EOF are finite, yet, the applicability to practical scenarios featuring mixed states is unclear.\\


\subsection{Recent developments}


\subsubsection{Detecting HD entanglement in practice}

{\noindent}Despite the elegance of the definitions of the ED and EOF [see Eqs.~(\ref{eq:Schmidt number}) and (\ref{eq:EOF})], carrying out the minimization over all pure-state decompositions is in general not feasible even if information about the full density matrix is available. The development of suitable ways to estimate the Schmidt number is therefore still an active field, with first witnesses put forward in 2001~\cite{SanperaBrussLewenstein2001}. 
At the same time, tomographic reconstructions of the density matrix are costly, requiring measurements in $(d+1)^2$ global product bases $\{\ket{m}\ket{n}\}_{m,n}$ or $d^2(d+1)^2$ global filter settings in single-outcome measurements, where a detector only registers one outcome~$m$ for each setting, selected with a suitable filter, see, e.g.,~\cite[Table~2]{FriisVitaglianoMalikHuber2019}. 
A common strategy therefore is to perform a number of suitable local measurements on the state $\rho$ that may not be sufficient for a full reconstruction of the density matrix but provide partial information that can be used to place lower bounds on quantities like the ED or EOF.

A simple example of such a strategy is to estimate the fidelity $\mathcal{F}(\rho,\ket{\psi})=\bra{\psi}\rho\ket{\psi}$ between the state $\rho$ produced in the lab and a fictitious pure target state $\ket{\psi}=\sum_{m=0}^{d-1}\lambda_m \ket{m,m}$, which is often taken to be the maximally entangled state $\ket{\Phi^+}$ for which $\lambda_m=1/\sqrt{d}$ for all~$m$. 
For any state $\rho$ with Schmidt number at most~$k$, the fidelity satisfies the relation~\cite[Supplementary Material, Sec. C]{FicklerEtAl2014}
\begin{align}\label{fidelity}
\mathcal{F}(\rho,\ket{\psi})    &\leq\, \sum\limits_{m=0}^{k-1}\lambda_m^2\,, 
\end{align}
which takes the simpler form $\mathcal{F}(\rho,\ket{\Phi^+})\leq k/d$ if the target state is maximally entangled. 
A violation of this inequality hence implies a Schmidt number of at least~$k+1$. 
At the same time, estimating the fidelity only requires $d+1$ global product basis measurements, or alternatively $d^2(d+1)$ global filter settings, and, as shown in~\cite{Bavaresco2018}, measurements in as few as two carefully chosen product bases are sufficient for providing lower bounds on the fidelity and corresponding lower bounds on the Schmidt number. 

Although it can sometimes be advantageous to select target states and corresponding subsequent measurement bases according to the results of measurements in an initially chosen basis~\cite{Bavaresco2018}, fixing the target state to $\ket{\Phi^+}$ leads to measurement bases that are mutually unbiased, which has the benefit that the data can be used not only for obtaining lower bounds on the Schmidt number, but also to lower-bound the EOF, see~\cite{Erker2017} and~\cite[Supplemental Material Sec. S.IV.]{Bavaresco2018}. Using this technique, it was possible to certify values such as an ED of~$29$ and $4$~ebits of EOF for a local dimension $d=31$, or an ED of~$55$ (but only $1.9$~ebits of EOF) for $d=97$ encoded in the discretized transverse-spatial pixel-mode basis of photons~\cite{HerreraValenciaSrivastavPivoluskaHuberFriisMcCutcheonMalik2020}. 
Meanwhile, criteria for the estimation of the fidelity of high-dimensional systems can also be extended to certain classes of multipartite states, see, e.g.,~\cite[Supplemental Material Sec. S.VI.]{Bavaresco2018} and~\cite{Cobucci2024}. 
Nevertheless, such advances have to be seen in the context of experiments with addressable local Hilbert-space dimensions exceeding $d=2.6\times10^5$ using a computer-controlled digital micromirror of $512\times512$ pixels and a photon-counting detector~\cite{SchneelochTisonFantoAlsingHowland2019}, indicating that a vast gap of opportunity is yet to be closed.

Fidelity-based criteria have been successfully employed in situations with considerable noise~\cite{EckerHuber2019}, but it is generally expected that they provide good estimates for high-fidelity sources but perform less well for low-fidelity sources. 
An alternative route towards Schmidt-number detection with local measurements is to directly use the correlations in $M\geq2$ pairs of bases $\{\ket{m\suptiny{0}{0}{(i)}}\subtiny{-1}{0}{A}\}_{m}$ and $\{\ket{{n\suptiny{0}{0}{(i)}}^{*}}\subtiny{-1}{0}{B}\}_{n}$, represented by the quantity
\begin{align}
    \mathcal{S}_{d}\suptiny{1}{0}{(M)}(\rho) = \sum\limits_{i=1}^{M}\sum\limits_{m=0}^{d-1} \langle m\suptiny{0}{0}{(i)},{m\suptiny{0}{0}{(i)}}^*|\nr\rho\,|m\suptiny{0}{0}{(i)},{m\suptiny{0}{0}{(i)}}^*\rangle,
    \label{eq:correlations}
\end{align}
where $m,n=0,\dots,d-1$ indicate the measurement outcomes, $i=1,2,\ldots,M$ label different measurement bases, and the asterisk denotes complex conjugation with respect to the first basis, so that $\ket{{n\suptiny{0}{0}{(1)}}^{*}}\subtiny{-1}{0}{B}=\ket{n\suptiny{0}{0}{(1)}}\subtiny{-1}{0}{B}$. 
When the $M$ bases on both sides are \emph{mutually unbiased}, all separable states satisfy $\mathcal{S}_{d}\suptiny{1}{0}{(M)}(\rho)\leq 1+(M-1)/d$ such that larger values detect entanglement~\cite{SpenglerHuberBrierleyAdaktylosHiesmayr2012}. In~\cite{Morelli2023} it was shown that for all states with Schmidt number $k$, one has $\mathcal{S}_{d}\suptiny{1}{0}{(M)}(\rho)\leq 1+k(M-1)/d$, a violation of which hence implies an ED of at least $k+1$. 
Although the number of density-matrix elements $\langle m\suptiny{0}{0}{(i)},{n\suptiny{0}{0}{(i)}}^*|\nr\rho \nr|m\suptiny{0}{0}{(i)},{n\suptiny{0}{0}{(i)}}^*\rangle$ that appear in $\mathcal{S}_{d}\suptiny{1}{0}{(M)}(\rho)$ is lower (i.e., requiring those matrix elements with $m=n$) than that appearing in the fidelity bounds~\cite{Bavaresco2018} (for the same number $M$ of bases), setups that estimate these elements from coincidence clicks nevertheless need to take data for all combinations of detector settings (all pairs of $m$ and $n$) to correctly determine the normalization. In addition, the $M$ bases are still required to be mutually unbiased. The latter restriction can be relaxed to allow arbitrary bases $\{\ket{m\suptiny{0}{0}{(i)}}\}_{m}$, in which case every state of Schmidt number at most $k$ satisfies~\cite{LiHuberFriis2025}
\begin{align}
\mathcal{S}_{d}\suptiny{1}{0}{(M)}(\rho)    &\leq\,
k\,\tfrac{M-\mathcal{T}}{d}\,+\,\mathcal{T},
\end{align}
where $\mathcal{T}:= \min\{\lambda, M\}$, with $\lambda := \bigl(\bigl[1+2d\sum_{i\neq j}G^{i,j}\bigr]\suptiny{0}{0}{1/2}+1\bigr)/2\geq 1$, is a function of the overlaps of the chosen basis vectors, i.e., 
$G^{\, i,j} := 1-(d+1)c^{\,i,j}_{\mathrm{min}} + \frac{1}{d}\sum_{m,n}|\langle m\suptiny{0}{0}{(i)}|n\suptiny{0}{0}{(j)}\rangle|^4$ 
with $c^{\,i,j}_{\mathrm{min}}=\min_{m,n} |\langle m\suptiny{0}{0}{(i)}|n\suptiny{0}{0}{(j)}\rangle|^2
$, which reduces to the criterion from~\cite{Morelli2023}, that is, $\mathcal{T}=1$ for mutually unbiased bases, i.e., when $|\langle m\suptiny{0}{0}{(i)}|n\suptiny{0}{0}{(j)}\rangle|^2=1/d$ for all~$m,n$ and for all~$i\neq j$. 

Approaches based on fidelity estimates or correlations in multiple bases can be complemented by criteria for entanglement detection based on covariance matrices of sets of carefully chosen observables~\cite{GuehneHyllusGittsovichEisert2007}. Here, the intuition is that while mean values of tensor-product operators do not reveal much about entanglement, certain higher-order moments, in particular, variances and covariances, are not compatible with separable states. Such considerations have been extended to the detection of ED from covariances~\cite{LiuFadelHeHuberVitagliano2024} and related inequalities for testing the ED via the quantum Fisher information matrix~\cite{DuLiuFadelVitaglianoHe2025}. The technique for verifying high-dimensional entanglement from covariance matrices can also be employed in settings with randomized measurements~\cite{LiuHeHuberGuehneVitagliano2023}, which has allowed the experimental detection of three-dimensional entanglement in two-photon states with local state-space dimension five~\cite{LibLiuShekelHeHuberBrombergVitagliano2025}. However, one should note that the random selection of measurement directions can be demanding in terms of flexibly adjusting experimental setups. 

Determining lower bounds for the Schmidt number has also been studied for formally infinite-dimensional continuous-variable (CV) systems. In particular, Gaussian ED witnesses based on quadrature measurements have been put forward in~\cite{ShahandehSperlingVogel2013}. Conceptually, these can be compared to the covariance-matrix criteria mentioned above, as the ED witnesses in~\cite{ShahandehSperlingVogel2013} can be expressed in terms of the first and second moments of{\textemdash}in this case{\textemdash}quadrature operators like $(\hat{a}_{n}+\hat{a}_{n}^{\dagger})/\sqrt{2}$ and $-i(\hat{a}_{n}-\hat{a}_{n}^{\dagger})/\sqrt{2}$, where $\hat{a}_{n}$ and $\hat{a}_{n}^{\dagger}$ are the creation and annihiliation operators for the $n$th mode. For CV systems such criteria are supplemented by tools for the detection of ED from phase-space quasi-probability distributions~\cite{LiuGuoFadelHeHuberVitagliano2025}. 


\subsubsection{Relaxing trust in Schmidt number detection}

{\noindent}Entanglement-detection methods such as those described above assume that the experimenter has full control over the measurement devices. This is an idealisation towards which experiments can only aspire. Therefore, several frameworks have been developed that allow for the detection of entanglement and ED while relaxing the trust in the measurements. The strongest form of these methods is called device-independent (DI), i.e.,~to make no assumption on the inner workings of the measurements~\cite{Brunner2008}. DI detection of entanglement needs the violation of a Bell inequality and its extension to Schmidt-number detection needs violations of sufficiently large magnitude, which is challenging to realize. Moreover, certain HD entangled states cannot be detected in this way, not even with access to arbitrary measurements and any number of copies~\cite{HirschHuber2020}.

In view of that, compromises have been proposed. One instance is the steering scenario \cite{Wiseman2007}, in which only Alice is viewed as a black box while Bob uses a fully trusted measurement device. Schmidt-number criterion have been developed in this framework \cite{Designolle2021}. The simplest witness uses two mutually unbiased bases, and a quantity as in Eq.~(\ref{eq:correlations}) but with Alice's measurement now seen as a free variable. Stronger criteria have been developed, relying either on more measurements~\cite{Designolle2022, Designolle2025} or semi-definite programming~\cite{Gois2023, Alessandro2025}. An experiment based on OAM used a physical dimension of $31$ and certified a Schmidt number of at least $15$~\cite{Designolle2021}. Another experiment using time-bin encoding with dimension $24$ and certified Schmidt number $9$~\cite{chang2023experimental}. 

Another route to proposing a compromise with DI is based on the entanglement-assisted prepare-and-measure scenario. The difference with the steering scenario is that neither Alice nor Bob is trusted, but Alice performs no measurement and instead relays her subsystem over a quantum channel to Bob, who then measures both subsystems jointly. Here, the assumption used in steering, namely a fully trusted Bob, is replaced with an assumption about the channel{\textemdash}most commonly its dimensionality. Hence, no part of the experiment must be fully characterized. This type of approach is reminiscent of dense coding, which too can be used to detect Schmidt numbers in a noise-robust way~\cite{Moreno2021}. However, dense coding requires complex entangled measurements for Bob which have not been implemented beyond a pair of qubits. However, recent schemes have shown that noise-robust Schmidt-number detection can be achieved by only applying product measurements~\cite{Bakhshinezhad2024}. Using this approach, a system employing eight photonic paths was used to demonstrate a maximal Schmidt number~\cite{Miao2025}.


\subsubsection{High-dimensionality of single quantum systems}

{\noindent}High-dimensionality is also relevant in scenarios that do not feature entangled states but only a single quantum system. The question of testing the dimensionality of a set of states has been discussed extensively, while notions of dimension for channels and sets of measurements have also been defined more recently. These developments are presented below.


\subsubsection{Sets of states}

{\noindent} While a single quantum state has no dimension per se, one can ask what is the minimal dimension of a set of states $\mathcal{E}=\{\rho_x\}_x$. The dimensionality of the set $\mathcal{E}$ can be lower bounded in a device-independent manner \cite{Gallego2010}. Consider a prepare-and-measure scenario, where one party receives a classical input $x$ and sends state $\rho_x$ to another party who performs a quantum measurement $\{M_{b|y}\}$ based on a classical input $y$. This gives rise to quantum correlations of the form $p(b|x,y)=\Tr\left(\rho_x M_{b|y}\right)$, from which a lower bound on the dimension of the set $\mathcal{E}$ can be inferred. This was first observed experimentally for quantum systems up to dimension four by using multiple degrees of freedom in photons, specifically polarization and orbital angular momentum encoding \cite{Hendrych2012} and polarization and spatial mode encoding \cite{Ahrens2012} respectively. Theoretical methods were then developed  for witnessing arbitrary dimensions, based on state discrimination~\cite{Brunner2013} and on random access coding~\cite{Tavakoli2015}. General methods for computing bounds on the correlations achievable in a given dimension were proposed in~\cite{Navascues2015} and made computationally scalable in~\cite{Rosset2019}. In this picture, two independent uses of a qubit would still count as a four-dimensional system, but refined tests that are sensitive to such parallel use were later proposed~\cite{Cong2017} and tested experimentally~\cite{Aguilar2018}. 

Usually, it is assumed that the preparation and measurement devices are permitted to exploit a shared classical random variable. Two complementary approaches have been studied. In the first, the devices are viewed as independent. This leads to greatly amplified quantum advantages~\cite{Hayashi2006} that have been exploited for tests of dimension based on nonlinear criteria~\cite{Bowles2015, Vicente2017, Tavakoli2021indep}. In the second, the shared classical variable is replaced with an entangled state~\cite{Pauwels2021, Pauwels2022}. This leads to amplified quantum correlation advantages but dimension-scalable criteria are presently limited to specialised cases~\cite{Bakhshinezhad2024}. 


\subsubsection{Channels}

A quantum channel, such as an optical fiber, describes the evolution of a quantum system, mapping an input state to an output state. Formally, it is represented by a completely positive and trace-preserving map $\Lambda$. Every channel has a Kraus decomposition $\Lambda(\rho)=\sum_{\lambda} K_\lambda\,\rho\, K_\lambda^\dagger$, with $\sum_{\lambda}K_\lambda^{\dagger}K_\lambda=\mathds{1}$. A specific Kraus decomposition corresponds to a specific realisation of the channel. The dimensionality of a channel can also be captured by a Schmidt number. The latter pertains to a realisation of a channel that coherently transmits quantum information between the smallest possible subspaces. Formally, it is defined as $\min\max_\lambda \text{rank}(K_\lambda)$, where the minimisation is over all Kraus decompositions of $\Lambda$.

Interestingly, there is a strong connection between channels and bipartite quantum states, which is useful for testing the dimensionality of a channel. The Schmidt number of a channel $\Lambda$ is equal to the Schmidt number of its corresponding Choi state~\cite{chruscinski2005partiallyentanglementbreakingchannels}, i.e., the bipartite state defined as $J_\Lambda=(\Lambda\otimes\mathds{1})\bigl[ \ket{\Phi^+}\!\!\bra{\Phi^+}^{\top_{\!B}}\bigr]$ with $\ket{\Phi^+}=(1/\sqrt{d})\sum_m|m,m\rangle$ and where $\top_{\!B}$ indicates the transposition with respect to the basis $\{\ket{m}\}_m$ on subsystem $B$.
This allows one to use the techniques from entanglement theory to build witnesses for the channel Schmidt number. One simply pulls a Schmidt-number witness, cf. Eq.~(\ref{eq:correlations}), from the Choi picture to the prepare-and-measure scenario:
\begin{align}
  \mathcal{S}_{d}\suptiny{1}{0}{(M)}&=\,\sum\limits_{i=1}^{M}\sum\limits_{m=0}^{d-1} \Tr\bigl(
    \ket{m\suptiny{0}{0}{(i)}}\!\!\bra{m\suptiny{0}{0}{(i)}}\otimes 
  |{m\suptiny{0}{0}{(i)}}^{*}\rangle\!\langle{m\suptiny{0}{0}{(i)}}^{*}| \,
  J_\Lambda\bigr)\nonumber\\
&=\,\frac{1}{d}\sum_{i,m} \Tr\bigl(\Lambda\big[\ket{m\suptiny{0}{0}{(i)}}\!\!\bra{m\suptiny{0}{0}{(i)}}\big]\ 
    |{m\suptiny{0}{0}{(i)}}\rangle\!\langle{m\suptiny{0}{0}{(i)}}|\,
    \bigr),
\end{align}
where $J_\Lambda$ is the Choi state and we have used the properties of the maximally entangled state. Hence, MUBs can be used to lower bound the Schmidt number $k$ via $(\mathcal{S}_{d}\suptiny{1}{0}{(M)}-1)/(M-1)\leq k/d$. 
In such detection scenarios, one trusts the input states and the output measurements. One can also relax these assumptions. For example, steering-based Schmidt-number witnesses only require trust on the input states. Both trusted and steering-based detection have been reported in~\cite{Engineer2025} using the transverse-spatial degree-of-freedom of photons and commercial multi-mode fibers. The highest reported dimensionalities in this experiment were $k=59$ (fully trusted, local dimension $d=131$) and $k=9$ (partially trusted, local dimension $d=29$) for a 2 meter-long fiber supporting 200 modes.


\subsubsection{Sets of measurements}

{\noindent}The genuine dimensionality of POVMs is based on excluding low-dimensional simulation models. It can be argued that a single POVM is effectively one-dimensional, as it is a quantum-to-classical map. By extension, any set of measurements that can be simulated with a single measurement, i.e., is jointly measurable~\cite{Guhne_2023JMreview}, is one-dimensional. 
To quantify the dimensionality of a set of measurements, one says that a collection $\{M_{a|x}\}$ of measurements is $k$-simulable~\cite{Ioannou_2022}, i.e., effectively $k$-dimensional, if
\begin{align}
\Tr[\,\rho \,M_{a|x}] = \sum_\lambda \Tr \bigl[ K_\lambda\,\rho\, K_\lambda^\dagger\, N_{a|x,\lambda} \bigr] \quad \forall\,\rho\,.
\end{align} 
Here $\{N_{a|x,\lambda}\}$ are POVMs and $\text{rank}(K_\lambda)\leq k$. If such a simulation model does not exist, the collection $\{M_{a|x}\}$ is genuinely (at least) $k+1$-dimensional. It is worth noting that deciding the dimensionality of POVMs is mathematically equivalent to the task of detecting Schmidt number of a bipartite state in a steering scenario~\cite{Jones_2023}. This allows one to use the methods from steering-based Schmidt number detection as witnesses for dimensionality of POVMs.

As an alternative dimensionality measure, one can define Schmidt number for bipartite measurements in a similar manner as for states~\cite{egelhaaf2025certificationquantumnetworksusing}. Here one calculates the Schmidt number of each individual POVM element and takes the maximum thereof. However, high-dimensional bipartite measurements are challenging with current experimental techniques due to the need of additional photons for their implementation. Such measurements have nevertheless been implemented in, e.g., two qutrit systems in the form of a fully-resolving Bell-state measurement \cite{Luo2019,Hu2020}.


\subsubsection{HD-QKD} \label{sec:Theory_HDQKD}

{\noindent}The use of HD systems for applications in quantum communication is an important research direction~\cite{Cozzolino2019}, which has already been explored in the early days of quantum information. Indeed, $d$-dimensional systems (qudits) offer an increased information capacity of $\log_2(d)$ compared to qubits, which are limited to one bit per carrier. Formally, this leads to a clear advantage via enhanced rates. In practice, however, this potential advantage must be carefully weighed against the increased technical challenge of implementing qudits compared to qubits. This is well illustrated by protocols such as quantum teleportation and dense coding, which have been generalized early on to arbitrary dimensions (via the natural extension of the Pauli algebra to the Clifford one), yet have been barely explored experimentally~\cite{Luo2019}, mainly due to the challenge of implementing generalized Bell-state measurements~\cite{bianchi2025}.

In contrast, quantum cryptography is an area where the use of HD systems has potential to provide an exploitable advantage in practice. Beyond the increase in information capacity, HD protocols may also take advantage of the significantly increased noise and loss-robustness of qudit resources compared to qubit ones. Notably, entanglement becomes increasingly robust to noise when considering systems of increasing dimension. Moreover, the implementation of these protocols relies on the distribution of HD entanglement followed by few local measurements, or even in simpler prepare-and-measure setups, hence avoiding the hurdle of complex joint measurements. 

This potential was already recognized more than 20 years ago, with the first protocols for quantum key distribution tailored to HD systems~\cite{Bechmann1999, Bechmann2000, Cerf2002}. The core idea of these protocols is to extend the standard qubit protocols to qudits using mutually unbiased bases (MUB) as a generalization of the Pauli bases. The first class of protocols uses a pair of MUB, thus generalizing the BB84 protocol. The second class uses a complete set of $d+1$ MUB (for $d$ being a prime power), extending the 6-state protocol. The security of these protocols has been investigated for different classes of attacks. The optimal individual attacks are based on cloning machines~\cite{Bruss2002, Cerf2002}, and security proofs against the most general (coherent) attacks have been derived~\cite{Cerf2002, Sheridan2010}. Both classes of protocols offer significant advantages over qubit protocols. In particular the noise robustness increases with the dimension~$d$, enabling QKD in noise regimes that are inaccessible to qubits. Second, even at lower noise, the secret key rate is enhanced. For example, considering protocols with dimension $d=11$ allows one to tolerate an error rate up to $Q=26.2\%$ (for 2 bases) and up to $Q=33,3\%$ (for $d+1=12$ bases) while qubit protocols are limited to $Q=12.6\%$. In addition, given a fixed error rate of $Q=5\%$, the asymptotic key rate increases by a factor of~$5$. 

Novel concepts in quantum cryptography were crucial for the development of full security proofs of HD-QKD, applicable to real experiments. First, entropic uncertainty relations enabled security proofs in the finite-size regime~\cite{Sheridan2010, Tomamichel2011}. Second, the technique of decoy states promoted abstract proofs to practical ones, applicable to implementations based on weak laser pulses~\cite{Zhang2014, Canas2017_HDDistib_MCF}. On the computational side, Semi-Definite Programming (SDP) techniques have been developed for bounding key rates~\cite{WangPrimaatmaja2019, Doda2021, Araujo2023}. This is relevant for HD-QKD, as these tools can in principle be applied to protocols with arbitrary states and measurements, though their performance is usually limited by the computational complexity of dealing with systems of increasing dimension. This problem can be overcome for specific protocols via an analytical formulation of the dual SDP problem, leading to finite-size and composable security proofs~\cite{Kanitschar2025, Kanitschar2025b}. Another hurdle for HD-QKD is that the presence of noise significantly increases the requirements in terms of error correction, hence reducing the key rate. Methods have been proposed to address this issue, based on the simultaneous use of many lower-dimensional subspaces~\cite{Doda2021}, or via Low-Density Parity-Check (LDPC) codes for high-dimensional alphabets~\cite{Mueller2023}. Finally, QKD protocols tailored to weak laser pulse implementations have also been generalized to HD systems, such as Differential-Phase-Shift~\cite{Stasiuk2023} and Coherent-One-Way~\cite{Sulimany2025}.

First proof-of-principle experiments for HD-QKD protocols based on MUBs have been reported, encoding qudits into transverse-spatial degrees of freedom of light based on spatial modes~\cite{Walborn2006, Etcheverry2013} and orbital angular momentum~\cite{Groeblacher2006, Mirhosseini_2015, Sit:17, Wang2019}. Alternatively, it was proposed to use temporal degrees-of-freedom to encode qudits into photonic time-bins~\cite{Ali-Khan2007, broughamSecurityHighdimensionalQuantum2013, Zhong2015}. This approach has several advantages, such as enabling the creation of very high-dimensional states in principle, which are well-suited to optical-fiber implementations. The time-basis readily provides an ideal measurement for establishing the key. The main challenge consists in performing measurements in some conjugate basis, for bounding the adversary’s information on the key. Schemes based on time-bins and phase encodings~\cite{Islam2017_a, Vagniluca2020PRAppl_d4QKD_2det}  have been effectively demonstrated for low dimensions ($d=4$), but their scaling to higher dimensions is challenging. An alternative is to consider measurements based on Franson (unbalanced Mach-Zehnder) interferometers for verifying coherence between different (typically neighbouring) time-bins \cite{fransonBellInequalityPosition1989}. Approaches based on hyperentanglement have also been explored, e.g., combining time-bins and polarization for free-space intra-city demonstrations exhibiting increased noise robustness~\cite{Bulla2023}. Furthermore, implementations based on path encoding have been reported, investigating increased noise robustness~\cite{Hu2021} and practical solutions based on photonic integrated chips combined with multicore optical fibres~\cite{Ding2017}. More generally, integrated photonic solutions are attractive in practice, and have been demonstrated for OAM~\cite{Zhang2025}, time-bins~\cite{yuQuantumKeyDistribution2025} and frequency-bin encodings~\cite{Tagliavacche2025_QKD}. 

HD systems also open interesting perspectives for cryptographic protocols with relaxed levels of trust. HD systems can in principle facilitate the implementation of device-independent QKD, via weaker requirements in terms of input randomness~\cite{Huber2013} and detection efficiency \cite{Vertesi2010}. The use of HD systems in the measurement-device-independent~\cite{Dellantonio2018} and twin-field~\cite{Mueller2025} QKD protocols has also been discussed. Finally, HD systems enable secure QKD in the time-lock model~\cite{Vyas2020}, combining concepts from classical and quantum cryptography.


\subsection{Challenges and outlook}

\subsubsection{Multipartite multidimensional entanglement}

{\noindent}Although the situation of entanglement theory is already complicated in bipartite high-dimensional systems, the real challenge for our understanding of future quantum networks lies in unraveling the structures of \emph{multipartite, multi-dimensional} entanglement. 
Since there is no equivalent of the Schmidt decomposition in multipartite systems, the situation is complicated from the outset, but a possible generalization of the ED to multiple systems is to start with a \emph{Schmidt rank vector}~\cite{HuberDeVicente2013, HuberPerarnauDeVicente2013} $\vec{r}\subtiny{0}{0}{\rm{S}}$ for pure states, which collects the Schmidt ranks for all bipartitions of the system in non-increasing order. For the transition to mixed states one may then formally apply a minimization over all decompositions for each bipartition as in Eq.~(\ref{eq:Schmidt number}), but the relations (similar to monogamy constraints for qubits~\cite{CoffmanKunduWootters2000, OsborneVerstraete2006}) between Schmidt numbers for different bipartitions are non-trivial~\cite{CadneyHuberLindenWinter2014}, making the practical characterization of this vector challenging. However, inspired by the covariance-matrix techniques mentioned before, methods for detecting Schmidt-number vectors that amount to checking systems of inequalities aided by linear programming have recently been put forward~\cite{LiuHeHuberVitagliano2025}.

Another challenge in the context of multipartite high-dimensional quantum information processing concerns the question of state convertibility. 
Seminal results~\cite{Nielsen1999, Vidal1999, JonathanPlenio1999} on \emph{single-copy} state conversion within the typical paradigm of local operations and classical communication (LOCC)~\cite{ChitambarLeungMancinskaOzolsWinter2014} give us tools to assess when \emph{bipartite} states can be converted into each other. However, the situation becomes more complicated even for three qubits~\cite{DuerVidalCirac2000, AcinAndrianovTarrach2000}, with infinitely many inequivalent classes even under stochastic LOCC (SLOCC) in general~\cite{VerstraeteDehaeneDeMoorVerschelde2002}, and with transformations between almost all multipartite (pure) states being impossible via LOCC~\cite{SauerweinWallachGourKraus2018}. 
The severe restrictions for multipartite state conversion via SLOCC~\cite{DeVicenteSpeeKraus2013, SpeeDeVicenteSauerweinKraus2017, DeVicenteSpeeSauerweinKraus2017, SauerweinWallachGourKraus2018} also hold (limited) potential for designing optimized protocols using side information to improve noise resistance~\cite{MorelliSauerweinSkotiniotisFriis2022}. 

However, access to multiple copies brings with it yet more complications in terms of multi-copy activation phenomena: Multiple copies of fully inseparable biseparable states (mixtures of partition-separable states, by definition not genuinely multipartite entangled, but entangled across all bipartitions) can be genuinely multipartite entangled (GME) when considering all local subsystems jointly~\cite{YamasakiMorelliMiethlingerBavarescoFriisHuber2022}. This is indeed always the case for some number of copies~\cite{PalazuelosDeVicente2022}, and in all dimensions~\cite{BaksovaLeskovjanovaMistaAgudeloFriis2025}, which gives us a glimpse of the importance of considering such multi-copy activation. Consequently, experimental tests of GME activation have already been conducted: In~\cite{ZhangGuehnePan2025}, two copies of individually biseparable three-photon states have been jointly processed via LOCC to produce a single three-photon GME state, whereas\cite{StarekGollerthanLeskovjanovaMethTirlerFriisRingbauerMista2025} reports the preparation of two copies of individually biseparable states of three trapped ions each, whose joint six-ion state is detected as GME by a suitable witness. 
Yet, recent work~\cite{WeinbrennerEtAl2024, VillegasAguilarEtAl2024} also suggests that much is still to be learned about activation phenomena. For one thing, different SLOCC classes of genuine multipartite entanglement can be reached even from biseparable states, further calling into question the paradigm of SLOCC for network problems. Second, some activated GME cannot be compressed to the single-copy Hilbert space, making it more difficult to access practically~\cite{WeinbrennerEtAl2024}. 


\subsubsection{Robust and efficient tests for HD entanglement}
\label{sec:TheoryTestsForHDEntanglement}
{\noindent}A key goal is to develop certification methods for HD systems that are based on minimal assumptions and at the same time efficiently implementable. The commonly used assumption of fully trusted measurements (within the device-dependent model) opens up a vulnerability in entanglement detection, as in practice the laboratory measurements do not perfectly align with the mathematical model describing them. Recent works have proposed to model such imperfections based on the fidelity between the lab observable and the target observable~\cite{Morelli2022}. Such imprecise measurements can have a significant detrimental impact on entanglement detection~\cite{Rosset2012, Morelli2022} and it motivates the development of entanglement criteria that are robust to generic but small imperfections. In~\cite{Tavakoli2024} it was shown how tests of entanglement and Schmidt number in steering scenarios can be corrected when imprecision is allowed in the otherwise trusted device. It remains an open challenge to build theoretical methods and experimentally robust scenarios that fully address such imperfections.

\begin{figure}
    \centering
    \includegraphics[width=0.9\linewidth]{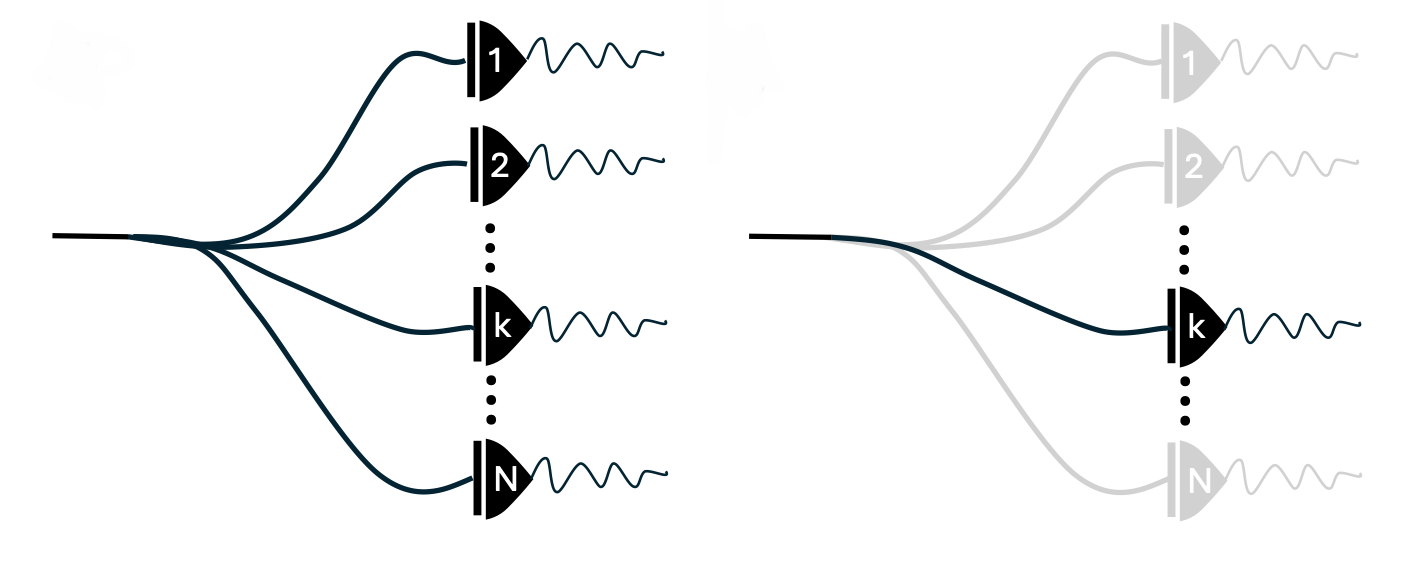}
    \caption{(Left) Genuine $N$-outcome measurement. (Right) Binarised implementation of $N$-outcome measurement based on $N$ independent ``click or no-click'' measurments.  Picture taken from \href{https://doi.org/10.1103/PhysRevA.111.042433}{Phys. Rev. A 111, 042433 (2025).}
    }
    \label{fig:Schmidtnumber}
\end{figure}

The certification of HD entanglement in (partially) device-independent scenarios, such as via HD Bell inequalities, HD steering tests, or entanglement-assisted prepare-and-measure scenarios, is also an interesting avenue for future research. Recent works have shown encouraging first results, but a number of key challenges still have to be addressed. Among these is the issue of loopholes that can plague the results of certain experiments, and thus require additional assumptions. It is of course desirable to remove all these assumptions by closing the corresponding loopholes. Besides the well-known locality and detection loopholes, experiments based on HD systems can be subjected to a new type of loophole, originating from the use of multi-outcome measurements. The latter are often challenging to implement in practice, and it is common that photonic implementations only emulate multi-outcome measurements through a collection of binary, “click or no-click” measurements by comparing the frequency of clicks from individual settings with the total number of clicks in all settings to infer the statistics of the multi-outcome measurement. While implementing a multi-outcome measurement through such a binarization procedure is legitimate in the device-dependent setting, this no longer the case when trust assumptions are relaxed. For instance, experimental tests of Bell inequalities in HD systems can lead to false positives when multi-outcome measurements are replaced with their binarisations, opening a \textit{binarisation loophole}~\cite{Tavakoli2025b}. 

Recently, by realizing multi-outcome measurements, experiments have begun to address the binarisation loophole by revisiting tests of HD Bell nonlocality~\cite{Dekkers2025} and also entanglement-assisted quantum communication~\cite{Miao2025}. However, the severity of the binarisation loophole varies over different experimental scenarios. For instance, in steering experiments, evidence has indicated that it is less severe~\cite{Tavakoli2025b}. In particular, highly noise-robust and loss-tolerant steering witnesses for HD systems have been developed based on binary measurements and demonstrated experimentally using discrete
transverse position-momentum~\cite{Srivastav2022}. 

Addressing the above challenges will require the development of novel theoretical tools tailored to current and near-future experimental capabilities. 
A first key challenge is to devise and implement robust methods for certifying Schmidt number addressing all relevant loopholes.


\subsubsection{High-dimensionality in single quantum systems}

{\noindent}Prepare-and-measure experiments are scenarios where a sender prepares states and sends them to a receiver who measures them. While entanglement has been widely studied both on its own and from a black-box perspective, studies of prepare-and-measure scenarios have been largely focused on the latter. This stems from applications in semi-device-independent quantum information, which view the sender and receiver as uncharacterised devices up to a mild assumption. Recently, the question has been raised whether quantum features in the prepare-and-measure scenario can be assessed on their own, without invoking the device-independent perspective. Models have been proposed for assessing the dimensionality of the set of states~\cite{Bernal2024} and whether they can be reproduced in classical models~\cite{Cobucci2025}. It is a largely an open question how to best  conceptualise the notion of dimensionality for preparation, transformation and measurement devices acting on single systems, and how these features can be detected in practice. 


\subsubsection{Quantum computing with HD quantum systems}

{\noindent}An increasing level of control has been been achieved over high-dimensional state spaces in many technologies that are promising contenders for the realization of quantum computers, including integrated photonic quantum technology~\cite{Wang2018, Chi2022}, superconducting circuits~\cite{ShlyakhovEtAl2018, Morvan2020}, trapped ions~\cite{Ringbauer2022, ZalivakoEtAl2025}, spin centres in semiconductor materials~\cite{SoltamovEtAl2019, KollerAstnerTissotBurkardTrupke2025}, and molecular platforms~\cite{ChizziniCrippaZaccardiMacalusoCarrettaChiesaSantini2022, MuminovEtAl2025}. This has motivated the study of a new paradigm of quantum computing based on qudits rather than on qubits, for reviews, see, e.g.,~\cite{WangHuSandersKais2020, KiktenkoNikolaevaFedorov2025}. 
Possible advantages of this approach include, for example, improved coherence~\cite{RingbauerEtAl2018} and richer entanglement structures~\cite{KraftRitzBrunnerHuberGuehne2018}, which can lead to benefits for error correction~\cite{WatsonCampbellAnwarBrowne2012, Campbell2014, LimLiuArdavan2023}.

Another prominent advantage lies in potential reductions in gate count~\cite{LanyonEtAl2009, NikolaevaKiktenkoFedorov2024a, KiktenkoNikolaevaXuShlyapnikovFedorov2020}. Entangling gates often feature lower gate fidelities compared to local gates, and can therefore lead to bottlenecks in many current devices. Consequently,  estimates of the expected performance of a quantum computation are often based on the number of required entangling gates~\cite{HaferkampFaistKothakondaEisertYungerHalpern2022}. 
Although this falls short of capturing all intricacies involved with the specific quantum circuits and their implementation in terms of hardware-native gates, the entangling-gate count can provide a good hardware-agnostic approximation. 
It is exactly this number of required entangling gates, which can be reduced when compressing qubit circuits to qudit architectures~\cite{GaoAppelFriisRingbauerHuber2023}, as some of the gates that would otherwise act as entangling gates between two or more qubits, can now be implemented on the $d$ levels of a single qudit.

However, to explore in detail how much of this potential advantage can be translated to real improvements in fidelities is a major open challenge. For one thing, addressing this challenge requires the design of hardware-specific native gate sets for qudits. Some progress has been reported in this direction, e.g., single~\cite{GaoKrennKyselaZeilinger2019} and two-qudit gates~\cite{GaoErhardZeilingerKrenn2019, GaoAppelFriisRingbauerHuber2023} have been proposed for photonic platforms, and a recent experiment has realized a heralded HD entangling gate for two qudits with $d=4$~\cite{LiuRenWanZhuChengWangWangXiHuberFriisGaoWangWang2024}. Similarly, trapped-ion platforms have been shown to support single~\cite{Ringbauer2022} and two-qudit gates~\cite{HrmoEtAl2023}, the latter in dimensions up to $d=5$. 
A second step towards meeting this challenge lies in determining suitable algorithms for compilation~\cite{MatoRingbauerHillmichWille2022, MatoRingbauerHillmichWille2023, NikolaevaKiktenkoFedorov2024b, MatoRingbauerBurgholzerWille2024} and transition~\cite{DrozhzhinNikolaevaKiktenkoFedorov2024, DrozhzhinKiktenkoFedorovNikolaeva2025}. 

Along with such basic questions of how to compose given unitaries into native qudit gate operations, comes the challenge of developing suitable methods and applications or to translate existing ones into qudit language. Some efforts have already been made in this direction, providing advances in the development of qudit error correction~\cite{GrasslKongWeiYinZeng2018}, a qudit stabilizer formalism~\cite{AignerMorRuizDuer2025}, a generalization of qubit graph~\cite{HeinDuerEisertRaussendorfVanDenNestBriegel2006} and hypergraph states~\cite{RossiHuberBrussMacchiavello2013} to tensor-edge hypergraph states for qudits~\cite{AppelHeilmanWertzLyonsHuberPivoluskaVitagliano2022}, measurement-based quantum computation with qudits~\cite{RomanovaDuer2025a}, a graphical calculus for multi-qudit computations with generalized Clifford algebras~\cite{Lin2025}, blind qudit quantum computation~\cite{RomanovaDue2025b}, and various quantum simulations with qudits~\cite{MethRingbauerEtAl2025, GavreevKiktenkoFedorovNikolaeva2025}.

\begin{backmatter}
\bmsection{Acknowledgments}
M.H.~acknowledges funding from the European Research Council (Consolidator grant ‘Cocoquest’ 101043705). 
M.H.~and N.F.~acknowledge funding from the Austrian Research Promotion Agency (FFG) through 
the project FO999914030 (MUSIQ) and the project FO999921415 (Vanessa-QC), funded by the European Union{\textemdash}NextGenerationEU. 
A.T.~was supported by the Knut and Alice Wallenberg Foundation through the Wallenberg Center for Quantum Technology (WACQT) and the Swedish Foundation for Strategic Research. R.U.~was supported by the Swedish Research Council (grant no. 2024-05341) and the Wallenberg Initiative on Networks and Quantum Information (WINQ). 
N.B acknowledges financial support from the Swiss State Secretariat for Education, Research and Innovation (SERI) under contract number UeM019-3. 
N.F.~acknowledges funding from the Austrian Science Fund (FWF) through the stand-alone project P 36478-N funded by the European Union{\textemdash}NextGenerationEU, 
as well as by the Austrian Federal Ministry of Education, Science and Research via the Austrian Research Promotion Agency (FFG) through the flagship project FO999897481 (HPQC) and the project FO999921407 (HDcode), funded by the European Union{\textemdash}NextGenerationEU. 
\end{backmatter}

\section{Synthesis and Future}
\author{Natalia Herrera Valencia\authormark{1},
Jacquiline Romero\authormark{29,30},
Micha\l{} Karpi\'nski\authormark{16},
Robert Fickler\authormark{20},
Will McCutcheon\authormark{1}}
\address{\authormark{1}Institute of Photonics and Quantum Sciences (IPAQS), Heriot-Watt University, Edinburgh, UK}
\address{\authormark{16}Faculty of Physics, University of Warsaw, Pasteura 5, 02-093 Warszawa, Poland}
\address{\authormark{20}Photonics Laboratory, Physics Unit, Tampere University, Tampere, FI-33720, Finland}
\address{\authormark{29}School of Mathematics and Physics, University of Queensland, Brisbane, 4072, Australia}
\address{\authormark{30}Australian Research Council Training Centre for Current and Emergent Quantum Technologies (QuTech), Brisbane, 4072, Australia}

\medskip

Many quantum systems in addition to photons are inherently high-dimensional and accommodate a qudit description. However, much of the research in quantum foundations, quantum information, and quantum technologies has centered on the qubit. This is perhaps because the qudit, although ubiquitous, is a double-edged sword. This can be attributed to two primary reasons: (1) The increased information capacity that comes with a qudit also means that more information is lost when the photon that carries the qudit is lost. (2) The higher information capacity also requires as many detector outcomes as there are dimensions (i.e. either $d$ detectors per basis, or $d$ time  slots in time-resolved detection). However, aside from these two basic constraints imposed by physics, the remaining challenges can be overcome with clever engineering and systems thinking. For example, recent experiments have demonstrated quantum key distribution using time-bin ququarts over 145~km of fibre \cite{Vagniluca2020PRAppl_d4QKD_2det}, and combining different photonic qudits (time-bin and path) has enabled a $\sim 50$~kbps secret key rate over 52~km of deployed fibre \cite{Zahidy2024:HDQKDAquila}.  The prizes for overcoming the challenges, as already highlighted earlier, are plenty: in quantum communications, higher information capacity, improved tolerance to noise, and resilience against certain classes of attacks; in quantum computing, reduced circuit depth and simplification of certain algorithms; in quantum sensing, improved sensitivity, multiplexed measurements, and robustness against some noise models. 

This roadmap has highlighted the challenges associated with photonic qudits and presented possible avenues for moving forward. The aim of this chapter is to identify common themes across the different degrees of freedom, and their applications to quantum technologies, to suggest possible collaborative developments to fully realise the potential of high-dimensional quantum information afforded by photons.

\subsection{Distribution: Shared Challenges and Opportunities}

The ability to distribute high-dimensional photonic quantum states in a robust and scalable manner is a central requirement for quantum communication and future quantum networking architectures. While the previous chapters of this roadmap have detailed the generation, measurement, and manipulation of high-dimensional states across different photonic degrees of freedom, their practical deployment ultimately hinges on how these states behave when transmitted through real-world channels, in particular optical fibre and free-space links that form the backbone of our existing communication infrastructure.

A recurring theme across high-dimensional distribution strategies is that transmission loss alone rarely defines the dominant system limitation. Instead, overall performance is often set by how distribution-induced distortions among modes defining the qudit interact with the complexity of performing coherent, high-dimensional measurements at the receiver. As a consequence, encodings that are comparatively straightforward to transmit (such as time-bin) may shift complexity toward detection and stabilisation, while encodings that are more challenging to distribute (such as spatial modes) can, in some cases, enable more direct or efficient measurement once successfully transmitted.

Importantly, these trade-offs are not unique to any single DoF. While different encodings offer distinct advantages in scalability, manipulation and practical deployment, all face constraints arising from transmission, stabilization, and interfacing. The following subsections examine how these constraints recur across time, frequency, path, and space, highlighting common limitations and design choices that emerge when distribution is treated as a system-level problem that cuts across different DoFs and motivates hybrid and conversion-based approaches.

\subsubsection{Infrastructure compatibility and system complexity}

In practice, the distribution of high-dimensional photonic states is constrained less by the abstract dimensionality of the encoding and more by the physical properties of the transmission medium. For most near- and mid-term quantum networks, optical fibre remains the dominant channel~\cite{LIU2025100551}. This imposes stringent requirements that immediately shape which degrees of freedom can be exploited directly and which require conversion, compensation, or additional control.

The choice of encoding determines not only where complexity is increased within the system, but also how readily it can be deployed at scale. Single-mode fibre-compatible encodings, such as time-bin, frequency-bin and pulse-modes, provide a genuine advantage for simplifying transmission by leveraging existing infrastructure. However, this advantage does not eliminate complexity, but rather shifts it. As efforts push towards larger accessible dimensional spaces, new development directions have emerged, including active interferometers, dense on-chip integration, and advanced detection methods like multi-outcome quantum pulse gates. While these approaches improve control, mode selectivity, or scalability in specific regimes, they also introduce additional architectural overhead, tighter requirements on phase stabilization, and increased noise and loss management. The result is a shift, rather than a removal, of system-level burden for large-scale deployment~(See Sections~\ref{sec:GenAndMeas_TimeBinChallengesAndOutlook}, \ref{sec:GenAndMeas_FrequencyChallengesAndOutlook} and \ref{sec:GenAndMeas_TemporalModeDetection} for focused discussions on challenges for time-bin, frequency, and temporal modes, respectively). 

In contrast, spatially structured encodings face a more fundamental incompatibility with single-mode fibers. Their inherent structure must either be preserved through specialized fibers or reconstructed through careful interfacing/conversion between generation and transmission. This typically demands precise mode-mixing and interference control (See \ref{sec:Applications_TransverseRecentDevelopments}), or additional mode-conversion stages and coupling strategies (see Sections \ref{sec:GenAndMeas_PathDistribution}). Over the past decade, significant effort has been devoted to the development of specialty optical fibers, driven primarily by the need to increase the capacity of classical telecommunication systems. At the same time, these advances have created new opportunities for implementing HD quantum communication and information-processing tools \cite{Xavier2020}. Nonetheless, the resulting resource overhead raises several open questions about when and over which distance scales the performance gain of high-dimensional encoding justifies the additional system burden.

\subsubsection{Phase stability as a unifying bottleneck}

Across otherwise distinct degrees of freedom, phase stability emerges as a unifying bottleneck for high-dimensional state distribution. Whether encoded in time bins, frequency bins, temporal modes, interferometric paths, or spatial channels, high-dimensional coherence relies on maintaining well-defined relative phases across an increasing number of modes.

While the physical origins of these phase errors differ, their impact on distributed high-dimensional states is strikingly similar, leading to basis-dependent decoherence, reduced interference visibility, and dimensionality-dependent performance degradation.

In time-bin and frequency encodings, this challenge is closely tied to interferometric stability and spectral phase fluctuations accumulated over long fibre links, often requiring active compensation schemes~\cite{Tagliavacche2025_QKD}. In contrast, for path and spatial encodings, phase instability is more strongly linked to environmental perturbations, differential path lengths, and mode-dependent propagation effects within fibres (Section~\ref{sec:GenAndMeas_TimeBinChallengesAndOutlook}) or free-space channels.

\subsubsection{Mode-dependent loss and non-uniform noise}

Unlike qubit encodings, where loss can often be treated as a binary event, high-dimensional systems are highly sensitive to mode-dependent loss. Rather than simply reducing the overall detection rate, imbalances that selectively affect subsets of modes or measurement bases directly distort the resulting state~(Section~\ref{sec:GenAndMeas_TransverseChallengesAndOutlook}) . Another related and shared challenge across high-dimensional distribution platforms is the presence of non-uniform noise, which, as pointed out in Section~~\ref{sec:GenAndMeas_TimeBinChallengesAndOutlook}, can become increasingly detrimental as the number of modes grows.

In fibre-compatible encodings, such effects may arise from imperfect interferometers, spectral filtering, or dispersion compensation, while in spatial and path-based schemes, they are often linked to coupling inefficiencies, mode-dependent projective measurements, cross-talk, and mode-selective attenuation during transmission. As pointed out in Section~\ref{sec:GenAndMeas_PathDistribution}, these non-uniform effects scale unfavorably with increasing dimensionality and can no longer be treated as small perturbations, emerging as a defining constraint on how high-dimensional states can be distributed, accessed, and benchmarked in practice.

The challenges discussed above place important limitations on how the promised advantages of high-dimensional photonic states manifest in realistic distribution scenarios. As pointed out in Section~\ref{sec:Theory_HDQKD} of the Theory chapter, beyond an increase in information capacity per photon, a central practical motivation for high-dimensional systems appears in quantum communications and cryptography because of the potential for enhanced robustness to noise and loss. A growing body of theoretical and experimental work has demonstrated that high-dimensional encoding can tolerate higher error rates and channel impairments under certain conditions~\cite{EckerHuber2019,Zhu2021,Srivastav2022}. In practice, however, this robustness does not arise in channel transmission alone. Rather, the effective robustness of a distributed high-dimensional state emerges from the interplay between state preparation, channel-induced distortions, and the techniques required to access arbitrary high-dimensional bases under realistic measurement constraints. As highlighted in Section~\ref{sec:GenAndMeas_TimeBinChallengesAndOutlook}, translating the theoretical tolerance to noise into a genuine system-level advantage therefore remains non-trivial, and depends critically on how distribution, detection, and control are jointly implemented.

\subsubsection{Conversion between degrees of freedom as a system-level solution}

Addressing the gap between theoretical robustness and practical performance motivates system-level strategies that decouple the choice of encoding for transmission from that used for state preparation and measurement. In this context, coherent conversion between photonic degrees of freedom has emerged as a powerful strategy for high-dimensional state distribution. Rather than relying on a single encoding to satisfy all requirements, hybrid approaches aim to exploit the strengths of different degrees of freedom at different stages of the network.

Recognised as a critical ingredient for scalable and modular quantum photonic architectures, the chip-to-chip distribution of quantum states relies on mapping path modes to independent fibres in a fibre-array, and more recently, has benefited from the coherent conversions between on-chip path encodings to polarization and spatial modes in few-mode or multicore fibres (see Section~\ref{sec:Applications_Path_CurrentDevelopments}). 

Alternative routes for transporting spatial or path qudits could also include conversion into polarization and/or time-bin encodings for long-distance fibre transmission, before being mapped back into formats optimised for detection or processing. However, experimental demonstrations for such conversion techniques have only operated at the qubit level (see Section~\ref{sec:GenAndMeas_PathDistribution}), and the generalized extension to higher-dimensional path-encodings may bring additional engineering burdens. 

In this manner, the performance of high-dimensional quantum networks becomes inseparable from the efficiency, stability, and scalability of mode conversion technologies, positioning them as a central enabling component rather than a secondary interface. 

Taken together, these challenges and considerations suggest that the distribution of high-dimensional photonic states should be viewed as a system-level optimisation problem, rather than a competition between degrees of freedom. Each encoding carries intrinsic strengths and limitations, but it is their interaction with transmission media, stabilisation requirements, and measurement strategies that ultimately determines network performance. Future progress will therefore depend on comparative benchmarking across platforms, realistic performance metrics that account for mode-dependent effects, and, as pointed out in Section~\ref{sec:Applications_PathChallengesAndOutlook}, the development of hybrid architectures that flexibly combine multiple degrees of freedom within a single network.

\subsubsection{High-dimensional quantum memories}
Along with the progress in distributing high-dimensional quantum states, another essential building block in quantum networks becomes important to be realized, namely, high-dimensional quantum memories that are essential for quantum repeaters.
In the following, we briefly review various physical implementations of quantum memories, however, limiting our discussion to multi-mode memory demonstrations allowing for the storage of a high-dimensional quantum state encoded through one or more degrees of freedom (DoF) of the photon.

In the time-energy domain, various demonstrations relied on implementing a quantum memory in rare-earth doped crystals, fibers,  and waveguides \cite{afzelius2009multimode} to show the storage of increasing number of temporal modes of light at the single-photon level \cite{ortu2022storage, zhang2023telecom}, entanglement between a memory and a photon \cite{laplane2017multimode, kutluer2017solid, kutluer2019time, seri2019quantum}, as well as entanglement between photon pairs where one has been stored and retrieved from the memory \cite{saglamyurek2011broadband, saglamyurek2015quantum, tiranov2017quantification, seri2017quantum, rakonjac2021entanglement, jiang2023quantum}. Recently, these ideas have also been extended towards integrated waveguide realizations  \cite{ zhang2023telecom} as well as to invoke single atoms as high-dimensional memories, which allows us to heralded entangled atoms serving as network nodes \cite{shalaev2025photonic}. 

In the spatial mode domain, rare-earth-ion-doped crystals acting as quantum memories have been studied for qutrits encoded in the OAM carrying modes \cite{zhou2015quantum}, however, the majority of the effort has been put towards the realization of high-dimensional quantum memories using cold atoms. Here, quantum states encoded in OAM of single photons and comparable faint light signals have been stored with increasing functionality over the last decade \cite{nicolas2014quantum,wang2021efficient, ye2022long, dong2023highly, yang2025efficient}.  Additionally, OAM entanglement has been successfully verified after retrieving it from two separate atomic quantum memories \cite{ding2015quantum, ding2016high}, which has been further extended to larger sets of modes through the use of OAM and path encodings \cite{ zhang2016experimental}. In a related approach, cold atom memories have been leveraged to demonstrate the storage of multi-modal quantum light in the form of different paths \cite{ pu2017experimental, parniak2017wavevector} including the high-dimensional entanglement between the photon and the memory’s spin wave excitation \cite{ li2020high}. 

In addition to storing quantum information encoded in a single DoF, there has also been a strong effort to improve the operation of these quantum memories through combining multiple DoFs . In atomic systems, quantum states encoded in spatial modes and polarization (vectorial light fields), have been stored in cold atom clouds \cite{ parigi2015storage} with extensions to path multiplexing \cite{ yang2025efficient} as well as the implementation in a warm Rubidium vapor cell \cite{ wu2025ai}. 
On the other hand, solid state quantum memories have been extended to multi-DoF operation through the usage of spectral, temporal and OAM modes \cite{ yang2018multiplexed} as well as spatio-temporal multiplexing \cite{ teller2025solid} with recent efforts pushing toward integrated solutions using laser-written waveguide memories for path-multiplexing of temporal modes \cite{ ou2025multichannel}.

While high-dimensional quantum memories have come a long way already, considering that first proof-of-principle demonstrations have only been demonstrated around a decade ago, there are still various challenges to tackle in all key parameters such as dimensionality, storage time, and efficiency. In addition, moving towards implementations that are easier to deploy in real-world network scenarios, such as integrated solutions, will become more important in the near future as long-distance high-dimensional quantum networks become available.  

\subsection{Detection: Arrays, Time-multiplexing and Unitaries}
A ubiquitous requirement of all photonics platforms is the ability to detect single photons, making the development of high-efficiency, low-jitter, low-noise single-photon detectors (SPDs) a shared challenge across all HD photonics platforms. Technologies exploiting multiple detectors, perhaps numbering in the millions, will require ready access to cheap solutions which can be practically and reliably integrated into devices and across networks. Current limitations in detector count rates and jitter are discussed in Sections~\ref{sec:GenPathSinglePhotonDetectors}. The specific requirements of integrating these detectors with different platforms is closely bound to the task of transforming from a given degree of freedom to one compatible with a given detector, and to transforming to the required measurement basis within that degree-of-freedom. 

An current attempt towards detection of HD states involves performing just single-outcome measurements, also known as binarised measurements.  Rather than perform all $d$ measurement outcomes simultaneously, sequentially projecting on to single outcomes in a click or no-click fashion, can enable the building up of statistics relevant to a general measurement whilst reducing the need for $d$-detectors. This binarization of measurements, whilst often being a practical first demonstration of methods, greatly reduces the power and security of measurements (Section~\ref{sec:TheoryTestsForHDEntanglement}). Nevertheless, single-outcome measurements are ubiquitous across platforms, from spectral filters in frequency bins, single-plane holography for transverse mode projections, non-linear temporal filtering for closely separated time-bins, right through to the first realizations of the quantum pulse gate (Section \ref{sec:PulseModesTemporalModeDetection}). Again however, these methods are limited and in each case movement toward general multi-outcome measurements are in development. These require either (a) $d$ detectors (whether fiber/waveguide-coupled or array pixels) or b) a single detector with $d$ distinct arrival times.

\subsubsection*{(a) $d$ Detectors} Where $d$ detectors are available, realising a desired measurement requires bringing each of the distinct measurement outcomes efficiently onto a unique detector. 
Most commonly, these detectors are fiber-coupled, and the challenge consists of transforming from a given encoding platform to the path degree of freedom with each of the path modes contained in different fibers. 
More recently, detectors coupled to integrated waveguides have required similar coupling strategies to distinct waveguide modes (Sections.~\ref{sec:GenPathSinglePhotonProcessing} and \ref{sec:AppTransverseToWaveguides}). Another promising route to increasing the number of available detectors is through the use of single-photon detector arrays, where again, realising generic measurements requires conversion to the free-space path/pixel modes of the array. 
Some straightforward examples included the use of spectral filters or diffraction gratings to convert frequency-bin to path, with dichroic mirrors easily enabling single-outcome measurements. In addition, off-the-shelf wavelength-division-multiplexers allow converting many closely separated freq-bins (the ITU grid) to separate fiber modes (Section.~\ref{sec:AppFreqbinQKD}and \ref{sec:GenAndMeasFreqBins}). Despite easy access to the canonical basis of frequency-bins for bin separations larger than single-photon spectrometer resolution, generalized measurements of frequency-bin require active control such as electro-optic modulation (Section.~\ref{sec:AppFreqbinQKD} and~\ref{sec:GenAndMeasFreqBins}). Other examples include the transformation of arbitrary transverse-spatial modes to arrays of fibers or array pixels. Here again, whilst single-outcome measurements can couple one spatial mode to a single-mode fiber with a single modulation plane, multi-outcome measurements require multiple passes of the modulation elements to realize general transformations (also known as a multi-plane light converter) from transverse modes to the detectors (Sections.~\ref{sec:GenTransverseParallelDetection} and \ref{sec:AppTransverseMillionsOfPixels}). 
Whilst access to increasing numbers of SPDs is becoming more routine~\cite{Deng2023} and SNSPDs mature into larger arrays~\cite{Venza2025,gao2025pixelscamerascalingsuperconducting} maintaining cryogenic temperatures required becomes more challenging due to the thermal load of the electronic read-out lines, and moving from direct readout towards parallel row-column readout becomes necessary~\cite{Allmaras2020}. 
One would hope that path encoding in integrated platforms overcomes some of these limitations (Section \ref{sec:GenPathSinglePhotonDetectors}) and yet reaching the requirements --- "detectors must evolve into detector arrays\cite{fleming2025high} with millions of pixels, operating at high speed, low noise, and with minimal jitter\cite{venza2025research}" --- remains far away from present day technologies (Section \ref{sec:AppTransverseMillionsOfPixels}).

\subsubsection*{(b) Time-multiplexed single detector} Using a single detector that associates different arrival times to the different measurement outcomes has the distinct advantage that the number of detectors does not need to increase with $d$. However, the temporal resolution or time interval is required to increase with $d$.
Such a measurement directly corresponds to the canonical basis of the time-bin encoding, however realizing generalized measurements in the time-bin platform remains a difficult challenge (See sections \ref{sec:appTimeBin} and \ref{sec:OSTD}). 
Another platform that routinely uses this measurement strategy is frequency-bin in which highly dispersive elements such as long fibers or chirped fibre Bragg gratings act as a time-of-flight spectrometer, using the dispersive Fourier transform principle. Here, different frequency components acquire different delays, thereby converting the frequency-bin basis to time-bins for detection (Sections \ref{timebin-recent} and~\ref{sec:AppFreqFreqToTime}). Again, generalized measurements require active electro-optic modulation, which is a non-trivial extension from this canonical basis measurement (Section~\ref{sec:GenAndMeasFreqBins}). 
The accessible dimension of time-multiplexed detection is limited by the detector jitter, which despite vast improvements, still prohibits very small time-bin resolution being realized (Section~\ref{sec:GenPathSinglePhotonDetectors}). Reaching very small time delays (sub 20ps) such as those realised in integrated optical delay lines requires active switching generally via nonlinear effects such as cross-phase modulation (Sections~\ref{sec:GenAndMeasTimeBinXPMSwitch}) or gating using nonlinear three wave mixing \cite{MacLean:2018}, though these methods are often restricted to single-outcome measurements.
An alternative approach towards improving the time-bin resolution is by means of temporal imaging, where dispersive propagation is combined with quadratic temporal phase modulation (implemented either electro-optically, or by three or four-wave mixing, see section \ref{sec:OSTD}) to magnify the temporal separation of the time bins to a resolvable level \cite{Mittal:2017, Joshi:2022picosecond, Horoshko:2025}.
The use of time-multiplexed detection highlights a key advantage of HD encoding over qubit encoding due to the limitations imposed by detector dead-time in quantum communication applications. By encoding in higher-dimensional qudits, the data rate available per photon, $log(d)$, can lead to clear advantageous for increasing $d$, Sections.\ref{sec:GenTimeDetectorDeadTime} and \ref{sec:AppTimeHDQKDLargerRates}.

\subsubsection{General Unitary Operations for Measurement}
An overarching requirement for general measurements in all platforms entails going beyond the canonical basis measurements, which may be readily available for the likes of path encoding, well-spaced time-bins, and frequency-bins. To realize true universal measurements, one needs both a canonical basis measurement as well as a unitary transformation to the detection basis, for which each platform encounters its own challenges. As one example, the multi-outcome quantum pulse gate is able to transform from an entire chosen set of pulse modes to the set of output spectral bins before being subsequently de-multiplexed to individual detectors. Significant work is being carried out in a similar direction for other photonic degrees-of-freedom, as discussed in detail in the application sections above. Given the clear propensity to convert between degrees of freedom for both detection and transportation, it seems likely that a variety of degrees of freedom will need to be simultaneously utilised for full functionality of future quantum networks. 

\subsubsection{Optical Space-Time Duality} \label{sec:OSTD}

The time-frequency and spatial encodings can be conceptually linked owing to optical space-time duality (OSTD) \cite{Kolner:2002}. OSTD relies on mathematical equivalence between the equation governing paraxial spatial diffraction and the equation describing propagation of an optical pulse in a medium with group velocity dispersion, in the slowly-varying envelope approximation. The Fourier-conjugate spatial variables---position and transverse momentum---are analogous to time and frequency (or, alternatively, frequency and time). This creates an analogy between spatial path encodings (sections \ref{Sec:GenAndMeas_Path}, \ref{Sec:App_Path}) and time-bin or frequency-bin encodings (sections \ref{sec:GenAndMeas:Time_Bin}, \ref{sec:appTimeBin}, and \ref{sec:GenAndMeas_Freq_Bin}, \ref{Sec:App_Freq_Bin}, respectively) , as well as an analogy between transverse spatial modes (e.g.\, Hermite-Gaussian modes, see sections \ref{Sec:GenAndMeas_Spatial_Modes}, \ref{Sec:App_Spatial_Modes}) and field-orthogonal temporal modes discussed in section \ref{sec:PulseModes}. The temporal optics techniques originating from OSTD are well established in classical optics \cite{Howe:2006,Torres:2011spacetime, Salem:2013}. During the last decade, their potential for photonic quantum information processing has started to be recognized \cite{Donohue:2016, matsuda2016deterministic, karpinski_bandwidth_2017, Mittal:2017, Fan:2019, Joshi:2022picosecond, zhu2022spectral, Srivastava2023erecting, sosnicki_interface_2023}, with early progress summarized in Karpi\'nski et al.\ \cite{karpinski2021control}.  

Devices and techniques from the spatial domain can find analogies in the time-frequency domain. For example, a spatial lens, viewed in the Fourier optics picture, imparts a transverse quadratic phase on a signal. In analogy, a time-lens can be realized by applying quadratic time-varying phase modulation to a pulse. The time lens, preceded and followed by propagation of the pulse in media with group delay dispersion implements temporal imaging: a fast time-dependent signal can be magnified in time to enable its detection by a slower detector. Another analogy is the dispersive Fourier transform spectrometer, also known as time-of-flight spectrometer, cf.\ section \ref{timebin-recent}, where the frequency of the photon is mapped to time by propagation in a medium with large group delay dispersion \cite{Avenhaus:2009,Davis:2017}. This is fully analogous to detecting the spatial spectrum of a signal by propagating to the far field.

Whereas the mathematical basis of the spatial and time-frequency encodings is fully analogous, there is an important practical difference. In the case of spatial experiments, one can easily detect signals in both conjugate bases: position and transverse momentum, with high resolution. This can be done by going from the near field to the far field using a lens and propagating by one focal length. In the time-frequency case, one typically operates only in the time-resolved regime, where the frequency spectrum cannot be measured (as in the case of ns-duration time bins with unresolvable sub-GHz spectral features) or in the frequency-resolved case (as in the case of frequency bins with spectral spacings of tens of GHz, which exhibit temporal beatings at 10-picosecond timescales, below the resolution of single-photon detectors). As of now, there is no efficient experimental approach to perform the time-frequency analogy of the near-to-far field transition, from a directly detectable time scale to a frequency resolvable regime. Further development in single-photon-compatible techniques to implement high-resolution time lenses, as well as development of highly dispersive, low-loss media is needed to achieve this goal \cite{karpinski2021control}. As of now this disparity between detection regimes leads to the distinct consideration and development of time-bin and frequency-bin encodings, whereas in the spatial case the near and far fields are often treated equivalently.

\subsection{Integration and Material Platforms}
Throughout this roadmap, integration has been identified as a requirement for realising applications of high-dimensional quantum photonics at a larger scale.  The trajectory of conventional general-purpose computers, which went from the room-scale ENIAC (Electronic Numerical Integrator and Computer) of the 1940s to the hand-held devices of the 2000s like the iPad, is a useful example that shows how integration can massively increase impact. ENIAC filled a room because it had very limited integration. In contrast, an iPad has benefited from many advances in CMOS technology which integrated logic, memory, control, and interconnect layers within \emph{one} semiconductor fabrication platform.  The challenge with quantum photonics (not just in high-dimensional quantum photonics) is that the use of different material platforms is unavoidable. Minimising the number of material platforms and interfaces between them will be crucial.  While conventional computers followed simple Moore’s law scaling for decades, the trajectory for quantum photonics is likely to be less straightforward. 

\subsubsection{Hybrid and Heterogeneous Integration}
Integrating hundreds of components in one chip, thereby achieving the generation,  manipulation, and measurement of qudits in one chip has been demonstrated most notably using path-encoding  \cite{Wang2018, Li2025},  in a monolithic silicon-on-insulator platform. However, the dream of realising a quantum photonic system-on-chip (SoC)  is unlikely to be monolithic because different material characteristics are needed for the different  functionalities of a quantum SoC: (1) deterministic or multiplexed on-chip sources, (2) large  interferometric networks, (3) on-chip filtering and routing,  (4) on-chip detectors, and (5) integrated classical control. Barring the discovery of a platform that is ideal for these different functionalities, either hybrid or heterogenous integration will be the future. 

Hybrid integration, where separately fabricated components are combined, is an important first step towards a quantum SoC. For example, recent work \cite{Kues2017} has coupled an InP  gain medium with Si$_3$N$_4$ platform to demonstrate for the first time a fully-on-chip frequency-entangled qudit source. Aside from needing different materials for different functionalities, it is also apparent that different photonic degrees of freedom are more suitable for some material platforms than others. While minimising propagation loss is important for all, the ease of generation and measurement of quantum states depends heavily on the photonic qudit used \cite{Kaur2025}. Time-bin encoding benefits from the fast electro-optic control in TFLN, but is also suitable for silicon-based platforms for applications where fast phase control is not critical. Frequency-encoding requires the generation of frequency combs, spectral manipulation via resonators, and dispersion control, also favouring TFLN and silicon-based platforms. Integration in path-encoding is almost exclusively done on silicon-based platforms, which afford a high density of waveguides. Transverse-spatial-mode-encoding is similar in its requirements to path-encoding, but in addition, is more sensitive to sidewall roughness and scattering, hence favouring platforms like silica and glass. Hybrid integration is a nearer-term solution to producing quantum SoCs because the technologies for fabricating each material independently is well-developed and each component can be optimised separately. Techniques for butt- and edge-coupling, and flip-chip and wire bonding are also mature. The flexibility that comes from a modular approach lowers fabrication risks, although the footprint that results is likely larger than that achieved with heterogenous integration.

Heterogenous integration is an approach that combines different materials at a wafer (or material) level, leading to a single platform from which one can carve out the quantum SoC.  Because the materials are bonded together, there is no optical or mechanical alignment necessary. Coupling losses and footprint will also be reduced. However, there is much process engineering necessary for this approach because combining different materials in one platform often leads to mismatch in material properties (e.g.\ thermal). These need to be managed to maintain reproducibility.  Another complication, at least while SNSPDs are the most efficient detectors for single photons, is the cryogenic temperature at which a heterogenous chip would need to operate. Heterogenous integration will become more important as the conversion between different photonic qudits becomes necessary.

\subsubsection{Performance scaling}
Performance scaling in high-dimensional quantum photonics is more complicated, in comparison for example, to conventional classical computers that followed Moore's law and  Dennard's law for decades. The simplicity of Moore's law (that the  number of transistors on a microchip doubles roughly every two years), and Dennard's law (as transistors get smaller, their power density stays constant) rests on the capacity to engineer one monolithic platform---silicon. On-chip quantum photonics is unlikely to be on a monolithic platform, hence the performance scaling will be different. In conventional computers, the reduction in transistor size has largely driven the increase in performance. In stark contrast,  it is impossible to name just one characteristic that would define the scaling in quantum photonics. The density-driven progress in classical computing is unlikely to be seen in quantum photonics because loss and coherence play such important roles. Scaling will be governed not just by simply shrinking components, but by  
coordinated improvements in material platforms to achieve low loss while maintaining coherence.  At least the number of coherently controlled modes has been increasing significantly, from the few-mode devices of the early 2010s to the large-scale devices that commercial roadmaps point to \cite{Alexander2025}.

\subsubsection{High-dimensional quantum photonics: epilogue}
It is a fascinating fact that life on earth is encoded in an alphabet consisting of four letters---the four DNA bases, G, A, T, C.  While we will never know how Nature stumbled upon this four-letter alphabet, it is most likely that $d{=}4$ is the result of a compromise between expressivity, robustness to error, and biochemical complexity.  In a similar vein, although
photonic quantum information is physically very different from biological information, it is clear from this roadmap that achieving the full potential of high-dimensional photonic quantum information for future technologies will take a  holistic, systems-approach. As demonstrated by the many works highlighted in this roadmap, progress towards a future where photonic qudits figure more significantly has accelerated in recent years. This momentum will continue only through a concerted effort and (where necessary) technological integration with other physical quantum systems, especially because photons underpin the interconnection and measurement of nearly all quantum architectures.

\begin{backmatter}
\bmsection{Acknowledgments}
J.R. thanks the Australian Research Council Training Centre For Current and Emergent Quantum Technologies (QuTech, IC240100012) for funding support and the Qudits@UQ team for reading an early draft of this roadmap article.
W.M. and N.H.V. acknowledge support from the UK Engineering and Physical Sciences Research Council (EPSRC) (EP/W003252/1) and European Research Council (ERC) Starting Grant PIQUaNT (950402). SPW was supported by by Chilean Fondo Nacional de Desarrollo Científico y Tecnológico (FONDECYT) Grant No. 1240746, ANID -- Millennium Science Initiative Program -- ICN17$_-$012, and ANID Anillo Project ATE250003.
\end{backmatter}

\bibliography{references}
\end{document}